\documentclass[aps,pre,twocolumn,superscriptaddress,a4]{revtex4}
\usepackage{times}
\usepackage[amssymb,thickqspace,cdot]{SIunits}

\usepackage{amssymb,amsmath} 
\usepackage{color}
\usepackage{graphicx}

\usepackage[colorlinks=true,linkcolor=blue,linktocpage=true,citecolor=blue,urlcolor=blue,bookmarksopen=true,bookmarksopenlevel=1,bookmarksnumbered=true]{hyperref}

\usepackage{scrpage}
\usepackage{float}
\usepackage{dsfont}
\usepackage{bbm}

\newcommand{\myref}[1]{(\ref{#1})}

\begin{document}

\preprint{}

\title{Density Reduction and Diffusion in Driven 2d-Colloidal Systems Through Microchannels}

\author{P.\ Henseler}
\email[]{Peter.Henseler@uni-konstanz.de}
\affiliation{Universit\"at Konstanz, Fachbereich f\"ur Physik, 78457 Konstanz, Germany}
\author{A.\ Erbe}
\email[]{Artur.Erbe@uni-konstanz.de}
\affiliation{Universit\"at Konstanz, Fachbereich f\"ur Physik, 78457 Konstanz, Germany}
\affiliation{Universit\"at Konstanz, Zukunftskolleg, 78457 Konstanz, Germany}
\author{M.\ K\"{o}ppl}
\affiliation{Universit\"at Konstanz, Fachbereich f\"ur Physik, 78457 Konstanz, Germany}
\author{P.\ Leiderer}
\affiliation{Universit\"at Konstanz, Fachbereich f\"ur Physik, 78457 Konstanz, Germany}
\author{P.\ Nielaba}
\affiliation{Universit\"at Konstanz, Fachbereich f\"ur Physik, 78457 Konstanz, Germany}
%\thanks{}

\date{\today}

\begin{abstract}
The behavior of particles driven through a narrow constriction is investigated
in experiment and simulation. The system of particles adapts to the confining
potentials and the interaction energies by a self-consistent arrangement of the
particles. It results in the formation of layers throughout the channel and of
a density gradient along the channel. The particles accommodate to the density
gradient by reducing the number of layers one by one when it is energetically
favorable. The position of the layer reduction zone fluctuates with time while the
particles continuously pass this zone. The flow behavior of the
particles is studied in detail. The velocities of the particles and their
diffusion behavior reflect the influence of the self-organized order of the
system. 

%The position and motion of defects
%is determined by the position and motion of the layer reduction zone. 
\end{abstract}
%68.60.Bs, 46.70.Hg, 95.10.Fh}

\maketitle

%%% Introduction
%\input{LTpaper-Intro-v7}
\section{Introduction}

Pedestrians in a pedestrian zone~\cite{Helbing01}, ants following a trail to
food places and many other systems of interacting entities, which are moving in
opposite directions to each other, show a prominent feature, namely the
formation of lanes along the direction of their motion. This formation of lanes
has been studied theoretically for colloidal particles in 3 dimensions
\cite{Rex05, Rex07, Rex07b} as well as in 2 dimensional systems
\cite{Chakrabarti03, Chakrabarti04, Dzubiella02}. These examples indicate that
flow of particles can have a substantial influence on the structure formation
of a system of interacting particles. Experimental studies on such systems have
not been performed up to date, first hints of a lane formation transition could
be seen in a 3-dimensional system of oppositely charged colloids driven in
opposite directions by application of an external electric
field~\cite{Leunissen05}. Studies of people in panic (for example trying to
escape from a building) show the influence of constrictions on such moving
ensembles. 

A system of 2-dimensionally confined moving colloidal particles also resembles
the classical analogon of a quantum point contact in mesoscopic
electronics~\cite{Wees88,Wharam88} or in metallic single atom
contacts~\cite{Scheer98,Dreher05,Pauly06}. These contacts exhibit transport in
electronic channels due to quantization effects.  Such quantum channels can be
seen as similar to the  layers in the macroscopic transport, since both occur
due to the interaction of the particles with the confining potential. A
classical version of a similar scenario can be built on a liquid helium
surface, which is loaded with charges. For such a system the formation of
layers has been reported as well~\cite{Glasson01}. The change of the number of
such  layers in the vicinity of a constriction has been predicted from Langevin
dynamics simulations of Yukawa particles~\cite{Piacente05}. 

In biological systems the transport of interacting particles through narrow
constrictions is of high importance for many processes, for example for the
size selectivity of transport in ion channels~\cite{Roth05}. The complexity of
such systems allows only to make simplified statements on the underlying
physics governing such phenomena. Experimentally easily accessible model
systems can reveal many of the underlying processes. In the context of
micro-fluidics and ``lab-on-a-chip'' devices one is interested in
non-equilibrium transport and mixing phenomena on the microscopic scale
\cite{Squires05}.

In this paper we present a 2-dimensional system of moving,
superparamagnetic particles. The interaction energies between the
particles and therefore the effective temperature of the system can be set by
application of an external magnetic field. The phase behavior of these
particles in 2 dimensions has been studied
extensively~\cite{Zahn99,Zahn03,Keim04,Eisenmann04}. In addition to
this, it has been shown that confinement of these particles in a
narrow channel leads to the formation of layers, in order to conform
to the boundaries set by the hard
walls~\cite{Haghgooie04,Haghgooie06}. The number and the stability of
these layers change as the density or the interparticle interactions
are varied. In this work we address the question how these layers
change when the particles are subject to a driven motion along the
channel. In order to investigate this moving state we first study the
properties of a static system using Brownian dynamics
simulations. Based on these results, the moving system is
characterized, and the results are compared to an experimental system
of superparamagnetic particles moving through a lithographically
defined channel.  

%%% Experimental Setup
%\input{LTpaper-Experiment-v7}
\section{Experimental Setup}

\begin{figure}[htb]
  \begin{center}
    \includegraphics[width=\columnwidth,clip]{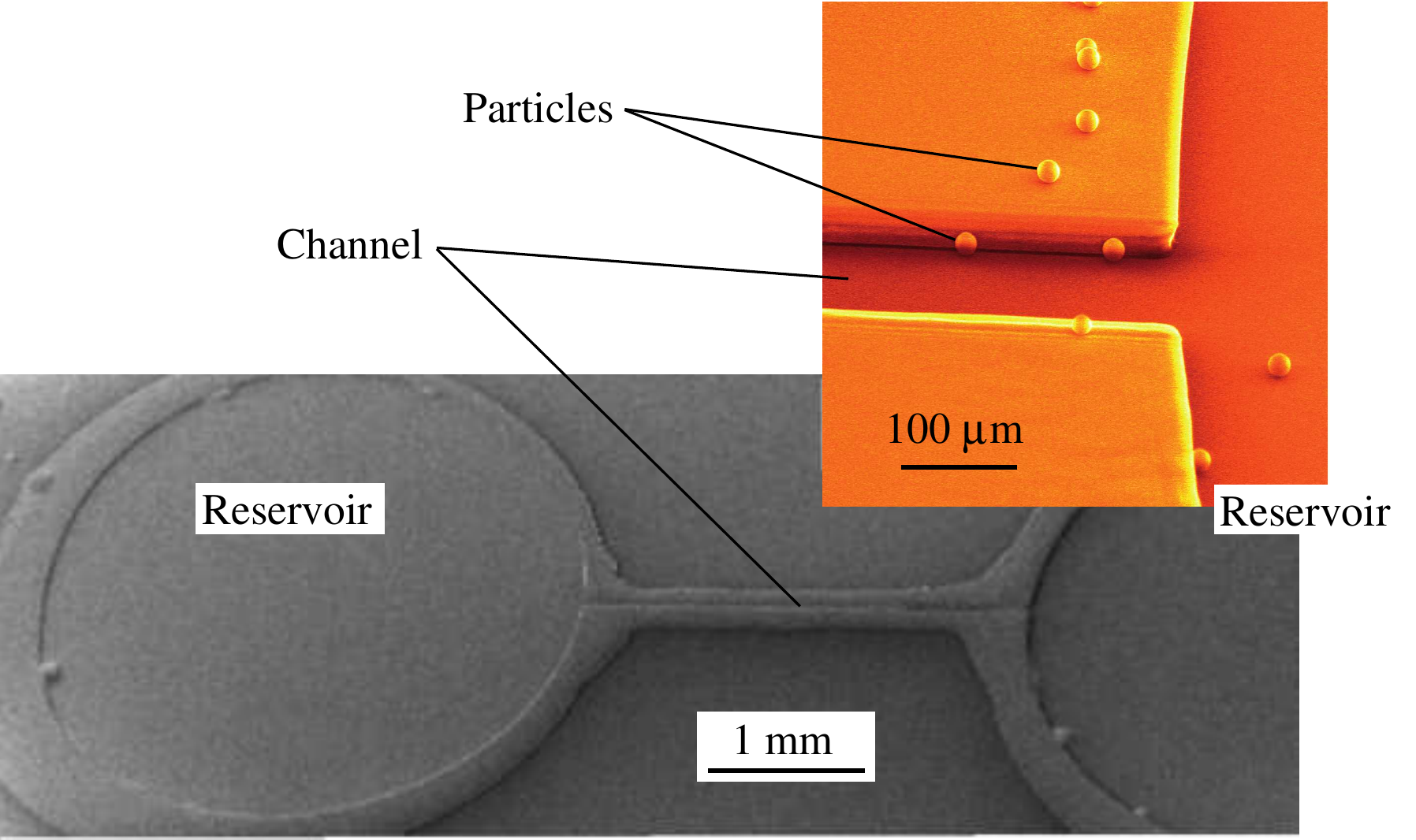}
  \end{center}
  \caption{SEM images of the full channel
  geometry connecting two reservoirs and an enlargement of the channel
  entrance/ exit region. Also some dried colloidal particles can be seen.
  During measurement the particles outside of the channel are removed so that
  they don't influence the particle transport within the
  channel.~\cite{Henseler06}}
  \label{fig:ChannelImagesExp}
\end{figure}

The particles are constricted to a narrow
channel connecting two reservoirs, which are defined on a substrate
using UV-lithography~\cite{Xia98}.  Images produced with a scanning electron
microscope (SEM) of such a channel setup are shown in
figure~\ref{fig:ChannelImagesExp}.
Channel geometries of various width and length have been produced. The
simulation results are compared to a channel being $\unit{60}{\micro\meter}$
wide, $\unit{2.7}{\milli\meter}$ long and having channel walls of about
$\unit{5}{\micro\meter}$ in height. The channel is filled with a suspension of
superparamagnetic particles of diameter $\unit{4.5}{\micro\meter}$ in water
(Dynabeads). Identical particles have been used previously and characterized
in~\cite{Bubeck}. A summary of the properties of these colloidal particles is
given in table~\ref{tab:Colloids}.
\begin{table}
  \centering
  \begin{tabular}{lll} \hline\hline
  	diameter & $\sigma$ & $\unit{4.55 \pm 0.1}{\micro\meter}$ \\
	mass density & $\rho_{\mathrm{colloid}}$ &
	$\unit{1.6}{\gram/\centi\meter^3}$ \\
	particle mass & $m$ & $\unit{(7.6\pm0.1)\cdot10^{-14}}{\kilo\gram}$ \\
	saturation magnetization & $M_0$ & $\unit{(5.7\pm0.4)\cdot
	10^{-13}}{\ampere \meter^2}$ \\
	effective susceptibility & $\chi_{\mathrm{eff}}$ &
	$\unit{7.5\cdot10^{-11}}{\ampere\meter^2/\tesla}$ 
%	at $T=\unit{295}{\kelvin}$ 
	\\ \hline
  \end{tabular}
  \caption{Particle properties of the Dynabeads used in the
  experiment.}
  \label{tab:Colloids}
\end{table}

Gravity confines the particles to the bottom surface of the channel due to the
density mismatch between the colloids and the surrounding water. An external
uniform magnetic field $\mathbf{B} = B\,\mathbf{\hat{z}}$ is applied
perpendicular to the bottom surface. As a consequence the colloids form a
monolayer in the $x$-$y$ plane with induced parallel dipole moments in
$z$-direction giving rise to a purely repulsive pairwise particle interaction.
The strength of the repulsive force at distance $r_{ij} = |\mathbf{r}_i -
\mathbf{r}_j|$ is given by 
\begin{equation}
	V_{ij}(r_{ij}) = (\mu_0/ 4\pi) {M^2}/{r_{ij}^3}
	\label{eq:DipolarPotentialExp}
\end{equation}
with the magnetic dipole moments $M = \chi_{\mathrm{eff}} B$ of the particles.
The importance of the pair-interaction can be characterized by the
dimensionless interaction strength 
\begin{equation}
  \Gamma = \mu_0 M^2 (\pi n)^{3/2}/({4\pi k_B T}),
  \label{eq:InteractionStrength}
\end{equation}
where $n$ denotes the (overall) particle number density, $k_B$ the Boltzmann
constant, $\mu_0 = \unit{1.257\cdot10^{-6}}{\volt\second/\ampere\meter}$ is the
magnetic permeability of free space and $T$ the temperature. For an unbounded
equilibrated 2D-system which forms a triangular lattice, the particle number
density can be written in terms of the lattice constant $\tilde{a}$ as
\begin{equation}
  n = \frac{2}{\sqrt{3}}\,\frac{1}{\tilde{a}^2}.
  \label{eq:NumberDensityTriangularLatt}
\end{equation}
So, $\Gamma = \left( \frac{2\pi}{\sqrt{3}} \right)^{3/\,2}
V_{ij}(\tilde{a})/\,(k_B T)$ is the mean dipolar interaction energy of
equation~\myref{eq:DipolarPotentialExp} in terms of the thermal energy.
Accordingly, the applied magnetic field $B$ which is connected to the
magnetization via $M=\chi_{\mathrm{eff}}B$ plays the role of an effective
inverse temperature.

The external magnetic field is the dominant magnetic field in this system as it
is obvious from the large particle separations in the video microscopy snapshot
of
Fig.~\hyperref[fig:channelsnapexperiment]{\ref{fig:channelsnapexperiment}(a)}
and the mutual induction between the colloids is negligible. Thermal and
magnetically induced fluctuations of the positions of the particles
perpendicular to the plane of inclination are less than 10\% of the particle
diameter and can be neglected.  Tilting of the whole channel setup induces
transport of the colloids from one reservoir into the other due to gravity. An
alternative driving method would be the application of an in-plane magnetic
field gradient.

Before starting experiments the system is set up exactly horizontal. The
particles are allowed to sediment to the bottom surface and arrange in the
equilibrium configuration within several hours. Before tilting the whole
apparatus the particles are either all confined in one reservoir~(by
use of laser tweezers) or uniformly
distributed along the channel and within both reservoirs. In the experiment an
inclination of $\alpha_{\mathrm{exp}} =0.6\degree$ is chosen, where the system
is in a gravitationally driven non-equilibrium situation, but not yet in the
regime of plug flow. This inclination results in an average particle drift
velocity $v_{\mathrm{drift}} \approx \unit{0.035}{\micro\meter/ \second}$. A
typical snapshot from the experiment of the particles moving along the channel
is given in figure~\ref{fig:channelsnapexperiment}(a). 

The particle trajectories are tracked with a video microscope. The repetition rate
of the video microscope setup is $\unit{10}{\second}$.  All experiments are
made at room temperature $T\approx\unit{295}{\kelvin}$.

In the experiment the number density of the colloids is defined as the number
of colloids divided by the area of the 2D channel within the field of view of
the video microscope accessible to the centers of the colloids.  This
dimensionless parameter $\Gamma$ was introduced by Zahn {\it et
al.}~\cite{Zahn99}, who studied experimentally the so-called KTHNY phase
transition in an unbounded two-dimensional equilibrium
system of superparamagnetic particles, to characterize the system state. They
found that for $\Gamma<\Gamma_i\equiv52.9$ the system behaves like a fluid, and
for $\Gamma>\Gamma_m\equiv60.9$ the system forms a triangular lattice. For the
$\Gamma$- values in between they observed the so-called KTHNY or hexatic phase.

In the experiments described below a magnetic field of strength
$B=\unit{0.24}{\milli\tesla}$ is applied, corresponding to $\Gamma \approx 72$
which is in the solid state region of the phase diagram.

%%% Simulation Details
%\input{LTpaper-Simulation-v8}
\section{Simulation Details}
We conduct Brownian dynamics (BD) simulations of a two-dimensional microchannel
setup in order to investigate the flow behavior of the colloidal particles
within the channel systematically for various parameter values of inclination,
overall particle density, and channel width. The equation of motion for an
individual colloidal particle is given by an overdamped Langevin equation.
This approach neglects hydrodynamic interactions as well as the short-time
momentum relaxation of the particles. Both approximations are fully justified
in the current experimental context. Typical momentum relaxation times are on
the order of \unit{100}{\micro \second} and therefore much shorter than the
repetition rate of the video microscopy setup~(\unit{10}{\second}) used in the
experiment. Thus the colloidal trajectories $\mathbf{r}_i(t) = (x_i(t),
y_i(t))$ ($i=1,\ldots,N)$ are approximated by the stochastic position Langevin
equations with the Stokes friction constant $\xi$
\begin{equation}
\xi \frac{d\mathbf{r}_i(t)}{dt} = -
\nabla_{\mathbf{r}_i} \sum_{i\neq j} V_{ij}(r_{ij}) + \mathbf{F}_{i}^
{\mathrm{ext}} + \mathbf{\tilde{F}}_{i}(t).
\label{Eq:PositionLangevin}
\end{equation}
The right hand side includes the sum of all forces acting on each particle,
namely the particle interaction, the constant driving force along the channel
$\mathbf{F}_{i}^ {\mathrm{ext}} = mg\sin(\alpha) \mathbf{\hat{x}}$ and the
random forces $\mathbf{\tilde{F}}_{i}(t)$. The latter describe the collisions
of the solvent molecules with the $i$th colloidal particle and in the
simulation are given by a Wiener process, i.e. by random numbers with zero
mean, $\langle \mathbf{\tilde{F}}_{i}(t) \rangle = 0$, and variance $\langle
\tilde{F}_{i \alpha}(t) \tilde{F}_{i \beta}(0) \rangle = 2k_B T \xi \delta(t)
\delta_{ij} \delta_{\alpha \beta}.$ The subscripts $\alpha$ and $\beta$ denote
the Cartesian components. The effective mass $m$ of the particles is
determined by the density mismatch between the particles and the
solvent.  These position Langevin equations are integrated
forward in time in a Brownian dynamics simulation using a finite time step
$\Delta t$ and the technique of Ermak~\cite{Ermak75, Allen}. 

Particles are confined to the channel by hard walls in $y$-direction and at
$x=0$ (channel entrance). These walls are realized both as ideal elastic hard
walls and as proposed in~\cite{Heyes93}, where a particle crossing the wall is
moved back along the line perpendicular to the wall until contact. Both
realisations result in the same flow behavior. Also we performed simulations
with the particles at the wall kept fixed. The channel end is realized as an
open boundary. To keep the overall number density in the channel fixed, every
time a particle leaves the end of the channel a new particle is inserted at a
random position (avoiding particle overlaps) within the first 10\% of the
channel, acting as a reservoir. A cutoff of $10\sigma$ was used along with a
Verlet next neighbor list~\cite{Allen}. Checks of particle overlaps
are included in the simulation, but for all ordered systems we never found two
overlapping particles.

Starting from a random particle distribution within the channel, we first
calculate an equilibrium configuration ($\mathbf{F}_{i}^ {\mathrm{ext}} =0$) of
a closed channel with ideal hard walls. Afterwards we apply to the
configuration of uniform density the external driving force and allow the
system to reorganize for $10^6$ time steps, before we evaluate the
configurations.  The time step $\Delta t = 7.5\cdot 10^{-5} \tau_B$ is used, with
$\tau_B = \xi \sigma^2/ k_B T$ being the time necessary for a single, free
particle in equilibrium to diffuse its own diameter $\sigma$.  We choose $\xi =
3\pi \eta \sigma$, with $\eta$ denoting the shear viscosity of the water. The
simulations are done with $2000-4500$ particles, for a channel geometry of $L_x
= 800\sigma$ and $L_y = (9-12)\sigma$, and $\chi_{\mathrm{eff, sim}} =
\unit{3\cdot10^{-11}}{\ampere\meter^2/ \tesla}$. Thus external magnetic fields
$B=\unit{0.1 - 1.0}{\milli\tesla}$ and a total particle density of
$n=0.4\sigma^{-2}$ correspond to $\Gamma \approx 21.34 - 2134$. 

%%% Equilibrium Properties of the Channels
%\input{LTpaper-Equilibrium-v8}
\section{Equilibrium Properties of the Channels}
\label{sec:EQPropertiesMicrochannel}

Equilibrated configurations of systems confined to a microchannel are used as
starting configurations for our analysis of the transport behavior. This
guarantees that at the beginning of the transport simulation the particles are
uniformly distributed over the whole channel. First, we compare some
results found for the 2D microchannels in equilibrium (the external driving
force is switched off) with the results of Haghgooie and
coworkers~\cite{Haghgooie04, Haghgooie05, Haghgooie06}.

During the equilibration process the channel beginning at $x=0$ and the channel
end at $x=L_x$ are either closed by ideal hard walls, or periodic boundary
conditions are applied in $x$-direction. By doing so, we assure that no
transport is initiated due to the boundary conditions used. The simulation
start parameters are chosen in such a way that they closely reflect the
situation of the experiment. In all simulations the area $L_x\cdot L_y$ is
defined as the region accessible to the particle centers.  This is the reason,
why in the following simulation snapshots the $y$-positions of the edge
particle centers coincide with the channel boundary. When comparing the channel
widths in the simulation to the widths of the channel in the experiment, one
has to add the particle diameter $\sigma$ resulting in $L_y^\mathrm{exp} = L_y
+ \sigma$, e.g.  a channel with $L_y = 10\sigma$ corresponds to a channel of
$L_y^\mathrm{exp} = 11\sigma = \unit{49.5}{\micro\meter}$ for the particles
used. The equilibration process is usually started from a uniform random
particle distribution over the whole channel. But to avoid a physical
instability of the starting configuration the particle separations are limited
to values greater than $0.7\sigma$. For very dense systems this initialization
method of course breaks down and we start from a hexagonally ordered
configuration. 

\subsection{Influence of the Confinement}

\begin{figure}[ht!b]
  \begin{center}
    \includegraphics[width=\columnwidth,clip]{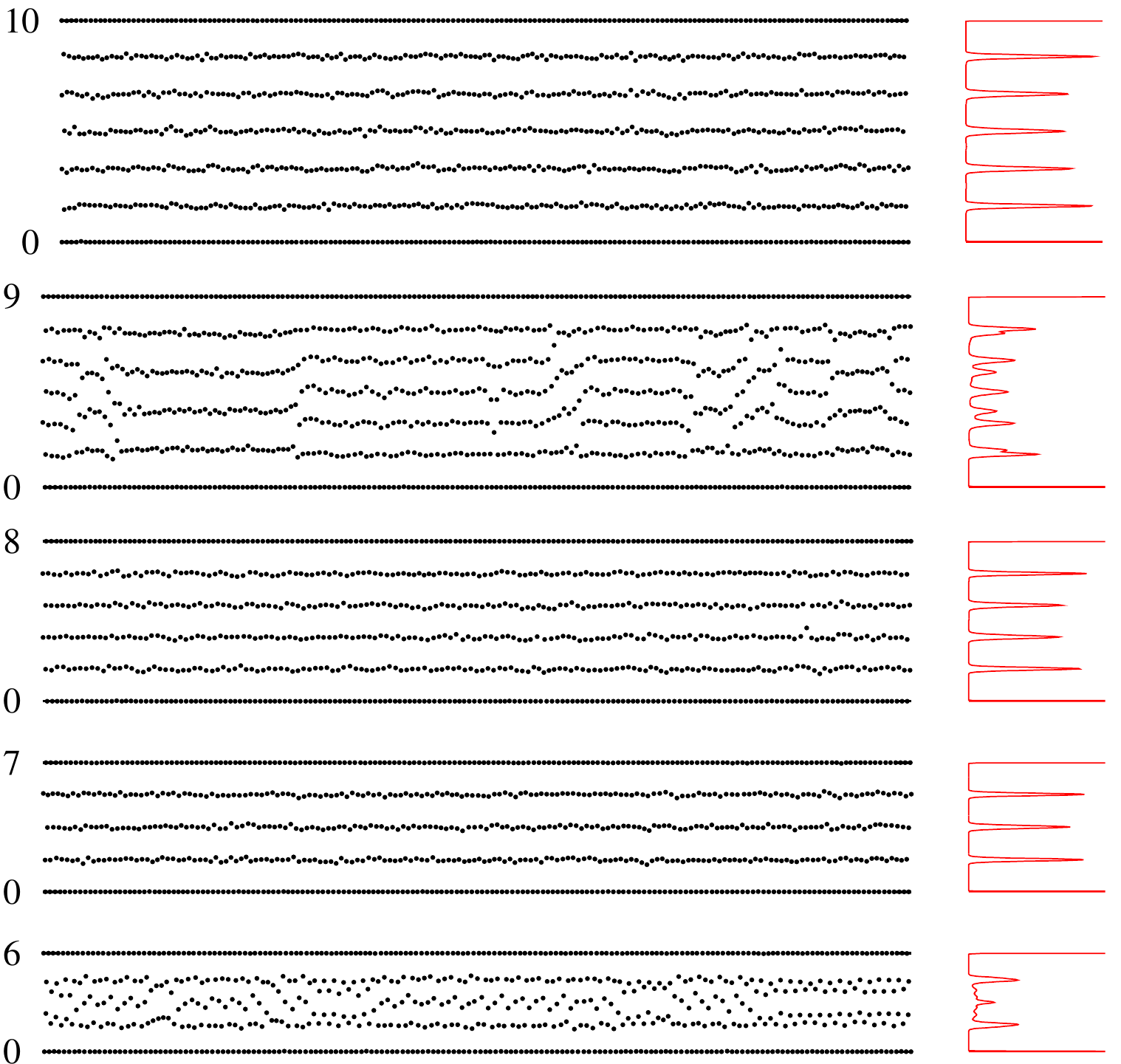}
  \end{center}
  \caption[Typical snapshots of equilibrated configurations and their averaged
  density profiles across the channel.]{Typical simulation snapshots of partitions with
  length $300\sigma$ of equilibrated configurations for a dipolar system
  ($B=\unit{0.5}{\milli\tesla}$, $\Gamma=533.74$) and a selection of channel
  widths ($10\sigma$, $9\sigma$, $8\sigma$, $7\sigma$, and $6\sigma$ from top
  to bottom). The channel widths are stretched by a factor of about $6.67$.
  All configurations have the overall particle density $n=0.4\sigma^{-2}$.  The
  red curves at the right of each configuration snapshot show averaged density
  profiles across the channel. For clarity reason, the large magnitude peaks at
  the walls have been truncated at a fixed peak height.}
  \label{fig:EQSnapshotsChannelWidth}
\end{figure}
The triangular lattice is the high density equilibrium configuration of an
unbounded 2D system. Here, we analyze how the confinement modifies the
resulting equilibrium configurations. We submitted simulation runs to determine
the equilibrium configuration in dependence of the channel width $L_y$ for a
superparamagnetic system with $B=\unit{0.5}{\milli\tesla}$ applied and the
global particle density $n=0.4\sigma^{-2}$ which corresponds to $\Gamma=533.74$
and is deep in the solid phase region. Typical snapshots of representative
parts of the equilibrium configurations being obtained are shown in
Fig.~\ref{fig:EQSnapshotsChannelWidth}. Shown are the regions
\mbox{$300\sigma\le x < 600\sigma$} of a channel with a total length of
$L_x=800\sigma$. Notice, that the channel widths are stretched by a factor of
about $6.67$. 

Obviously, whether an ordered or a perturbed configuration is formed strongly
depends on the channel width $L_y$.  For certain channel widths it is
energetically favorable for the system to arrange into what we call {\it
layers}. Right of configuration snapshots of
Fig.~\ref{fig:EQSnapshotsChannelWidth} the equilibrium density profiles
transverse to the channel walls are plotted. They are calculated by taking the
average over 2000 equilibrium configurations. For the channel widths
$L_y=7\sigma,\,8\sigma$, and~$10\sigma$ the peaks of these density histograms
are well separated and occur at almost regular spacing across the channel.
These properties are the signature of a well defined layered structure parallel
to the walls. For the channel widths $L_y = 6\sigma$ and $L_y=9\sigma$ the
system cannot equilibrate into such a single layered structure over the full
channel and only partial layering is visible in the configuration snapshots.
Such a confinement induced layering phenomenon is in agreement with the results
for liquid-dusty plasmas~\cite{Teng03} and the results of the simulations of
Haghgooie~\cite{Haghgooie04}.

The channel widths of $10\sigma$, $9\sigma$, $8\sigma$, $7\sigma$, and
$6\sigma$ correspond to the widths $6.80 R$, $6.12 R$, $5.44 R$, $4.76 R$, and
$4.08 R$ in units of $R=1.471\sigma$, which is the expected separation of
layers for the unbounded system. 
\begin{figure}[htb]
  \begin{center}
    \includegraphics[width=\columnwidth,clip]{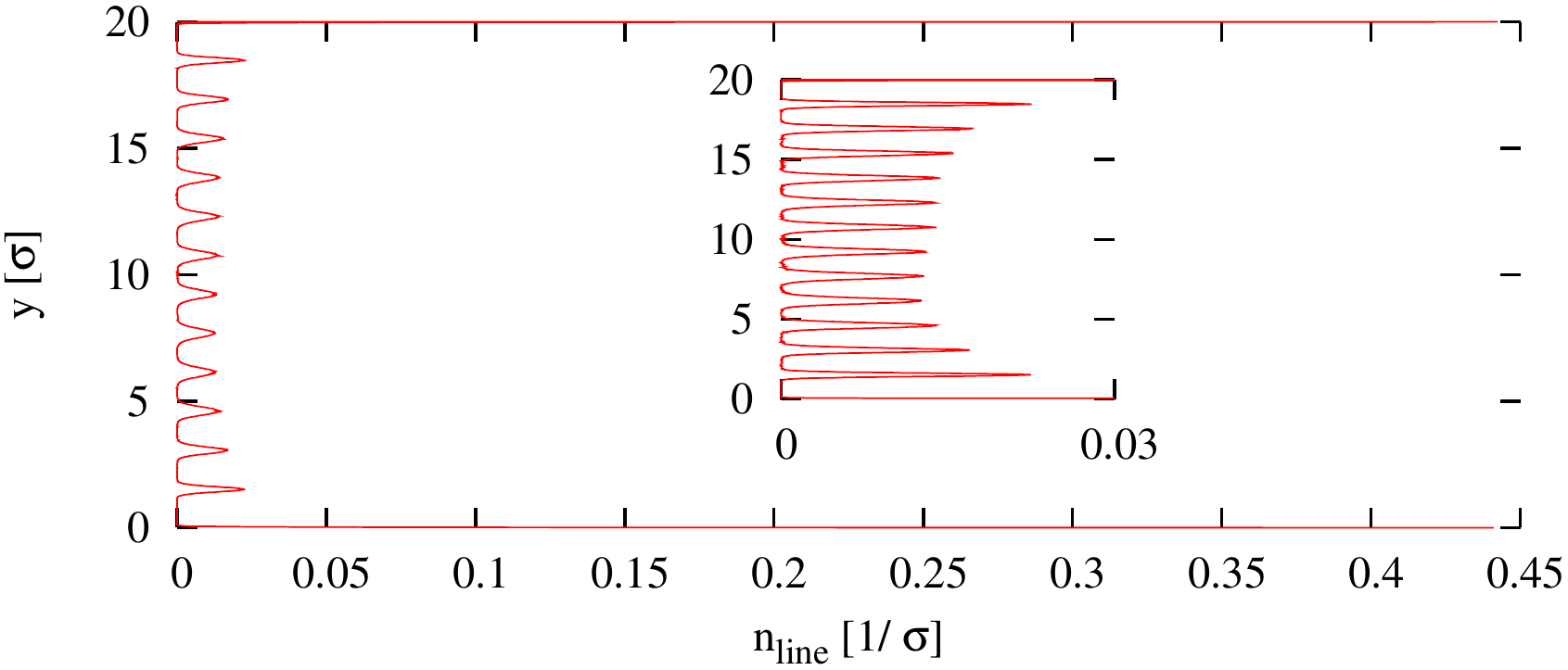}
  \end{center}
  \caption[Full density profile transverse to the walls for $L_y=20$ for the
  solid state.]{Simulation: Full density profile transverse to the confining walls for
  $L_y=20\sigma$ and $\Gamma=133.44$.}
  \label{fig:YHistY20}
\end{figure}
Even for wide channels of width $L_y=20\sigma$ a clearly boundary induced
layered structure occurs for a system at $\Gamma=133.44$. This is shown in
Fig.~\ref{fig:YHistY20}.

\subsection{Layer Order Parameter}

The number of layers forming within the channel can be identified by an
appropriate local order parameter.  We therefore divide the channel of width
$L_y$ into several bins in $x$-direction each containing $n_{\mathrm{bin}}$
particles and define for different number of layers $n_l$ the so-called
{\it layer order parameter}
\begin{equation}
  \Psi_{\mathrm{layer},\,n_l} = \left| \frac{1}{n_{\mathrm{bin}}}
  \sum_{j=1}^{n_{\mathrm{bin}}} e^{\mathtt{i} \,\frac{2\pi (n_l-1)}{L_y}
  \,y_j}\right|,
  \label{eq:LayerOrderParameter}
\end{equation}
which is unity for particles distributed equidistantly in
$n_l$ layers across the channel width starting at $y = 0$, and significantly
smaller for the non-layering case.

\begin{figure}[htb]
  \begin{center}
    \includegraphics[width=\columnwidth,clip]{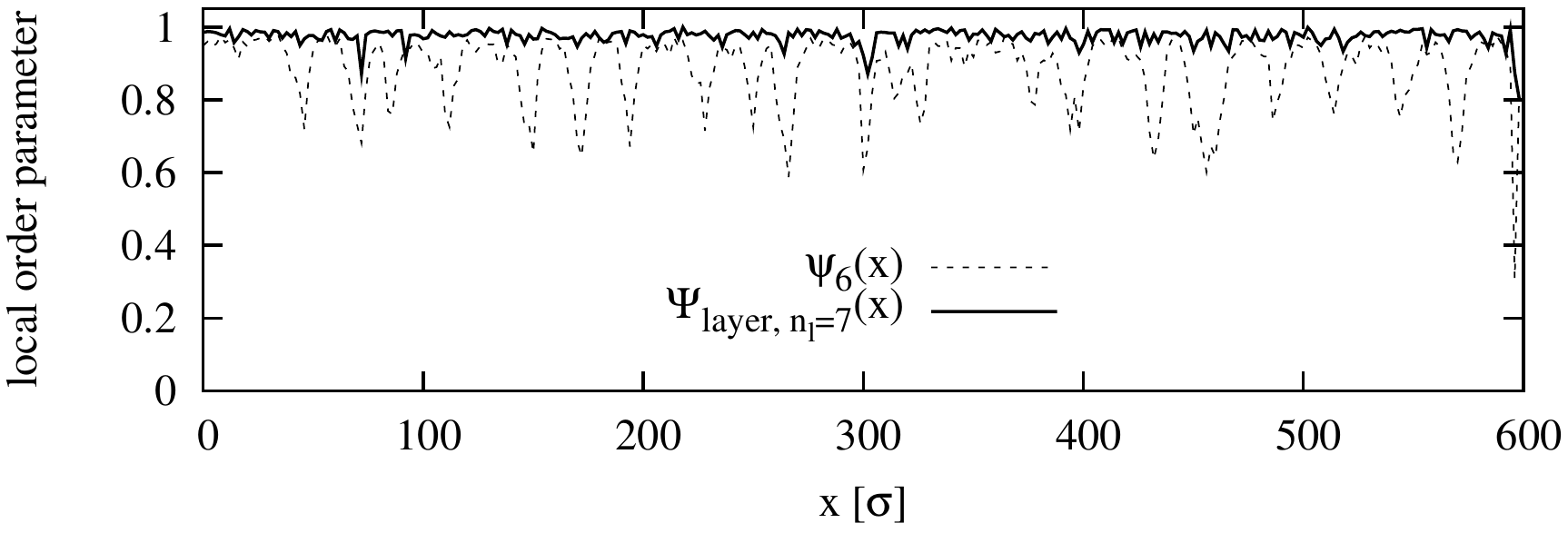}
  \end{center}
  \caption[Comparison of the local layer oder parameter and the local
  orientational order parameter.]{Simulation: Comparison of the local layer oder parameter
  $\Psi_{\mathrm{layer}, n_l=7}(x)$ for 7 layers and the local orientational
  order parameter $\psi_6(x)$ along the channel.}
  \label{fig:VglPsi6LOP}
\end{figure}
An exemplary comparison between the results of the local layer oder parameter
$\Psi_{\mathrm{layer}, n_l=7}(x)$ for 7 layers and the local orientational
order parameter% $\psi_6(x)$
\begin{equation}
	\psi_{6}(x_i) = \left| \frac{1}{N_b} \sum_{j=1}^{N_b}
	\mathrm{e}^{6\mathtt{i} \theta_{ij}} \right| ,
\end{equation}
which returns a measure of the orientational order based upon the distribution
of angles $\theta_{ij}$ (measured with respect to a fixed axis) of the lines
joining a particle $i$ with its surrounding $N_b$ neighbors, is shown in
Fig.~\ref{fig:VglPsi6LOP} for a channel of width $L_y=10\sigma$ and
$\Gamma=533.74$ as depicted in Fig.~\ref{fig:EQSnapshotsChannelWidth}. Both
order parameters have been averaged over 500 equilibrium configurations. The
equilibrium system consists of 7 layers which is indicated by
$\Psi_{\mathrm{layer}, n_l=7}(x)$ having values close to unity. The small
offset results from not fully equidistant peak separation. The distance between
the central layers is slightly greater than the distance between the wall layer
and the layer next to the wall. The local orientational order parameter
$\psi_6(x)$ has values greater than $0.6$, the signature of a nearly triangular
system, but exhibits several dips along the channel length. These are connected
to the occurrence of defects, {\it i.e.} bulk particles having $5$ or $7$
nearest neighbors instead of six and edge particles with $3$ or $5$ nearest
neighbors. The nearest neighbors of each particles are determined by a Delaunay
triangulation.

\begin{figure}[htb]
  \begin{center}
    \includegraphics[width=\columnwidth,clip]{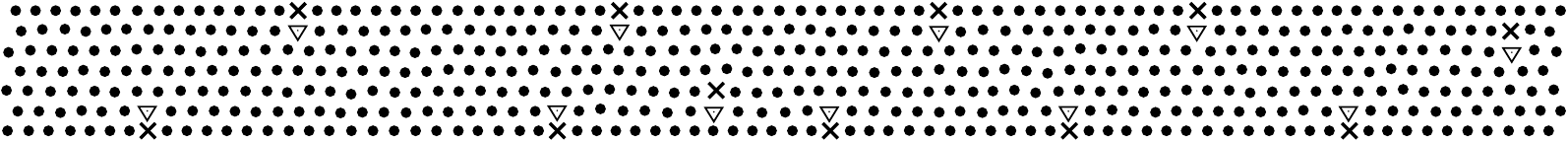}
  \end{center}
  \caption[Snapshot of a partition of an equilibrium defect
  configuration.]{Simulation: Snapshot of a partition ($450\sigma\le x \le580\sigma$) of
  the equilibrium defect configuration for the system with $L_y=10\sigma$ as
  shown in Fig.~\ref{fig:EQSnapshotsChannelWidth}. Full circles ($\bullet$)
  mark the bulk particles with 6 nearest neighbors and particle on the wall
  with 4 nearest neighbors, symbol $\mathbf{\times}$ corresponds to fivefold
  symmetry (or threefold if on the wall), and symbol $\triangledown$ to
  sevenfold symmetry (or fivefold if on the wall).}
  \label{fig:DefectsEQ}
\end{figure}
In Fig.~\ref{fig:DefectsEQ} all defects within a partition of the
equilibrated configuration are marked. For the layered system state the defects
always occur in pairs (forming a dislocation) and are located predominantly
close to the walls with quite a regular spacing. Due to the purely repulsive
nature of the particle pair-interaction the edge particles are pressed against
the confining ideal hard walls as it is obvious from the high peaks of very
small width at the boundary of Fig.~\ref{fig:EQSnapshotsChannelWidth}. These
defects along the walls are a consequence of a (slightly) higher line density
of the edge particles compared to the bulk layers. For example, for the system
with $L_y=10\sigma$ of Fig.~\ref{fig:DefectsEQ} the line density of the wall
layers is about $6$\% higher than of the nearest bulk layers. Edge layers have
only a single neighbor layer whereas bulk layers have two. Putting an
additional particle into a layer results both in stronger interaction within
this layer and of this layer with its neighboring layers. Thus, it is
energetically favorable for the system to have defects along the wall instead
within the bulk, because there the involved energy barrier is lower.

The appearance of dislocations along the wall was also seen
in~\cite{HenselerDiplom, Kong04}, where we systematically analyzed the
equilibrium configurations constricted within a circular hard-wall confinement
for dipolar and screened Coulomb pair interaction as function of the particle
number. In these systems the particles arrange in multiple circles and the
defects occur due to the bending of the lattice in presence of the curved
boundary. This is in contrast to the situation here, where the planar walls
give no need for the lattice to bend.

So, we can conclude that the layer order parameter is  more suitable than
$\psi_6(x)$ for the detection of layered structures and changes therein,
because it is insensitive to defects close to the wall.

\subsection{"Phase Diagram" of the Laterally Confined Dipolar System}

Two independent simulation parameters have a strong influence on the state of the
dipolar system laterally confined between two parallel ideal hard walls. These
are the wall separation $L_y$ and the dimensionless interaction strength
$\Gamma$. In the following we will compare these dependencies for our
simulation parameters qualitatively with the results of
Haghgooie~\cite{Haghgooie05}.

\subsection*{System State Dependency on the Channel Width}

\begin{figure}[htb]
  \begin{center}
    \includegraphics[width=\columnwidth,clip]{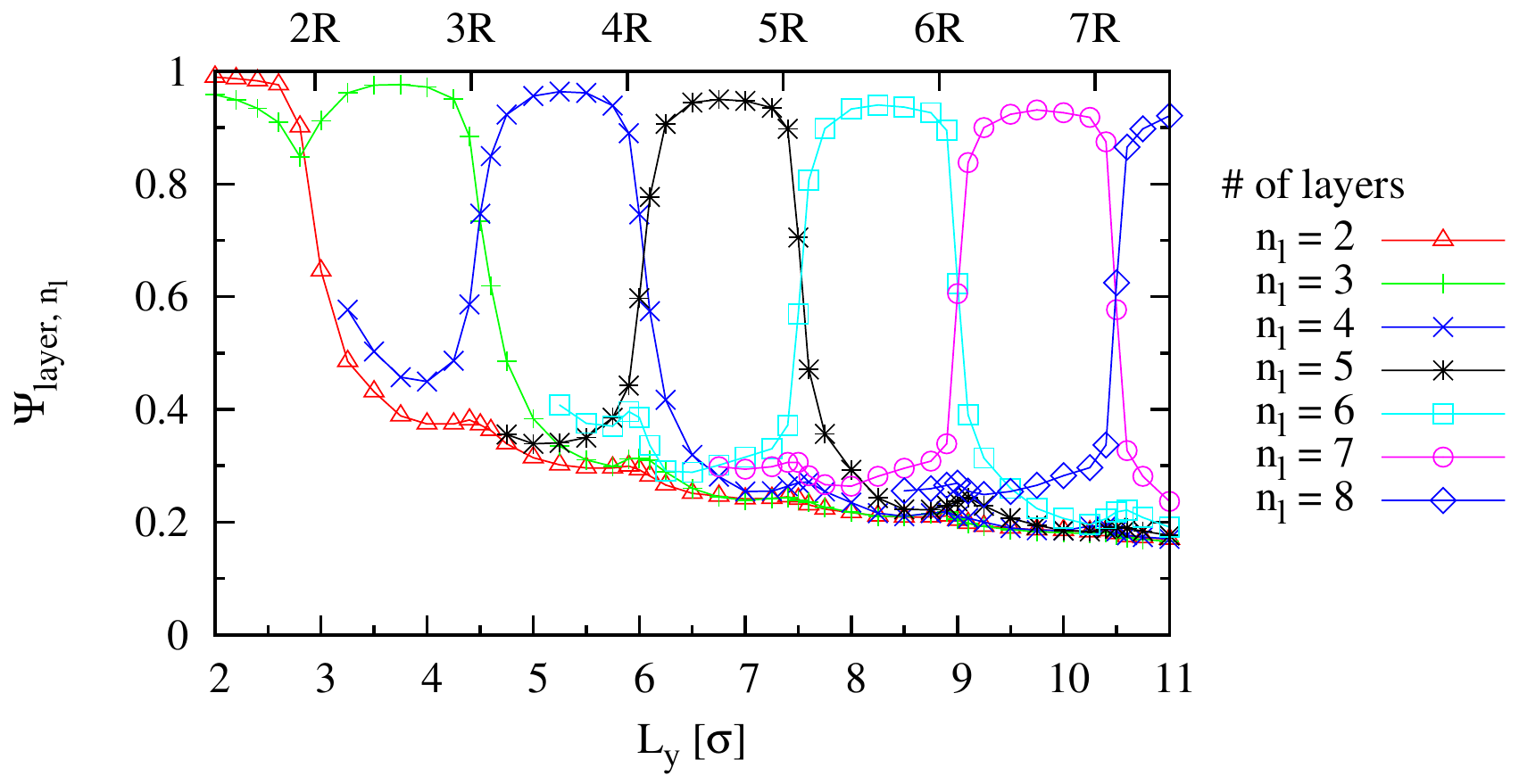}
  \end{center}
  \caption[The layer order parameter as function of the channel
  width.]{Simulation: The layer order parameter as function of the channel
  width. The simulation parameters are: $B=\unit{0.25}{\milli\tesla}$,
  $\Gamma=133.44$, $R=1.471\sigma$, $L_x=800\sigma$, and periodic boundaries in
  $x$-direction.}
  \label{fig:StateDiagramNLayerLy}
\end{figure}
The influence of the channel width on the system state is analyzed by examining
the behavior of the global layer order parameters
$\Psi_{\mathrm{layer}, \,n_l}$. The result is shown in
Fig.~\ref{fig:StateDiagramNLayerLy} for channel widths between $2\sigma$ and
$10\sigma$. The global layer order parameters as function of the channel widths
show for different number of layers $n_l$ distinct response regimes where their
values are close to one. On top of the graph we also indicated the channel width
in units of the length scale $R$.
Clearly, the change of the number of layers happens with a period of $\sim R$.
But for integer multiples of $R$ the system is not in a layered configuration,
but in the transition between two layered structures. This means that the
confinement induced optimal layer separation is smaller than the separation $R$
expected for the unbounded system. 

\begin{figure}[h!tb]
  \begin{center}
    \includegraphics[width=\columnwidth,clip]{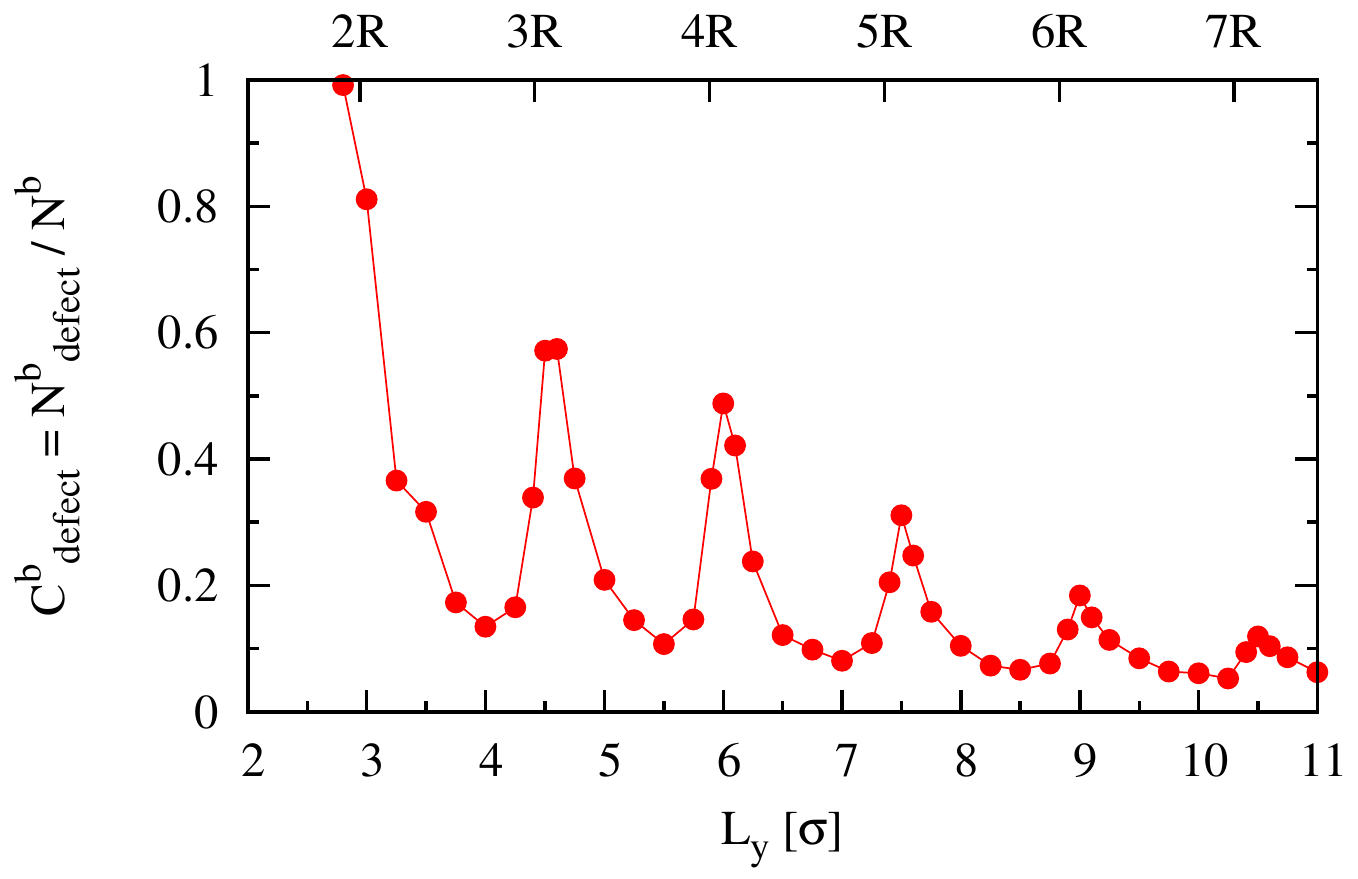}
  \end{center}
  \caption[The bulk defect concentration as function of the channel
  width.]{Simulation: The
  bulk defect concentration as function of the channel width for identical
  simulation parameters as in Fig.~\ref{fig:StateDiagramNLayerLy}.}
  \label{fig:EQDefectConcentrationLy}
\end{figure}

The above scenario can be confirmed by looking at the bulk defect concentration 
\begin{equation}
  C^b_{\mathrm{defect}} \equiv \frac{N^b_{\mathrm{defect}}}{N^b}
  \label{eq:BulkDefectCondentration}
\end{equation}
which is defined as the ratio of the number $N^b_{\mathrm{defect}}$ of bulk
particles with either more or less than six nearest neighbors and the total
number $N^b$ of bulk particles. All particles with a distance greater then
$0.5\sigma$ are defined as bulk particles. In
Fig.~\ref{fig:EQDefectConcentrationLy} $C^b_{\mathrm{defect}}$ is plotted as
function of the channel width for identical simulation parameters as used
above. The concentration of defects in the bulk shows an oscillatory behavior
with a period of $\sim R$. The peak positions indicate the channel widths where
the system can not equilibrate into a layered structure, and the positions of
the minima coincide with stable layer configurations. This behavior is in good
agreement with the results of Haghgooie as can be seen from taking slices of
constant $\Gamma_H$ in figure 6 of~\cite{Haghgooie05}.

\subsection*{Time Evolution of the Defect Configuration}

\begin{figure}[htb]
  \begin{center}
    \includegraphics[width=\columnwidth,clip]{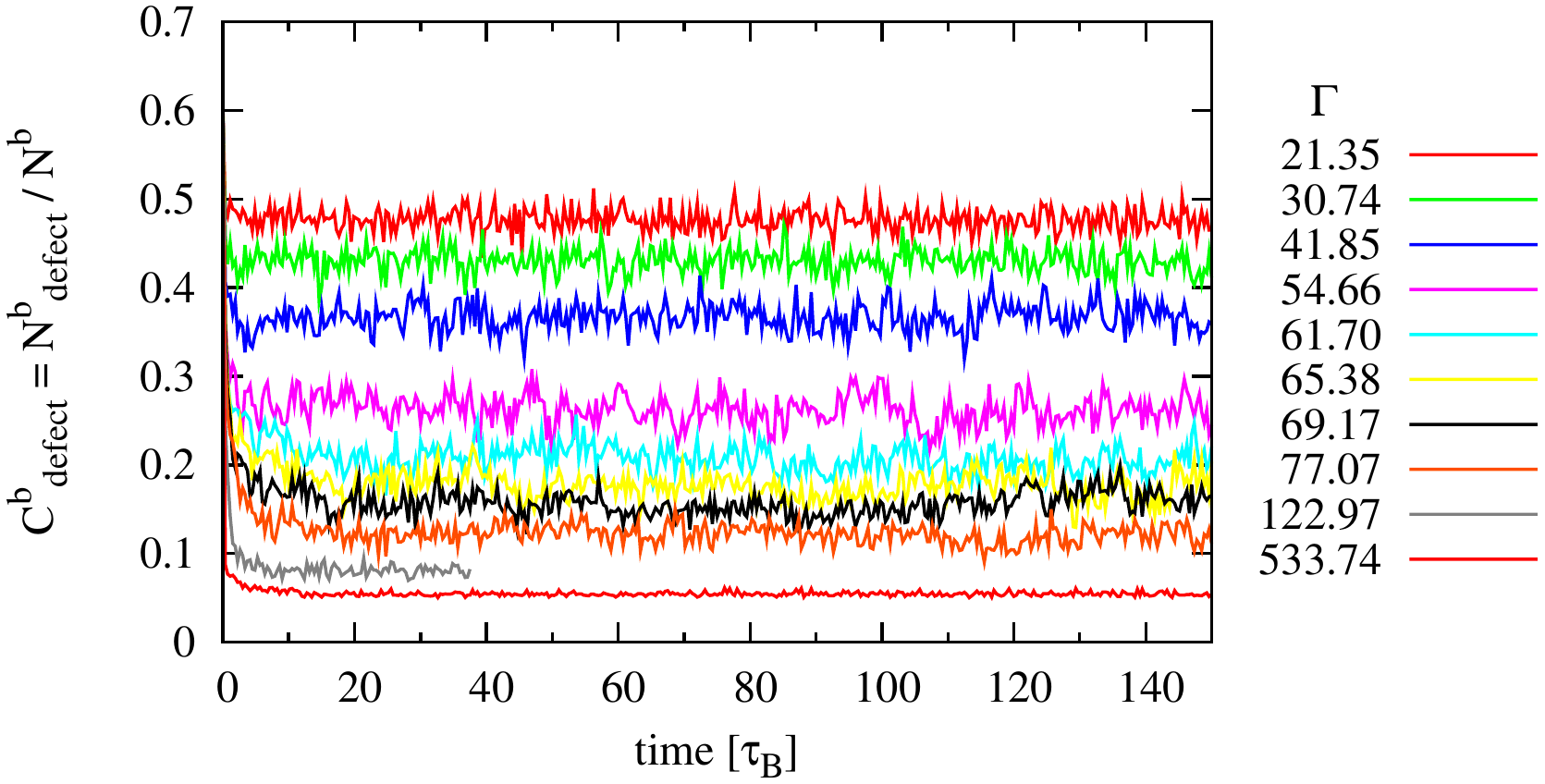}
  \end{center}
  \caption[Time evolution of the defect concentration.]{
  Simulation: Time evolution of the defect concentration $C^b_{\mathrm{defect}}$ for a
  channel with $n=0.4\sigma^{-2}$, $N=4000$, and $L_y=10\sigma$ kept fixed. The
  particle interaction strength is modified via the applied magnetic field $B$
  having values between $\unit{0.1}{\milli\tesla}$ and
  $\unit{0.5}{\milli\tesla}$.
  } 
  \label{fig:DefectConcentration}
\end{figure}
In Fig.~\ref{fig:DefectConcentration} the time evolution of the defect
concentration $C^b_{\mathrm{defect}}$ of the bulk particles during an
equilibration run is explicitly plotted for a selection of $\Gamma$ values for
a channel of width $L_y=10\sigma$. All runs are started from a random particle
distribution. After a time of $10~\tau_B$ the defect concentration remains
unchanged for all $\Gamma$ values. For $45.0 < \Gamma < 80$, {\it i.e.} for the
transition region between the liquid and the solid state, the equilibration
process is slower than for the other values. The fluctuations increase near the
phase boundary. These effects are consistent with the results of
Haghgooie~\cite{Haghgooie05} obtained for an unbounded system.

\subsection*{System State Dependency on the Interaction Strength}

\begin{figure}[h!t]
  \begin{center}
     \includegraphics[width=\columnwidth,clip]{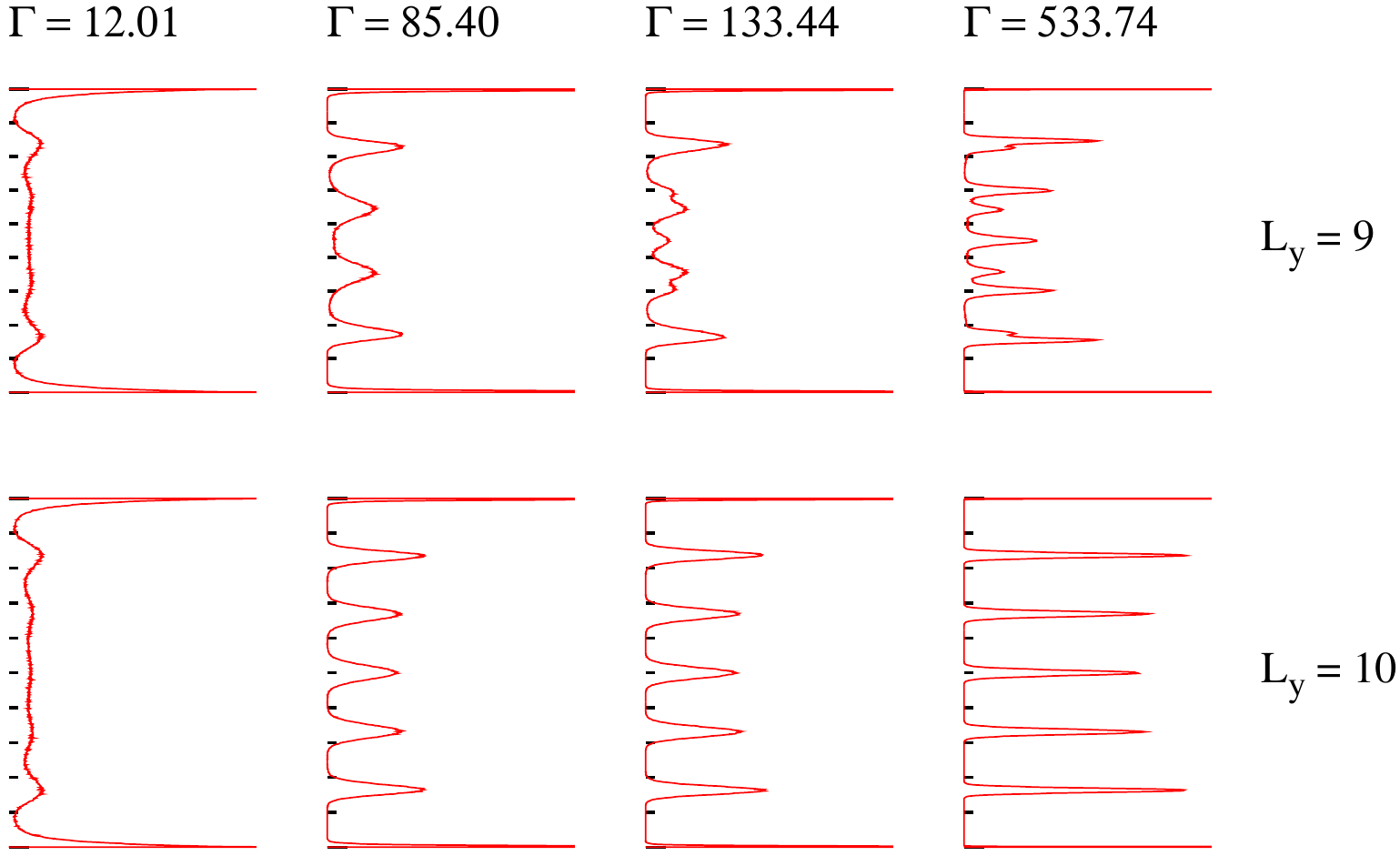}
  \end{center}
  \caption[Density profiles transverse to the walls for $L_y=9\sigma$ and $L_y
  =10\sigma$ in dependence of $\Gamma$.]{Simulation: Density profiles transverse to the
  walls for $L_y=9\sigma$ and $L_y =10\sigma$ in dependence of $\Gamma$. Again,
  the peaks at the walls are truncated for better clarity.}
  \label{fig:YHistVarB}
\end{figure}
In Fig.~\ref{fig:YHistVarB} we show density profiles transverse to the
confining walls for the two channel widths $L_y = 9\sigma$ and $L_y = 10\sigma$
at four values of $\Gamma$. On the left hand side both systems are liquid
whereas on the right hand side they are both in the solid state. These density
histograms are obtained by taking the average over 3500 configurations in
equilibrium.  The system characteristics are very different depending on the
$\Gamma$ value and the channel width $L_y$. For high $\Gamma$ values, where the
system is in the solid state, the density profile for the channel width
$L_y=10\sigma$ is sharply peaked at the positions of the seven layers. On
decrease of the interaction strength $\Gamma$ these peaks broaden and have a
Gaussian profile down to a value of $\Gamma\approx 65$. The central peaks show
greater broadening than the peaks at the wall, {\it i.e.} the system melts
first in the center of the channel. Even for low $\Gamma$ values as
$\Gamma\approx12.01$, where the unbounded system would be deep in the liquid
state, the particles at the wall are still relatively localized in their
$y$-positions. A clear density minimum between the colloids in the edge layer
and the colloids of the central region can always be identified. For the
channel width $L_y=9\sigma$ the
melting scenario is different. The peak profile is less pronounced for
$\Gamma=533.74$ and there is less order across the channel. A mixture between a
structure of 6 and of 7 layers is indicated by the positions of the peak
maxima. The structure of seven layers is favored more, because the peaks
connected to a structure of 7 layers are more pronounced than the remaining
peaks related to 6 layers. Decreasing $\Gamma$ again leads to a broadening of
the peaks and the structure with six layers becomes more favorable ($\Gamma =
133.44$). The unbounded system would be well in the solid state at this value
at this interaction strength. For $\Gamma=85.40$ only the peaks related to six
layers remain, and for $\Gamma=12.01$ no significant qualitative difference to
the situation for the channel of width $10\sigma$ exists.

\begin{figure}[h!tb]
  \begin{center}
    \includegraphics[width=\columnwidth,clip]{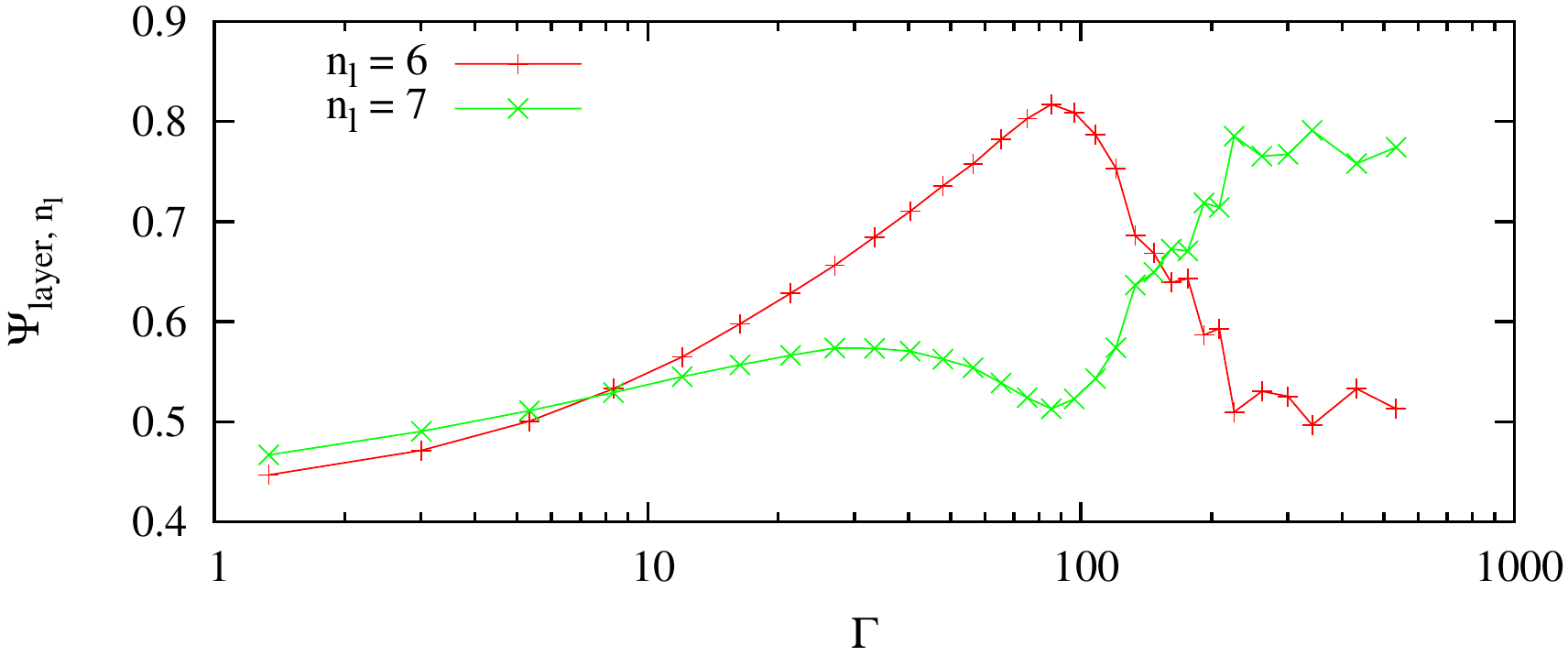}
  \end{center}
  \vspace*{-3.5ex}
  \caption[Global layer order parameters as function of the interaction
  strength for a channel of width $L_y =9$.]{Simulation: Comparison of the dependency on
  the interaction strength $\Gamma$ of the global layer order parameter
  $\Psi_{\mathrm{layer},\,n_l}$ with $n_l=6$ and $n_l=7$ for the channel width
  $L_y = 9$.}
  \label{fig:VariationNLayerY9}
\end{figure}
These changes of the peak characteristics of the density profile across the
channel of width $L_y = 9\sigma$ is also reflected in the behavior of the layer
order parameters in Fig.~\ref{fig:VariationNLayerY9} for $n_l = 6$ and $n_l =
7$ layers on variation of the interaction strength.
$\Psi_{\mathrm{layer},\,n_l=6}$ exhibits a maximum at about $\Gamma=90$, and
strongly decreases for higher $\Gamma$ values whereas the values of
$\Psi_{\mathrm{layer},\,n_l=7}$ increase to values of about $0.8$.

\begin{figure}[ht!b]
  \begin{center}
    \includegraphics[width=\columnwidth,clip]{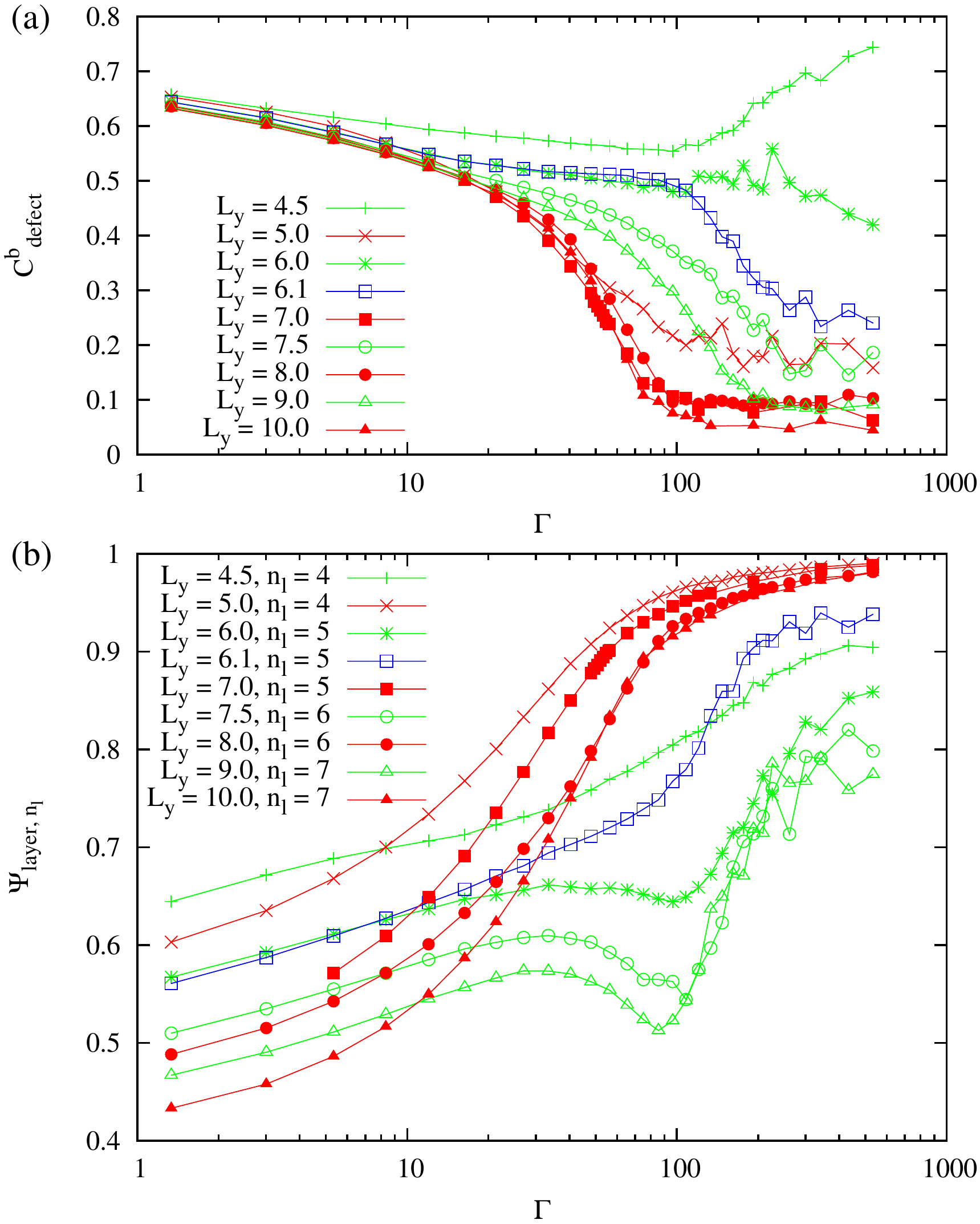}
  \end{center}
  \vspace*{-3.5ex}
  \caption[Order parameters in dependency of the dimensionless interaction
  strength for a selection of channel widths.]{Simulation: Order parameters in dependency
  of the dimensionless interaction strength $\Gamma$. In {\bf (a)} the bulk
  defect concentration $C^b_{\mathrm{defect}}$ and in {\bf (b)} the layer order
  parameter $\Psi_{\mathrm{layer},\,n_l}$ are shown for a selection of channel
  widths.}
  \label{fig:Nlane5DefectConcentrVarB}
\end{figure}
In Fig.~\ref{fig:Nlane5DefectConcentrVarB} the behavior of the bulk defect
concentration $C^b_{\mathrm{defect}}$ and the layer order parameter
$\Psi_{\mathrm{layer},\,n_l}$ on variation of $\Gamma$ are summarized for a
selection of channel widths. The curves are color coded depending on whether
the equilibrium configuration has a boundary induced layered structure (red
curves) or not (green curves). The blue curves are connected to the channel
width $L_y=6.1\sigma$, where the equilibrium system has a perturbed structure
with 5 layers as it can be deduced from \mbox{$\Psi_{\mathrm{layer},\,n_l=5} >
0.9$} and $C^b_{\mathrm{defect}}\approx 0.25$ for $\Gamma > 300$. Particles
changing between the central and its two neighboring layers perturb the 5
layers. 

The red curves for the defect concentration of bulk particles
$C^b_{\mathrm{defect}}$ in Fig.~\ref{fig:Nlane5DefectConcentrVarB}(a) show a
very similar behavior. All of them monotonically decrease up to $\Gamma \approx
100$ to values $<0.1$ and stay constant thereafter. Only for the small channel
width $L_y=5\sigma$ which has an equilibrium configuration of 4 layers the
final defect configuration is about $0.17$. Due to the small channel width the
defects being induced in the layers next to the edge layers (because of the
higher line concentration of the edge particles) are of greater influence. For
$L_y\ge6.0\sigma $ the blue and the green curves of
$C^b_{\mathrm{defect}}(\Gamma)$ also show a monotonic decay, but of varying
magnitude and the $C^b_{\mathrm{defect}}$ becomes constant at significantly
higher $\Gamma$ values than for the red curves. It is interesting to note, that
the curves for $L_y=6.0\sigma$ (green) and $L_y=6.1\sigma$ (blue) fall on top
of each other for $\Gamma<105$, but significantly diverge for $\Gamma>105$
where layers form for $L_y=6.1\sigma$ but not as strong for $L_y=6.0\sigma$
(cf.  Fig.~\ref{fig:Nlane5DefectConcentrVarB}(b)).

In Fig.~\hyperref[fig:Nlane5DefectConcentrVarB]{\ref{fig:Nlane5DefectConcentrVarB}(b)} the global
layer order parameter $\Psi_{\mathrm{layer},\,n_l}$ is plotted as function of
the interaction strength $\Gamma$. Shown are the functional dependencies of
$\Psi_{\mathrm{layer},\,n_l}$ for the parameter $n_l$ which have the maximum
value for $\Gamma>500$. The layer order parameters
$\Psi_{\mathrm{layer},\,n_l}$ increase monotonically to values greater than
$0.9$ for the systems connected to the red curves. The transition to the
layered structure takes place for $\Gamma<100$. 
In general,
Fig.~\hyperref[fig:Nlane5DefectConcentrVarB]{\ref{fig:Nlane5DefectConcentrVarB}(b)}
shows that in case of layer formation, larger values of $L_y$ require larger
$\Gamma$ values for layering. As observed before in
Fig.~\ref{fig:VariationNLayerY9} the $\Psi_{\mathrm{layer},\,n_l}$ have
non-monotonic behavior and the transition to the final state takes place for
$\Gamma> 110$, which is greater than for the layered structures. The highest
values of $\Psi_{\mathrm{layer},\,n_l}$ are less than $0.9$.

Piacente and coworkers~\cite{Piacente04} studied the structural, dynamical
properties and melting of a quasi-one-dimensional system of charged particles,
interacting through a screened Coulomb potential in equilibrium. This system is
related to our situation, but a different particle interaction potential is
used and the particles are confined in $y$-direction by a parabolic potential.
They also find a rich structural phase diagram with different layered
structures as function of the screening length $\kappa_D^{-1}$ and the electron
density $n_e$ of the system.

% Re-entrance phase behavior?
A re-entrant phase behavior, {\it i.e.} a melting process succeeded by a system
solidification and subsequent further melting, was observed for particle
confinement inside of a
circle~\cite{Bubeck99,Bubeck,HenselerDiplom,Schweigert00} or in static 1D
periodic light fields~\cite{Wei98,Bechinger01,Strepp,Strepp02} both in
experiment and simulation. For our planar wall confinement we do not find any
re-entrant behavior as function of the dimensionless interaction strength
$\Gamma$ (the inverse effective temperature). In
Fig.~\hyperref[fig:Nlane5DefectConcentrVarB]{\ref{fig:Nlane5DefectConcentrVarB}(a)}
the defect concentration of the bulk decreases monotonically with increasing
$\Gamma$ and thus gives no hint on a reentrant behavior. This observation again
is in agreement with the results of~\cite{Haghgooie05}.

For particles inside a disc shaped cavity the increase of radial fluctuations
is responsible for the re-stabilization of an ordered shell structure with
increasing temperature. 
In our case the influence of the
confining hard walls does not seem to have a similar effect on the particle
fluctuations in $y$-direction to give rise to a re-entrance behavior. We
conclude that the re-entrance phenomenon depends strongly on way of
confinement. It would be interesting to study the influence of the curvature of
the confinement on the melting scenario systematically. On
the other hand, a boundary induced reentrant behavior between different layered
structures is observed for increasing channel width $L_y$ (cf.
figures~\ref{fig:StateDiagramNLayerLy} and \ref{fig:EQDefectConcentrationLy}).

\begin{figure}[htb]
   \begin{minipage}[b]{0.15\columnwidth}
    {\bf (a)} 
    
    $\Gamma=56.38$
   \end{minipage}
   \hfill
   \begin{minipage}{0.8\columnwidth}
    \includegraphics[width=\columnwidth,clip]{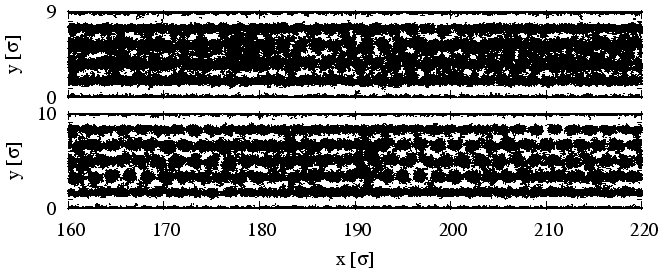}
   \end{minipage}

   \begin{minipage}[b]{0.15\columnwidth}
    {\bf (b)} 
    
    $\Gamma=133.44$
   \end{minipage}
   \hfill
   \begin{minipage}{0.8\columnwidth}
    \includegraphics[width=\columnwidth,clip]{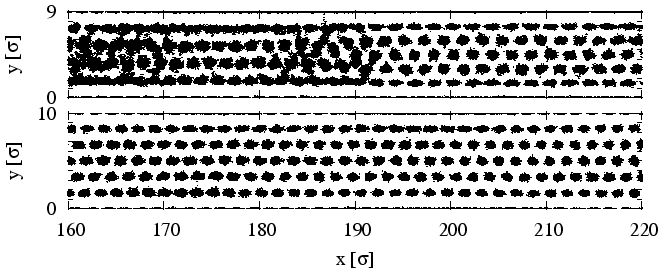}
   \end{minipage}
  \vspace{-2ex}
  \caption[Superimposed particle positions in equilibrium for two channel
  widths.]{Simulation: Superimposed particle positions in equilibrium for channel segments
  of length $60\sigma$ and widths $L_y=9\sigma$ and $10\sigma$ during the time
  interval $\Delta t = 15\tau_B$ \mbox{($\equiv2\cdot10^5$ BD steps)}. Shown is
  the situation {\bf (a)} in the fluid regime ($\Gamma=56.38$ or
  $B=\unit{0.1625}{\milli\tesla}$) and {\bf (b)}~in the solid state
  ($\Gamma=133.44$ or $B=\unit{0.25}{\milli\tesla}$). For the images of (b) the
  corresponding density profiles are given in Fig.~\ref{fig:YHistVarB}.}
  \label{fig:TrajectoryEQ}
\end{figure}
Figure~\ref{fig:TrajectoryEQ} illustrates the qualitatively different particle
mobilities due to the confinement according to their $y$-position for the two
channel widths $L_y=9\sigma$ and $L_y=10\sigma$. Shown is the overlay of $1000$
equilibrium configurations, which corresponds to a run of length~$\Delta t =
15\tau_B$, both for the fluid state
[Fig.~\ref{fig:TrajectoryEQ}(a)] and the solid state
[Fig.~\ref{fig:TrajectoryEQ}(b)]. We recognize already from
these superimposed snapshots that the walls affect the particle mobilities
transverse to the walls for both widths. The edge particles have a very low
mobility to move away from the confining walls they are pressed against. A
clear depletion zone exists between the edge and its neighboring layer.
Generally, the spreading of the particle positions in $y$-direction
increases with growing distance to the walls. When comparing the two widths
$L_y = 9\sigma$ and $L_y = 10\sigma$ we again realize that the layered system
of width $L_y=10\sigma$ is higher ordered with smaller spreading of the
particle positions. In the fluid state, shown in
Fig.~\hyperref[fig:TrajectoryEQ]{\ref{fig:TrajectoryEQ}(a)}, the particles still move
predominantly within the layers parallel to the walls. This boundary induced
layering effect is stronger for $L_y=10\sigma$ than for $L_y=9\sigma$.

The effect of the type of confinement on the ordering of a crystal confined to
stripes of finite width was analyzed using Monte-Carlo simulations by Ricci and
coworkers~\cite{Ricci, Ricci06, Ricci07}. In their case, the particle pair
interaction is given by the inverse power law $\propto r^{-12}$. They studied
the influence of ideal planar hard walls and structured walls obtained by
fixing the wall particles at separations they would have in a bulk system. Our
findings are in good agreement with their results.

\subsection{Diffusion Behavior}\label{sec:SFD}

For channels, which are small enough, so that the particles
cannot pass each other the diffusion behavior of the particles changes. The
sequence of the particles remains unchanged and the particles move in a {\it
single file} (SF). The long-time behavior of the mean-square displacement for
infinite long channels is predicted to be~\cite{Richards77, Fedders77}
\begin{equation}
  \langle \Delta x^2 \rangle = 2F\sqrt{t}.
  \label{eq:SFD1}
\end{equation}
Here $F$ is the single file mobility and $t$ the time.

Such a behavior is an example of {\it anomalous diffusion} or {\it non-Fickian
diffusive behavior}, which is characterized by the occurrence of a mean-square
displacement of the form $\langle \Delta \mathbf{r}^2 \rangle \propto
t^\alpha$, where $\alpha \neq 1$. The motion is called {\it sub-diffusive} for
the (anomalous) diffusion coefficient $0<\alpha<1$ and {\it super-diffusive} for
$\alpha > 1$. The phenomenon of single file diffusion (SFD) has received a lot
of attention in recent publications, especially after the experimental
observation of Wei~{\it et al.}~\cite{Hahn96, Wei00, Kollmann03, Lutz04, Lin05,
Demontis06, Marchesoni06, Taloni06, Taloni06a, Savelev06, Coupier07,
Coupier07a, Majumder07, Nelissen07}.

\begin{figure}[hbt]
  \begin{center}
    \includegraphics[width=\columnwidth,clip]{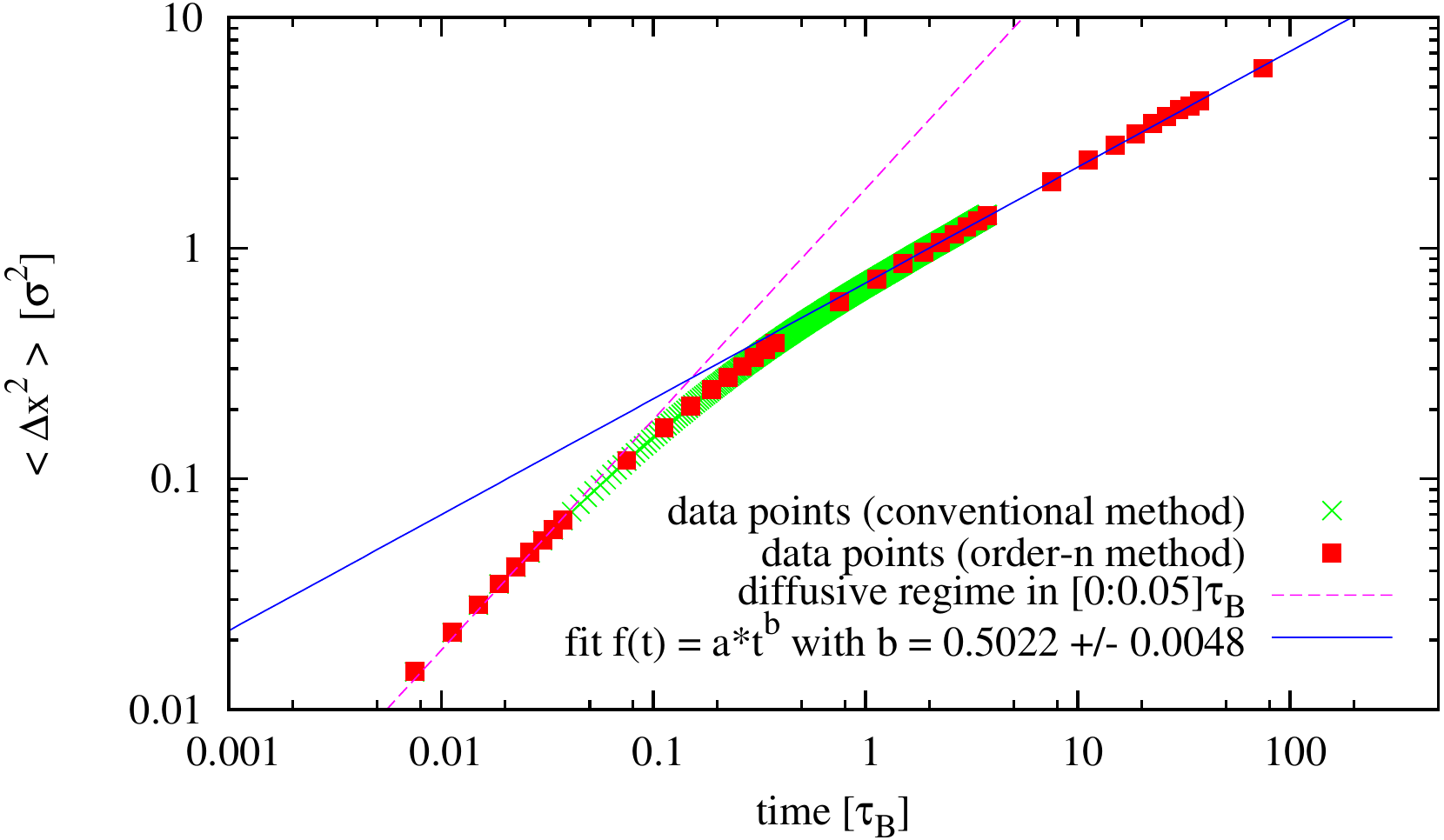}
  \end{center}
  \caption[Single file diffusion behavior for a channel setup with
  $L_y=0.5\sigma$.]{Simulation: Single file diffusion behavior for a channel setup with
  $L_y=0.5\sigma$, $n=0.8\sigma^{-2}$, $\Gamma=60.39$, and periodic boundary
  condition in $y$-direction in absence of any driving field. The two symbol
  types used for the calculated data points refer to two different ways of
  evaluation of the MSD.}
  \label{fig:SFD1}
\end{figure}
In Fig.~\ref{fig:SFD1} we plot the mean square displacement (MSD) as a function
of the simulation time in a double logarithmic graph. The data points are
obtained from a simulation run of a channel having ideal hard walls, the width
$L_y=0.5\sigma$, and periodic boundary condition in $x$-direction and no
driving field. Two different algorithms have been used to evaluate the MSD.
Both, the conventional analysis of the MSD (green crosses) and the so-called
{\it order-n algorithm} (red squares) to measure correlations being introduced
in~\cite{Frenkel} give identical results. At short times, {\it i.e.} at times
less than $0.1\tau_B$, the MSD increases $\propto t$, which is characteristic
for the ballistic movement of the particles. The dashed magenta line in
Fig.~\ref{fig:SFD1} has the slope $\alpha=1$ as it is the case for normal
diffusive behavior. Clearly, for $t\le0.1t_B$ the simulation data points fall
onto this curve. After the time $\tau_B$, which can be interpreted as the time
a particle needs to meet one of its nearest neighbors and to realize it cannot
overtake, the MSD approaches the square-root time dependency characteristic for
SFD. This is indicated in Fig.~\ref{fig:SFD1} by the solid blue line, which
is a fit of the function $f(t)=A\cdot t^\alpha$ with the two fit parameters $A$
and $\alpha$ to the data points with $t\ge \tau_B$. The resulting slope is
$\alpha=0.5022\pm0.0048$, which is in perfect agreement with SFD behavior. 

Now, the following questions arise: How does the longitudinal and transversal
particle diffusion behavior depend on the channel width $L_y$? How does the
transition take place from the single-file diffusion behavior for channel
widths where particle can not pass each other to the Fickian diffusion behavior
of bulk systems?  

\begin{figure}[thb]
  \begin{center}
    \includegraphics[width=\columnwidth,clip]{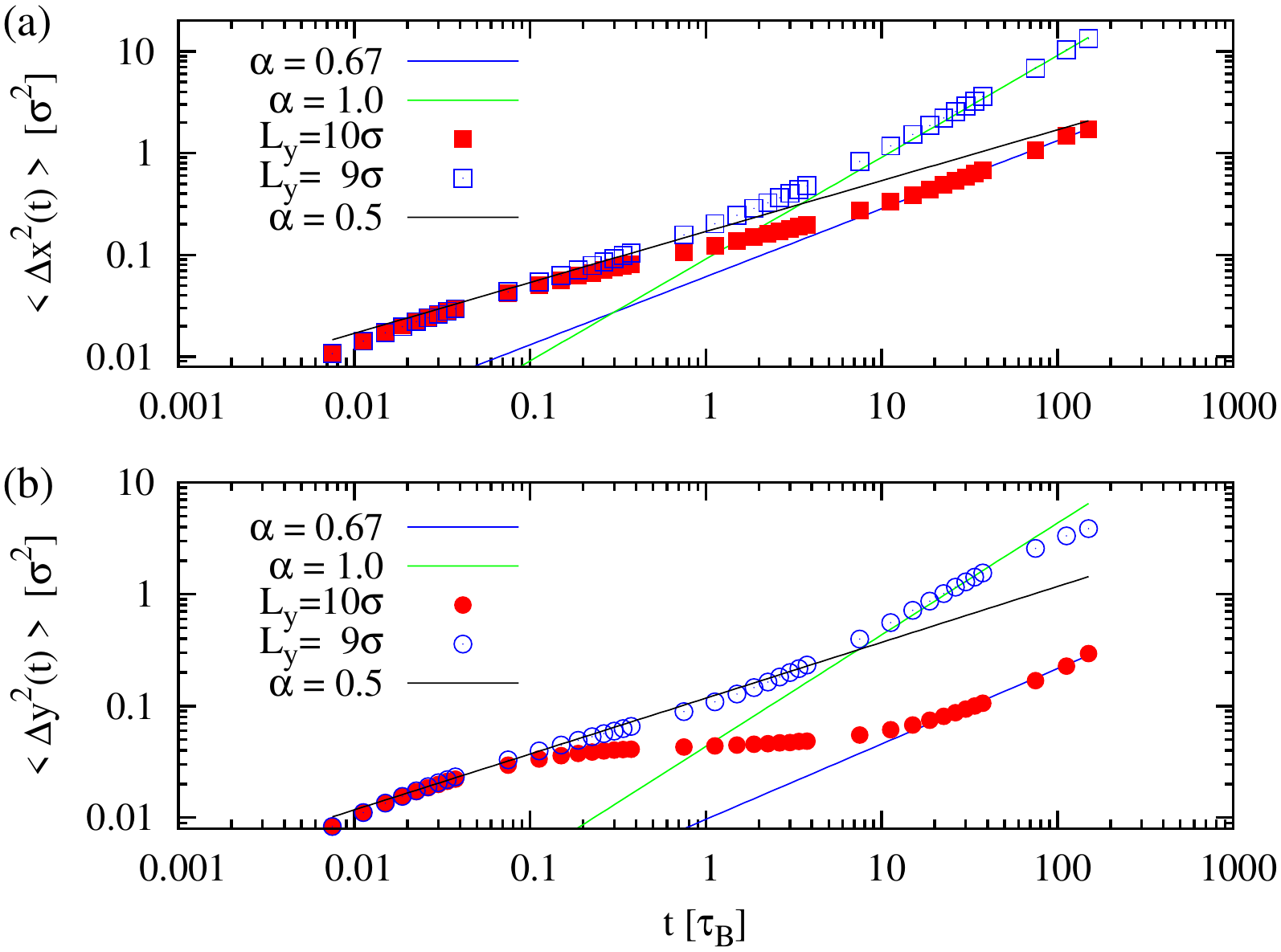}
  \end{center}
  \caption[Comparison of the mean-square displacement parallel and perpendicular
  to the confining channel walls for the two channel widths $L_y=9\sigma$ and
  $L_y=10\sigma$.]{Simulation: Comparison of the mean-square displacement {\bf (a)}
  $\langle \Delta x^2\rangle$ parallel and {\bf (b)} $\langle \Delta
  y^2\rangle$ perpendicular to the confining channel walls for the two channel
  widths $L_y=9\sigma$ and $L_y=10\sigma$. The simulation parameters are:
  $L_x=800\sigma$, $n=0.4\sigma^{-2}$, $B=\unit{0.2}{\milli\tesla}$, and
  $\Gamma=83.4$. The periodic boundary condition is applied in $x$-direction.}
  \label{fig:MSDVgl9and10}
\end{figure}
To give a first answer to these questions, we evaluated the time-dependency of
the mean-square displacement for the two channel widths $L_y=9\sigma$ and
$L_y=10\sigma$. The results of the MSD $\langle \Delta x^2\rangle$ parallel to
the channel walls and of the MSD $\langle \Delta y^2\rangle$ transversal to the
channel walls are shown in the
figures~\hyperref[fig:MSDVgl9and10]{\ref{fig:MSDVgl9and10}(a)}
and~\hyperref[fig:MSDVgl9and10]{(b)}.  Obviously, the dependency of the MSD on
time differs strongly for the two channel widths. The long time behavior of the
longitudinal MSD $\langle \Delta x^2\rangle$ is linear proportional to time $t$
for the width $L_y=9\sigma$, whereas for $L_y=10\sigma$ the long time behavior
scales with the exponent $\alpha \approx 0.67$. For times $t<0.4\tau_B$ the
longitudinal MSD scales approximately with the exponent $\alpha=0.5$. For the
transversal MSD $\langle \Delta y^2 \rangle$ the time dependency is similar
(cf.~Fig.~\hyperref[fig:MSDVgl9and10]{\ref{fig:MSDVgl9and10}(b)}), but for the
width $L_y=10\sigma$ the transversal MSD has a plateau for intermediate times.
This is due to the boundary induced formation of seven layers, which is not the
case for width $L_y=9\sigma$. Notice, that for either $\langle \Delta x^2
\rangle$ or $\langle \Delta y^2 \rangle$ the absolute values in the
intermediate and long time limit are smaller than it is the case with well
defined layers. 

The crossover from single-file diffusion with the exponent $\alpha=0.5$ to
Fickian diffusion with $\alpha=1$ in the bulk limit has recently been
analyzed theoretically by Mon, Percus and Bowles~\cite{Mon02, Bowles04, Mon06,
Mon07}. These authors present a phenomenological theory in terms of the hopping
time $\tau_{\mathrm{hop}}$, which is defined as the average time a particle
must spend before it can ``hop'' over (pass) its nearest neighbor in
longitudinal direction. They theoretically show that with increasing
transversal system size the diffusion constant will increase from zero
according to $D \propto (\tau_{\mathrm{hop}})^{-1/2}$, where
$\tau_{\mathrm{hop}}$ is a function of the pore radius 3D or the channel width
in 2D. They confirmed this predicted behavior by MC and MD simulations of hard
spheres within a pore and hard discs confined to a microchannel respectively.
In general, when particles are allowed to pass their neighbor particles, the
long time dynamics is given by Fickian diffusion. This can be understood by the
simple argument: After the mean time $\tau_{\mathrm{hop}}$ a particle passes
one of its nearest neighbors in either direction. Therefore the long time
diffusion behavior is given by conventional Fickian diffusion, but for times
less than $\tau_{\mathrm{hop}}$ the SFD behavior is expected. A similar
argument was already used in the context of a two chain lattice gas model of
Kutner {\it et al.}~\cite{Kutner84}.

%%% Transport Behavior of Colloids in Mircochannels
%\input{LTpaper-Transport-v8}
\section{Transport Behavior of Colloids in Microchannels} \label{ch:LayerReduction}

Now, we want to address the transport behavior of colloids confined to such
microchannels as described in the previous section. The colloids are driven by
the application of an external driving force $\mathbf{F}$ and thus form a
system in non-equilibrium. This driving can be of gravitational origin as in
our case, or due to the presence of an electrical or magnetic field or an
osmotic pressure difference between both channel ends. To match the experimental
situation closely, we will concentrate mainly on colloids with repulsive
dipolar pair-interaction driven by gravity.  First we introduce the effect of
dynamical rearrangement of the colloids during their flow along the channel. We
call this effect {\it layer reduction}.

\subsection{Layer Reduction}\label{sec:LayerReduction}
\begin{figure}[hbt]
  \begin{center}
  \includegraphics[width=\columnwidth,clip]{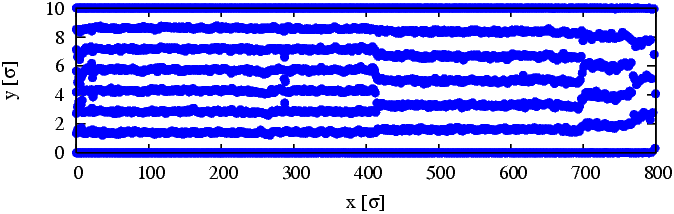}
\end{center}
\caption[Full channel snapshot from simulation for a channel with ideal hard
walls]{Simulation: Full channel snapshot for a channel with ideal hard
walls ($\Gamma = 533.74$, $\alpha=0.2\degree$) after $10^6$ BD simulation steps
having reached a stationary non-equilibrium state.  Note, that the scaling on
the $y$-axis is stretched by a factor of 20 compared to the $x$-axis scaling.}
\label{fig:SnapshotFullChannel}
\end{figure}

\begin{figure}[htb]
  \begin{center}
  \includegraphics[width=\columnwidth,clip]{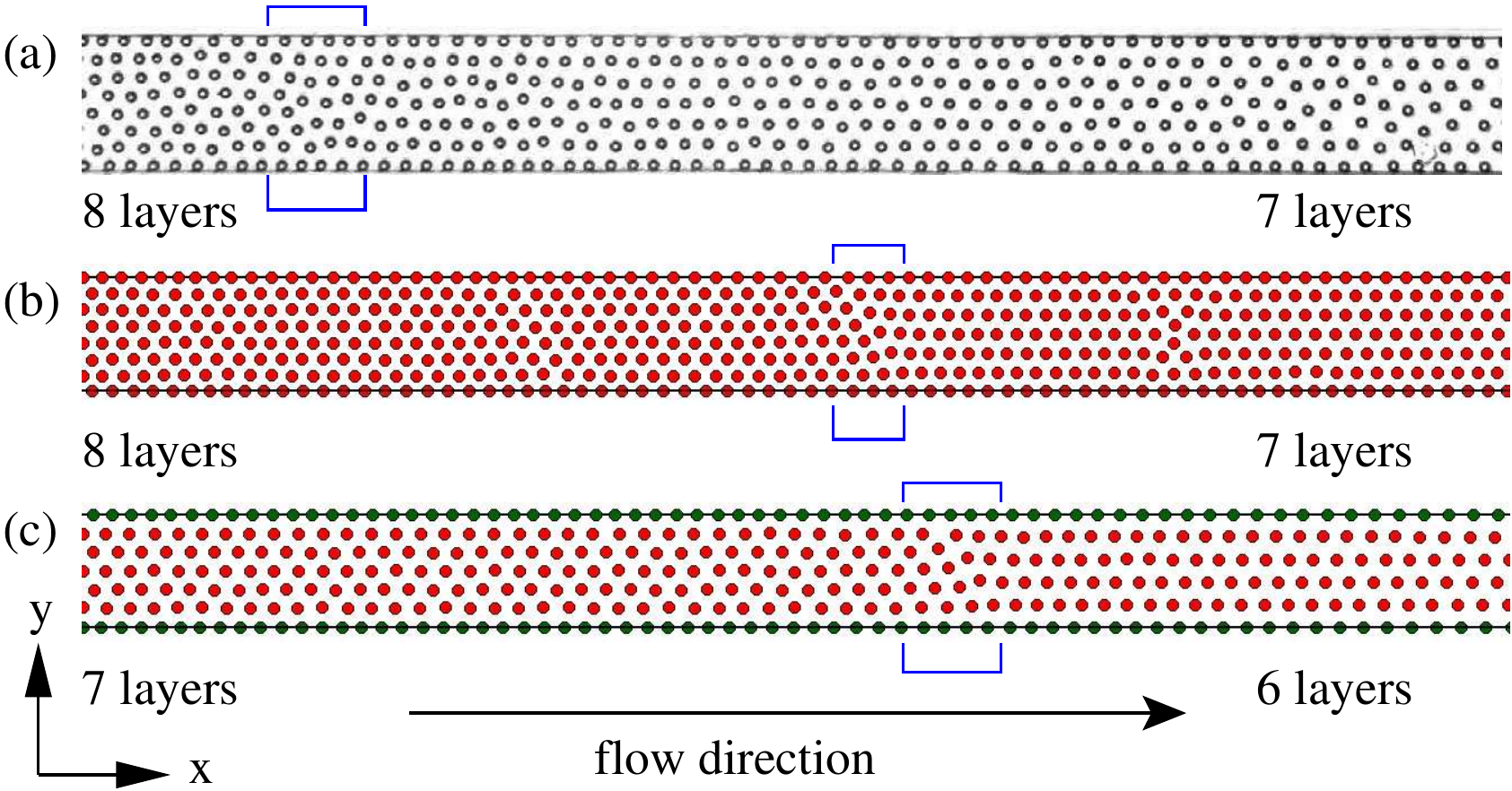}
\end{center}
\caption[Snapshots of the stationary non-equilibrium configurations around the
layer transition]{{\bf (a)} Experiment: Non reworked video microscopy snapshot of colloidal
particles moving along the lithographically defined channel. The channel
partition shown has the size $(\unit{692\,\times\,60}){\micro \meter} =
(153.8\times13.33) \sigma$, and the interaction strength is $\Gamma \approx
72$. {\bf (b)} Simulation: Snapshots for a channel with ideal hard walls
[$(\unit{573.3\times45}){\micro \meter} = (127.4\times10) \sigma$, $\Gamma =
640.5$], {\bf (c)} the same as in (b) with the particles at the walls (marked
green) kept fixed [$(\unit{573.3\times45}){\micro \meter}$, $\Gamma = 5026$].
The rectangles mark the region of the  layer reduction.}
\label{fig:channelsnapexperiment}
\end{figure}

A first impression of the particle arrangement under the influence of an external
driving field give the figures~\ref{fig:SnapshotFullChannel}
and~\ref{fig:channelsnapexperiment} which depict typical configuration
snapshots from simulation and experiment. The particles move along the channel
from left to right in the positive $x$-direction. 
The external magnetic field strength $B$, which is responsible for the strength
of the pair-interaction, and the overall particle number density $n$ are chosen
in such a way, that the confined equilibrium system is hexagonally ordered.
This is true also for the unbounded system under identical conditions.
Figure~\ref{fig:SnapshotFullChannel} is a representative snapshot taken in the
simulation of the full channel having the length $L_x = 800\sigma =
\unit{3.6}{\milli\meter}$. The first 10\% of the channel act as reservoir.  In
the experiment the channel length is $L_x = 444.4\sigma =
\unit{2.0}{\milli\meter}$. The strength of the constant driving force
$\mathbf{F}^{\mathrm{ext}} = F\mathbf{e}_x$ can either be specified directly or
by definition of the inclination $\alpha$ resulting in $F=mg\sin\alpha$. Under
the influence of external driving the particles still form layers.  Additionally we
observe, both in experiment and in simulation, a decrease of the number of
layers in the direction of motion~\cite{Henseler06, Henseler}. The
layer transitions are clearly visible in
Fig.~\ref{fig:SnapshotFullChannel}, where they are located at
$x\approx420\sigma$ from 8 to 7 layers, at $x\approx700\sigma$ from 7 to 6
layers, and at $x\approx 770\sigma$ from 6 to 5 layers.
	
The images of Fig.~\ref{fig:channelsnapexperiment} show in enlargement the
part of the channel near the region of layer reduction being marked by the
rectangle. The video microscope snapshot of
Fig.~\hyperref[fig:channelsnapexperiment]{\ref{fig:channelsnapexperiment}(a)}
is taken from the experiment~\cite{Henseler06}. The small white spots at the
particle centers allow for precise tracking of the particle trajectories with
the video microscope.  Similar snapshots we get from our BD~simulations with
either co-moving
(Fig.~\hyperref[fig:channelsnapexperiment]{\ref{fig:channelsnapexperiment}(b)})
or fixed edge particles
(Fig.~\hyperref[fig:channelsnapexperiment]{\ref{fig:channelsnapexperiment}(c)}).
In these two subfigures the filled circles represent the particles at their
real size relative to the channel width. For these highly ordered systems the
layer transitions take place on the scale of only a few particle diameters.

\subsection{Density Gradient along the Channel}
The simulation snapshots above are taken after a time long enough for the
system to reach a stationary non-equilibrium situation. Applying the external
driving force to the equilibrated channel configuration leads to the build-up
of a particle density gradient along the channel. This is an effect of the
chosen boundary conditions at the channel entrance and exit, which leads to a
pressure difference between both channel ends. After about $10^6$~BD time steps
this density gradient does not change any more, which is the signature of a
stationary state. The exact origin of the density gradient is given by
details of the particle-particle interactions in combination with the
driving force and will  be subject of a separate publication. 

\begin{figure}[tbh]
  \begin{center}
	\includegraphics[width=\columnwidth,clip]{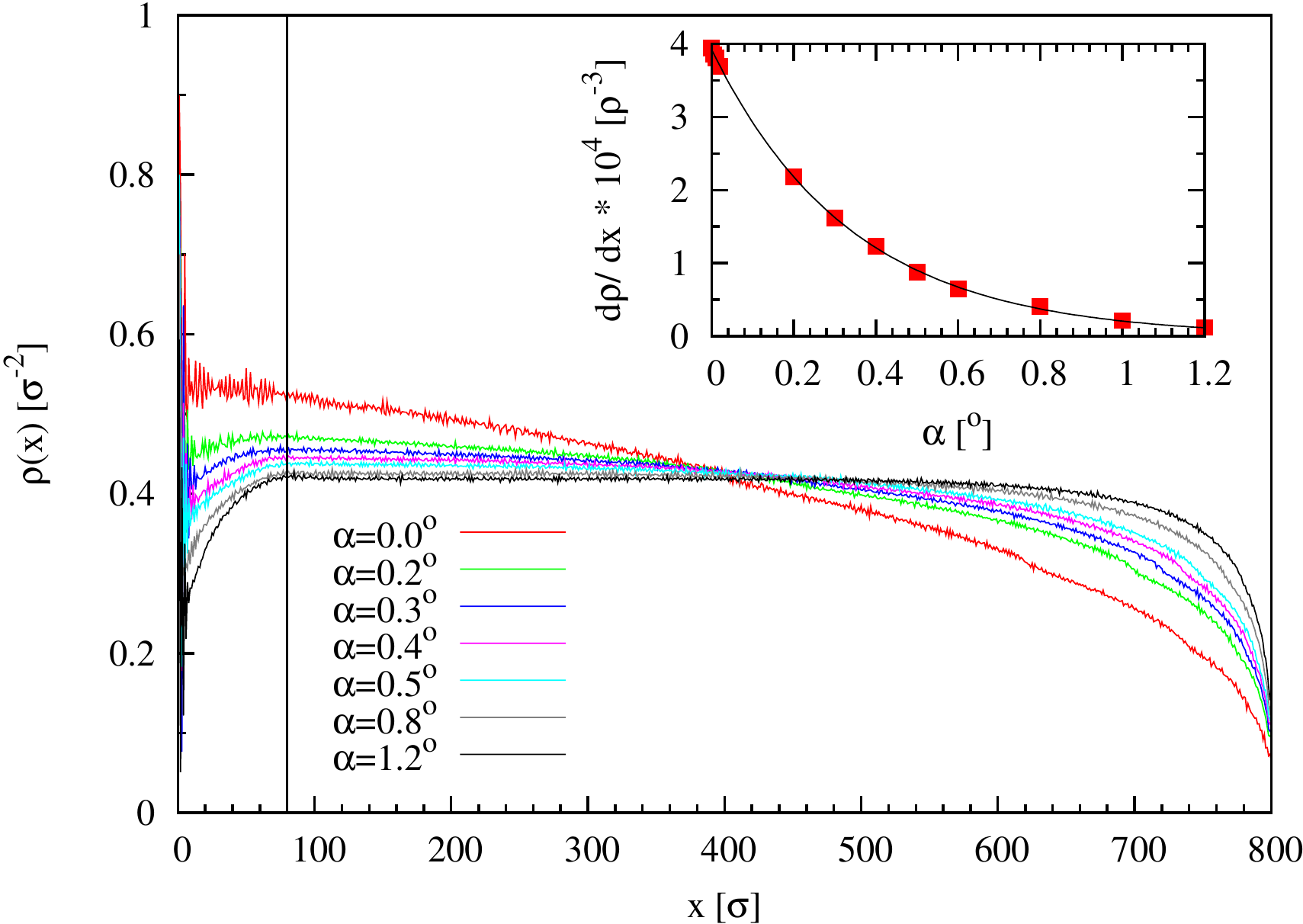}
  \end{center}
  \caption[Stationary non-equilibrium density profiles along the channel for
  several values of the slope $\alpha$]{Simulation: Stationary non-equilibrium density
  histograms along the channel for several values of the slope $\alpha$
  ($\Gamma=533.75$ and $n=0.4\,\sigma^{-2}$). The vertical line at $x=80\sigma$
  marks the right end of the reservoir, {\it i.e.}, the maximum $x$-value up to
  where particles are inserted randomly. The inset shows the density gradient
  in the interval $x\in[150,600] \sigma$ as a function of the inclination
  $\alpha$ being obtained from linear fits to the density histograms. Here the
  line connecting the data points serves as a guideline to the eye.}
  \label{fig:densitygradient}
\end{figure}
To study the robustness of the formation of the density gradient and its
connection to the  layer reduction in our system of gravitationally driven
particles, we performed simulations for a variety of inclinations $\alpha=0.0^\circ
- 10.0^\circ$ keeping the overall particle density fixed at $n =
0.4\,\sigma^{-2}$. The resulting stationary non-equilibrium density profiles
along the channel are shown in Fig.~\ref{fig:densitygradient}. They are
calculated from histograms of the $x$-positions of $1000$~configurations in
stationary non-equilibrium. A very significant decrease of the local density
occurs for $x>700\sigma$, which is caused by the open boundary at $x=800\sigma$.
The region $0\le x < 80\sigma$ acts as reservoir, where new particles are
inserted at random position whenever a particle drops out at the end of the
channel. To avoid unnecessary high perturbations due to random particle
re-insertion in the reservoir the channel is closed at $x=0\sigma$ by a
semipermeable ideal hard wall. Note, that with increasing inclination $\alpha$
a depletion layer forms within the reservoir which is the result of a greater
outflow than input of new particles.

All density profiles show a nearly linear density gradient in the interval
$x\in[150,600] \sigma$, which is maximal for $\alpha = 0^\circ$ (cf. Inset of
Fig.~\ref{fig:densitygradient}). Even at $\alpha = 0^\circ$ a (osmotic) pressure
difference between both channel ends exists for the boundary conditions used,
and a small particle flux is induced. For inclinations \mbox{$\alpha >
1.0^\circ$} the density gradient becomes almost zero. For these inclinations
the driving force dominates, and we find plug flow of the particles without
layer reduction. A decrease of the inclination (driving force) gives rise to an
increase of the density gradient. Under non-plug flow condition we find a
self-induced arrangement of the particles to a nearly hexagonal lattice and
the occurrence of layer reductions with the particles moving across. 

\subsection{Dynamical Properties}
\subsubsection{Drift Velocity}
\begin{figure}[htb]
  \begin{center}
    \includegraphics[width=\columnwidth,clip]{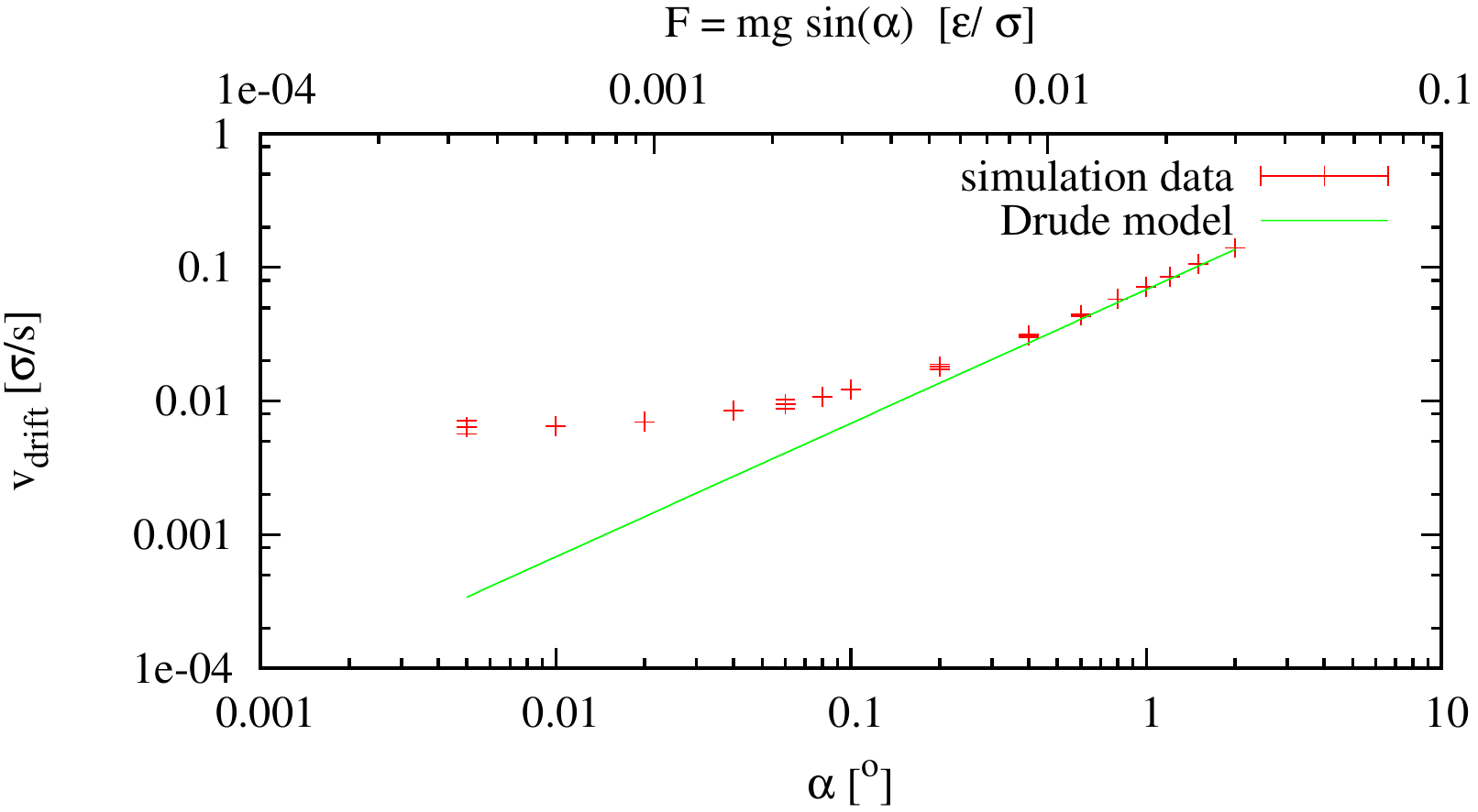}
  \end{center}
  \caption[Average particle drift velocity as function of the
  inclination.]{Simulation: Average particle drift velocity in the interval
  $x\in[200,\,770]\sigma$ as function of the inclination $\alpha$ or equivalently
  the driving force $\mathbf{F} = mg\,\sin(\alpha)\,\mathbf{e}_x$.
  ($L_x=800\sigma$, $L_y=10\sigma$, $n=0.4\sigma^{-2}$ and $\Gamma = 533.74$).
  The solid line is the expected drift velocity for non-interacting particles
  due to the external driving (Drude model).}
  \label{fig:AngleTotalDriftV}
\end{figure}

It is also interesting to study the average overall drift velocity as
function of the driving force. The result is shown in
Fig.~\ref{fig:AngleTotalDriftV}. 
For $\alpha>0.5\degree$ the particle flow
is dominated by the driving force. This is the regime of plug flow, where the
particles move with 
\begin{equation}
  \langle v_{\mathrm{drift}} \rangle_{\mathrm{Drude}} = \frac{mg}{\xi}\,\sin\alpha
  \label{eq:VDrude}
\end{equation}
as expected for non-interacting particles. Such a dependency was formulated by
P.~Drude~\cite{Ashcroft} for electrical conduction
to explain the transport of electrons in metals. For $\alpha < 0.5\degree$
the average drift velocity deviates from the expectation of the Drude model.
Interestingly, for these inclinations the particles move faster than expected.
The Drude model is based on a friction dependent mobility coefficient, only.
For inclinations $\alpha<0.2\degree$ the diffusion behavior of the particles
has to be taken into account, too. Therefore, the interplay of the small drift
and of the diffusion behavior gives rise to a change of the mobility in
$x$-direction.
\begin{figure}[hbt]
  \begin{center}
    \includegraphics[width=\columnwidth,clip]{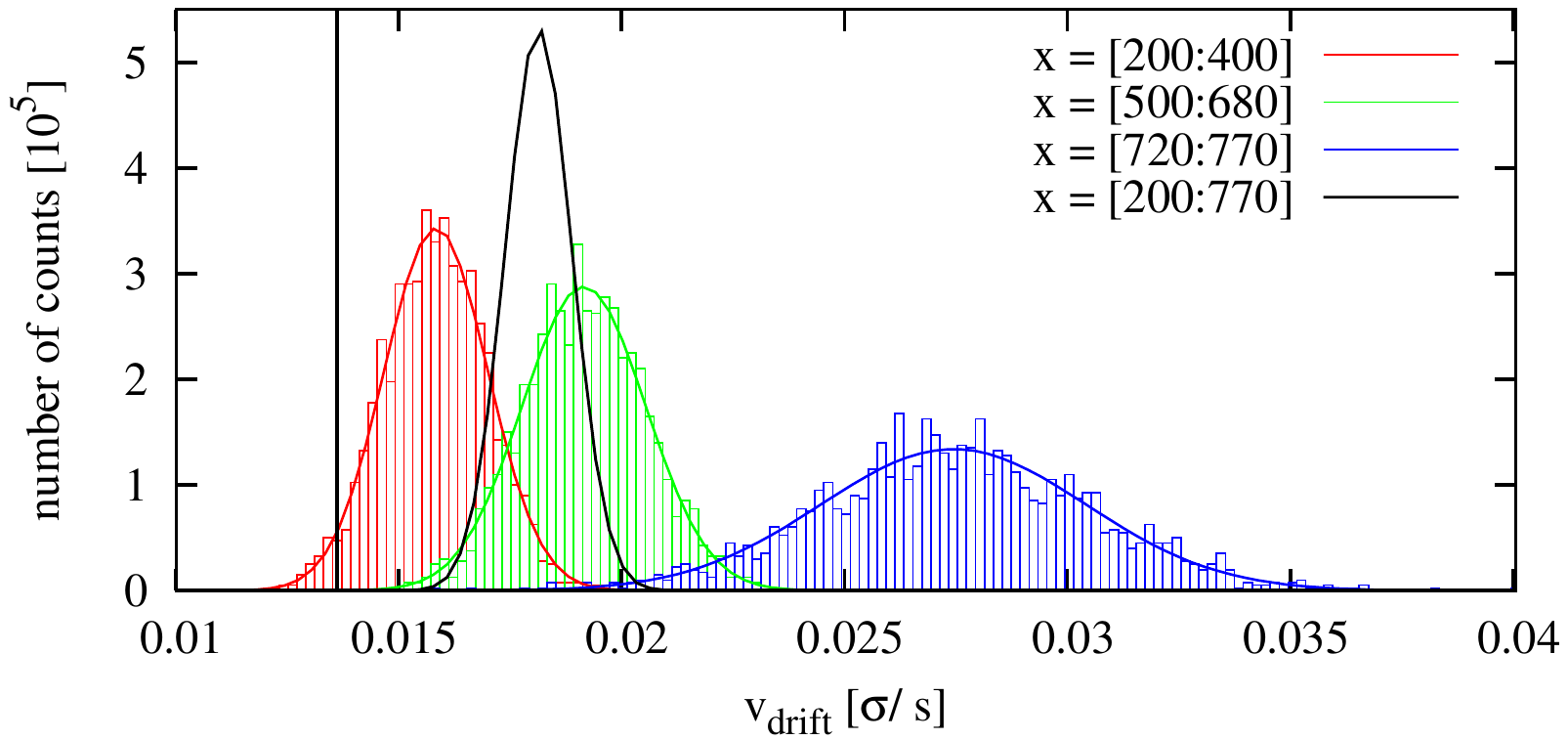}
  \end{center}
  \caption[Drift velocity histograms of the particles in different channel
  regions.]{Simulation: Drift velocity histograms of the particles in different channel
  regions. The vertical line marks the expected drift velocity for
  non-interacting particles according to the Drude model (cf.
  equation~\myref{eq:VDrude}).}
  \label{fig:DriftHist}
\end{figure}
\begin{figure}[t!hb]
  \begin{center}
    \includegraphics[width=\columnwidth,clip]{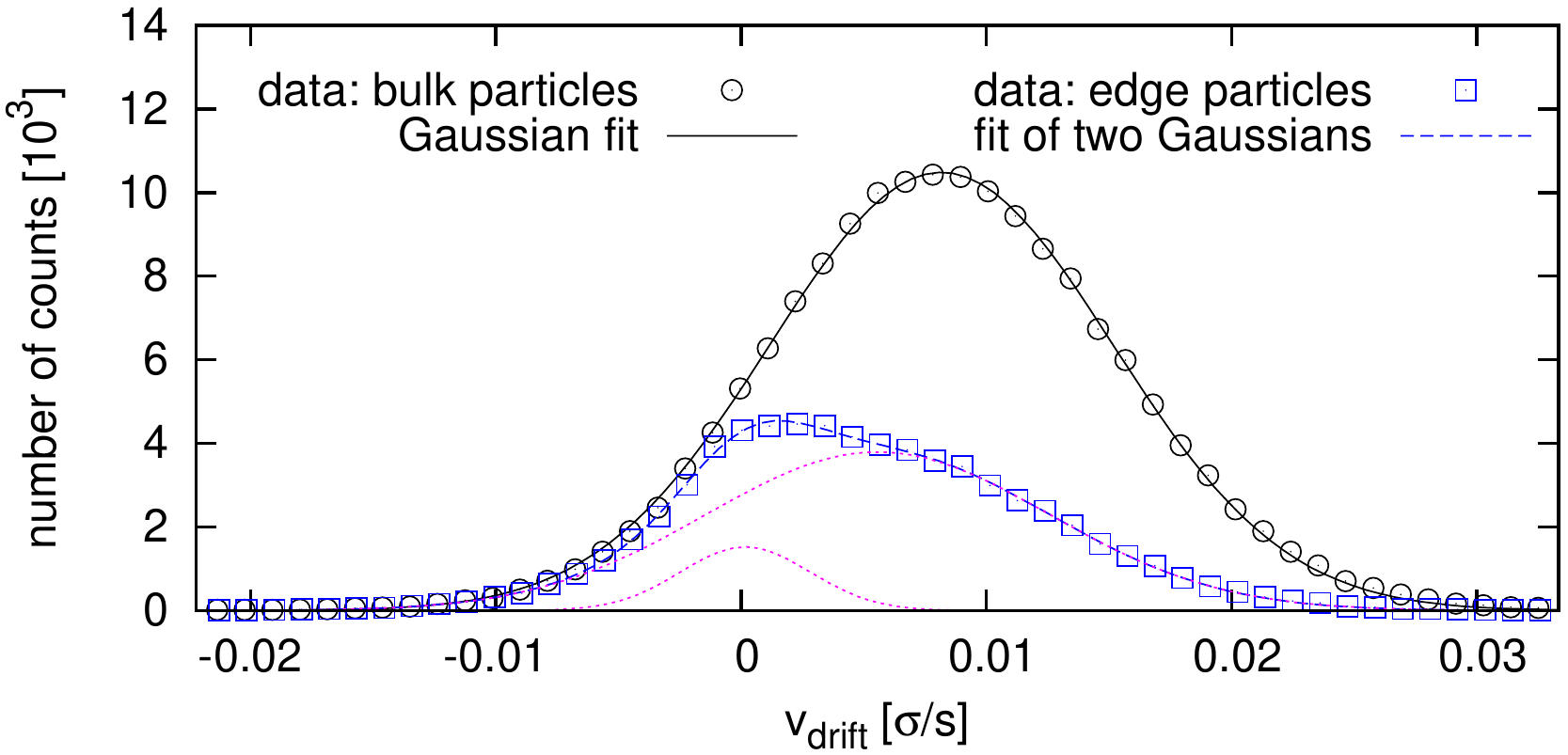}
  \end{center}
  \caption{Experiment: Drift velocity histograms of the particles in the full
  field of view. Shown are the histograms for the bulk and the edge particles,
  respectively. The data points of the bulk particles can be fitted well by a
  single Gaussian, whereas the edge particles need to be fitted by a
  superposition of two Gaussian functions. The results of the fits are
  indicated by the lines.}
\end{figure}

Particles moving along the channel get accelerated. This becomes obvious from
Fig.~\ref{fig:DriftHist}, where histograms of the drift velocity together
with Gaussian fits in different $x$-regions of the channel are
plotted. 
All hydrodynamic interactions are neglected. Generally, the
particle velocities $v_x$ in $x$-direction are normally distributed about the
average drift velocity.  For the angle $\alpha=0.2\degree$ the average drift
velocity is $\langle v_{\mathrm{drift}}\rangle \approx
\unit{0.081}{\micro\meter/\,\second}$. In the experiment an
inclination of $\alpha_{\mathrm{exp}} = 0.6^\circ$ was chosen, which results in
$\langle v_{\mathrm{drift}} \rangle \approx \unit{0.035}{\micro
\meter/\second}$. The velocities of the particles in the
experiment are lower as compared to simulations, possibly due to the
influence of hydrodynamic interactions. The comparison between edge
particles and bulk particles shows the effect of layer changes of the
particles on the velocities. As mentioned in
section~\ref{sec:EQPropertiesMicrochannel} dislocations are present
along the walls. These dislocations lead to an increased number of
layer changes for the edge particles. During the layer transition the
particles move in $y$-direction rather than $x$-direction; therefore
we expect to see a superposition of 2 velocity distributions in
$x$-direction, one centered around zero for particles changing layers
and one centered around the velocity of the particles in the edge
layer. Figure~\ref{fig:DriftHist} shows Gaussian fits for the bulk
particles and the edge particles. It is apparent that the velocity
distribution of the edge particles can be fit by a superposition of 2
Gaussian fits, resembling the particles changing lanes (around
\unit{0}{\micro\meter\per\second}) and moving
straight~(\unit{0.031}{\micro\meter/\second}). The different velocities
of bulk and edge layers are caused by the difference in density of the
layers. The behavior of bulk particles can be fit with a single
Gaussian, because the percentage of particles changing layers is much
lower than in the edge layer as the transition is confined to a small
region.

\subsubsection{Example Particle Trajectories}

\vspace*{-2.5ex}
\begin{figure}[htb]
  \begin{center}
    \includegraphics[width=\columnwidth,clip]{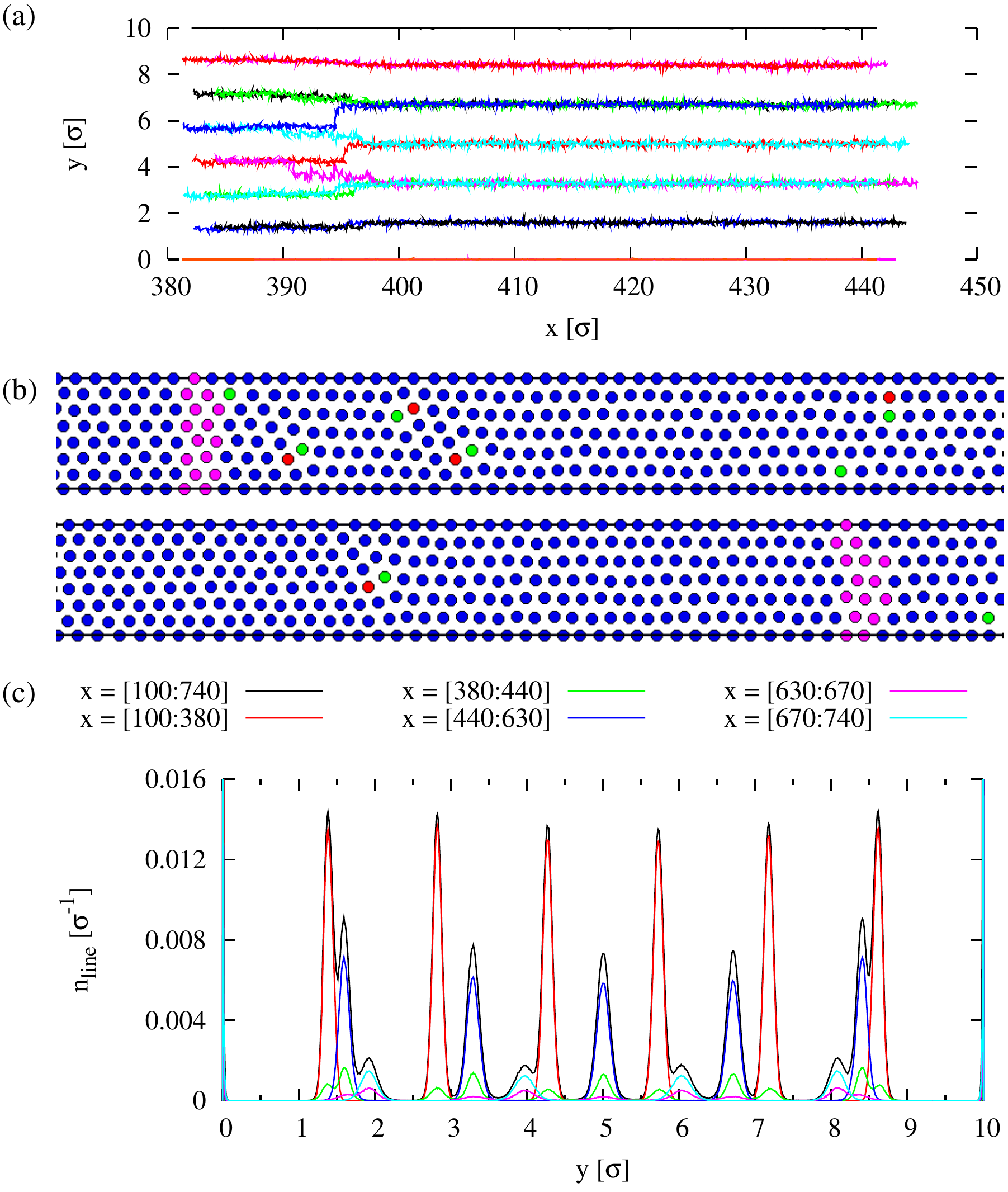}
  \end{center}
  \caption[Example particle trajectories of the particles crossing the layer
  reduction zone]{Simulation: {\bf(a)} Example particle trajectories which show the
  dynamical rearrangement of the particles crossing the layer reduction zone
  from 8 layers to 7 layers. Shown are the trajectories for the time interval
  $\Delta t=37.5\tau_B$ ($\equiv 5\cdot10^5$ BD steps). {\bf (b)} Corresponding
  snapshots of the starting and final configuration. The general color coding
  (see text) is used. Additionally, all the particles which trajectories are shown
  in (a) have been marked in magenta. {\bf (c)} Histograms of the $y$-positions
  within different $x$-regions evaluated for $1.5\cdot10^6$ BD steps. The peaks of
  the edge particles are truncated for clarity reason. The system parameters
  are identical to those of Fig.~\ref{fig:SnapshotFullChannel}.}
  \label{fig:ExampleTrajectories1}
\end{figure}
The particles flow across the layer reduction zone
(cf.~Fig.~\ref{fig:ExampleTrajectories1}), whereas the position of the layer
reduction zone almost remains unchanged. We show in
Fig.~\hyperref[fig:ExampleTrajectories1]{\ref{fig:ExampleTrajectories1}(a)}
representative particle trajectories for a selection of particles. These are
marked in magenta in the configuration snapshots
(Fig.~\hyperref[fig:ExampleTrajectories1]{\ref{fig:ExampleTrajectories1}(b)})
at their beginning and the final location of the trajectories. The trajectories
clearly show that we do not observe plug flow of a crystal, but rather a
dynamic behavior of particles moving in layers and adapting to the external
potential. The particles move a distance of about $60\sigma$ whereas the layer
transition stays located withing $x\in [390,\,400]\sigma$.

The edge particles are pushed against the ideal hard walls at $y=0\sigma$ and
$y=L_y=10\sigma$ by the repulsion of the inner particles of the channel. This
is the reason for their minimal fluctuations perpendicular to the flow
direction. The corresponding fluctuations of the non-edge layers are
significantly larger, and a small increase of the mobility in $y$-direction with
increasing wall separation is found. In the central region some particles
change very abruptly from one layer to another whereas others shift more
smoothly. The particles in the layers next to the edge layers only show a small
and smooth change in their $y$-position. In the regions with fixed number of
layers no particle transitions between layers are observed for our
simulation parameters. 

All particles are identical. In
Fig.~\hyperref[fig:ExampleTrajectories1]{\ref{fig:ExampleTrajectories1}(b)}
we just color coded
the particles according to the number of nearest neighbor particles they have.
Bulk particles with six nearest neighbors and all edge particles are marked
blue, whereas red particles have a fivefold symmetry and green particles have a
sevenfold symmetry of nearest neighbors. The actual number of nearest neighbors
is determined using a Delaunay triangulation. In the start configuration three
defect pairs (dislocations) are in the region of the layer transition form 8 to
7 layers, whereas in the final configuration the layer reduction position
is connected to a single dislocation. The slightly higher density of the edge
particles gives rise to the scattered green particles in the next edge layer.

For the same system we analyze the density profiles transverse to the walls
within several sub-regions along the channel. Therefore we evaluate
$1.5\cdot10^6$ BD steps corresponding to a time interval of $\Delta t \approx
122.5\tau_B$. The full density profile for $x\in[100,\,740]\sigma$ (black
curve) is a superposition of several profiles connected to distinct layering.
Highly ordered layer structures with sharply peaked density profiles occur for
eight layers in $x\in[100,\,380]\sigma$ (red curve), seven layers in
$x\in[440,\,630]\sigma$ (blue curve), and six layers in $x\in[670,\,740]\sigma$
(cyan curve). The $x$-regions in between are the layer transition regions.

\begin{figure}[h!tb]
  \begin{center}
    \includegraphics[width=\columnwidth,clip]{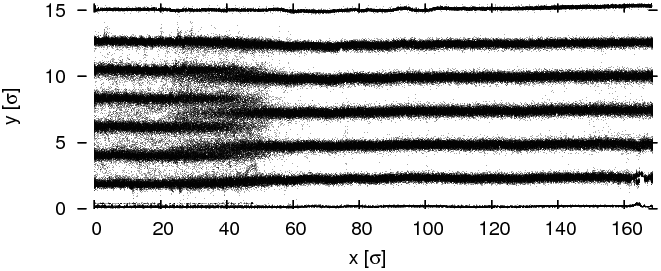}
    \includegraphics[width=\columnwidth,clip]{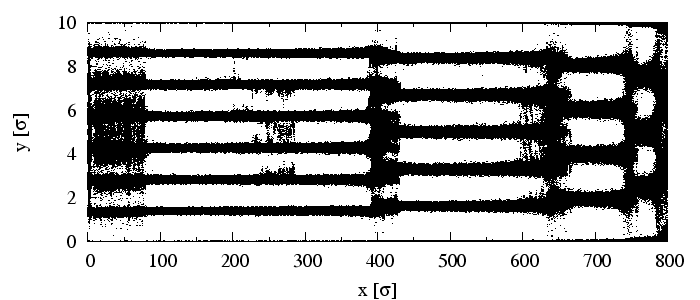}
  \end{center}
  \caption{Superimposed configuration snapshots: {\bf (Top)} of the experiment,
  {\bf (Bottom)} of the simulation for $1.5\cdot10^6$ BD steps, which
  corresponds to $\Delta t \approx 122.5\tau_B$. ($B=\unit{0.5}{\milli\tesla}$,
  $\Gamma=533.74$, $L_x=800\sigma$, $L_y=10\sigma$, $n=0.4\sigma^{-2}$,
  $\alpha=0.04\degree$)} \label{fig:AllTrajectories}
\end{figure}
In Fig.~\ref{fig:AllTrajectories} we explicitly plot the
superposition of 1911 video microscopy snapshots of the experimental
system and 
3000 configuration snapshots used for the density profile evaluation
above.
In the experiment the particles move on average $\langle \Delta x
\rangle \approx 670\sigma$. The layer reduction zone is confined in
the interval $x \in(5,50) \sigma$.  
In the simulation the particles have moved forward on
average the distance $\langle \Delta x \rangle \approx 202\sigma$, {\it i.e.}
more than a quarter of the channel length.  The layer transition positions
remain located within an interval of length $45\sigma$. The particles are
inserted at a random position in the region $x\in(0,80)\sigma$. Perturbations
of the configuration due to the random particle insertion heal after a few BD
steps. Therefore the configuration for $x>90\sigma$ is not influenced by this
particle re-insertion method.

\subsubsection{Defect Removal}
\begin{figure}[htb]
  \begin{center}
    \includegraphics[width=\columnwidth,clip]{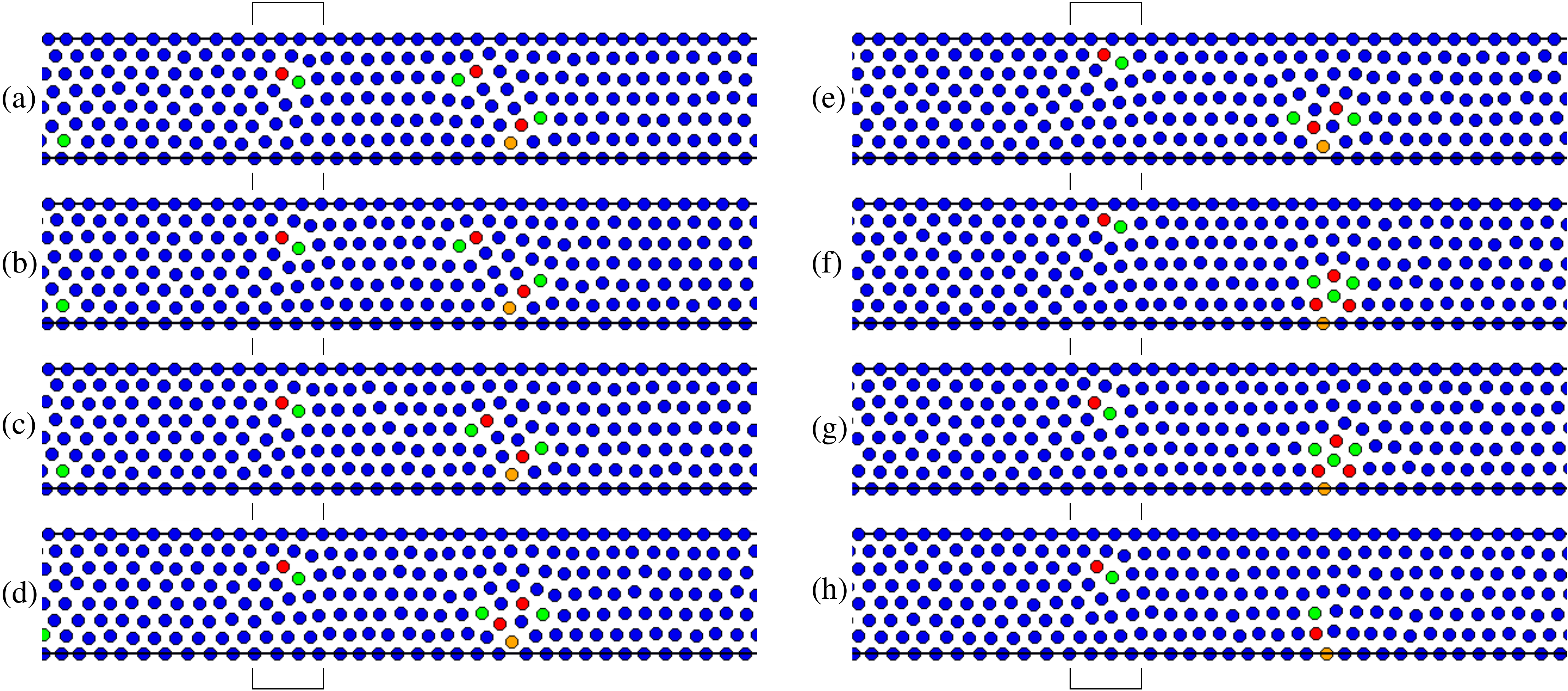}
  \end{center}
  \caption[Removal of a defect after the layer transition
  region]{Simulation: The sequence
  of configuration snapshots {\bf (a)} -- {\bf (h)} shows the process of a
  vanishing ``defect'' after the layer transition zone (marked by the black
  rectangle) due to the change of particle marked in orange into the edge
  layer. The snapshots have been taken every 500 BD time-steps, {\it i.e.}
  $\Delta t = 0.0375 \tau_B$.}
  \label{fig:DefectRemoval}
\end{figure}
Sometimes ``defects'' remain after the point of layer reduction, which vanish
on further flow. Here we call a defect a pair of particles having 7 and 5
neighbors respectively which disturb a given layer configuration. These can be
identified from dips they form in the local layer order parameter defined in
equation~\myref{eq:LayerOrderParameter} of the current configuration.
Generally, small density gradients along the channel give rise to a larger
number of defects than higher density gradients. This already is a hint on the
close connection of the occurrence of layer transitions to the local number
density. A defect can be neutralized by a particle changing into the edge
layer. Such a neutralization process of two defects is shown in the sequence of
configuration snapshots of Fig.~\ref{fig:DefectRemoval} taken every 500~BD
time-steps. The orange colored particle moves into the edge layer and thereby
removes the perturbation of the layered structure of 7 layers after the
position of the layer transition region marked by the black rectangle. In the
final snapshot~\hyperref[fig:DefectRemoval]{\ref{fig:DefectRemoval}(h)}
seven unperturbed layers remain. Recognize that again the $x$-position of the
layer reduction remains unchanged. 

\subsubsection{Diffusion Behavior}
\begin{figure}[htb]
  \begin{center}
    \includegraphics[width=\columnwidth,clip]{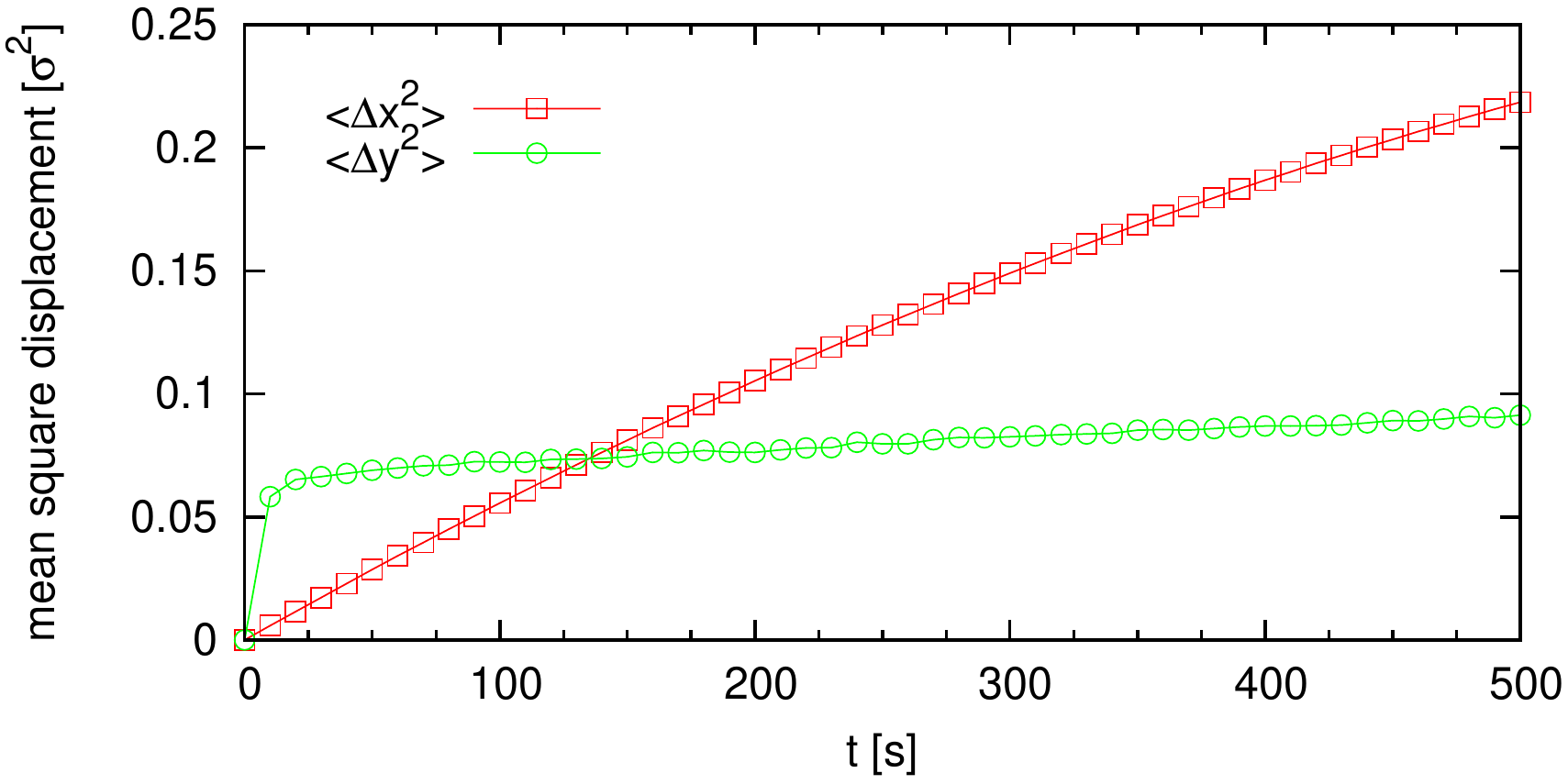}
  \end{center}
  \caption{Particle diffusion behavior in the experiment.}
  \label{fig:DiffExp}
\end{figure}
In section~\ref{sec:EQPropertiesMicrochannel} the diffusion behavior of
particles in the equilibrated channel was discussed. The question
arises whether a similar behavior can be observed for the driven
system. Figure~\ref{fig:DiffExp} shows the MSD in $x$- and
$y$-direction for one of the bulk layers in the experiment after the
driven motion of the system has been subtracted. In
$x$-direction we see a linear increase of the MSD at short times,
which starts to saturate at longer times. A clear transition to SFD,
as it was found in the simulation data for the equilibrated channel,
cannot be found. One reason can be given by the rather short times at
which the experimental MSD could be obtained. In $y$-direction we see
saturation at rather short time scales. This behavior is due to the
fact that the particles move in stable layers during most of the
experiment and are therefore restricted in their $y$-movement. A
similar behavior was shown in Fig.~\ref{fig:MSDVgl9and10} for a
channel width which induces stable layers. 

\subsection{Connection between the Layer Transition and the Density Gradient}
\begin{figure}[htb]
  \begin{center}
    \includegraphics[width=\columnwidth,clip]{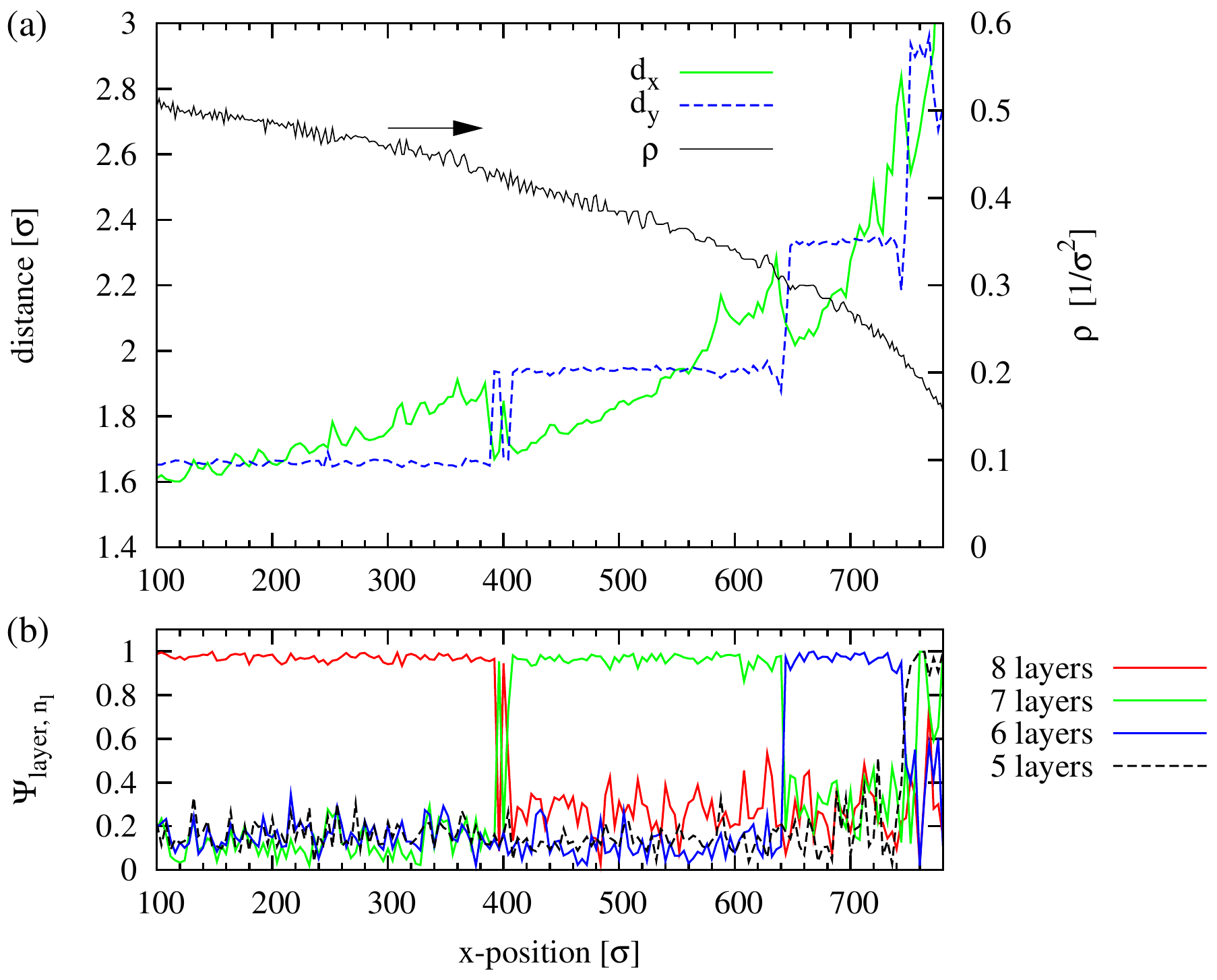}
  \end{center}
  \caption[Local lattice constants, particle density and layer order
  parameters.]{ {Simulation: \bf (a)} Local lattice constants $d_x$ and $d_y$ and local
  particle density $\rho$, {\bf (b)} Corresponding local
  layer order parameters $\Psi_{\mathrm{layer},\,n_l}$. The system parameters
  are: $L_x=800\sigma$, $L_y=10\sigma$, $n=0.4\sigma^{-2}$, and $\Gamma =
  533.74$.}
  \label{fig:SeperationDensity-jump310-0}
\end{figure}

The reduction of the number of  layers originates from a density gradient along
the channel. The local particle density $\rho(x)$ inside the channel is shown
in Fig.~\ref{fig:SeperationDensity-jump310-0} together with the local lattice
constants $d_x$ and $d_y$. The particle separations of neighboring particles in
$x$- and $y$-direction are used to calculate the local lattice constant $d$ of
the triangular lattice. Due to the density gradient along the channel, the
ordered structure is not in its equilibrium configuration at all points along
the channel. Thus the local lattice constant $d_x$, calculated from the
particle separations in $x$-direction, can deviate from the local lattice
constant $d_y$, calculated from the particle separations in $y$-direction and
multiplication with the factor $2/\sqrt{3}$. At the left end of the channel,
$d_x$ increases to larger values than $d_y$, indicating that the ordered
structure is stretched along the $x$-axis. At the position of the layer reduction
the system changes back to a situation, where $d_x$ is smaller than
$d_y$ by decreasing $d_x$ and increasing $d_y$ by about 20\%
simultaneously. These changes of separations compensate each other and result
in a continuous change in the local density at the position of the layer
reduction. The behavior of the system shows that the stretching of the ordered
structure before the layer reduction causes an instability towards decreasing
the number of layers. This decrease compresses the system along the
$x$-direction, but apparently lowers the total energy of the system.

\begin{figure}[htb]
  \begin{center}
    \includegraphics[width=\columnwidth,clip]{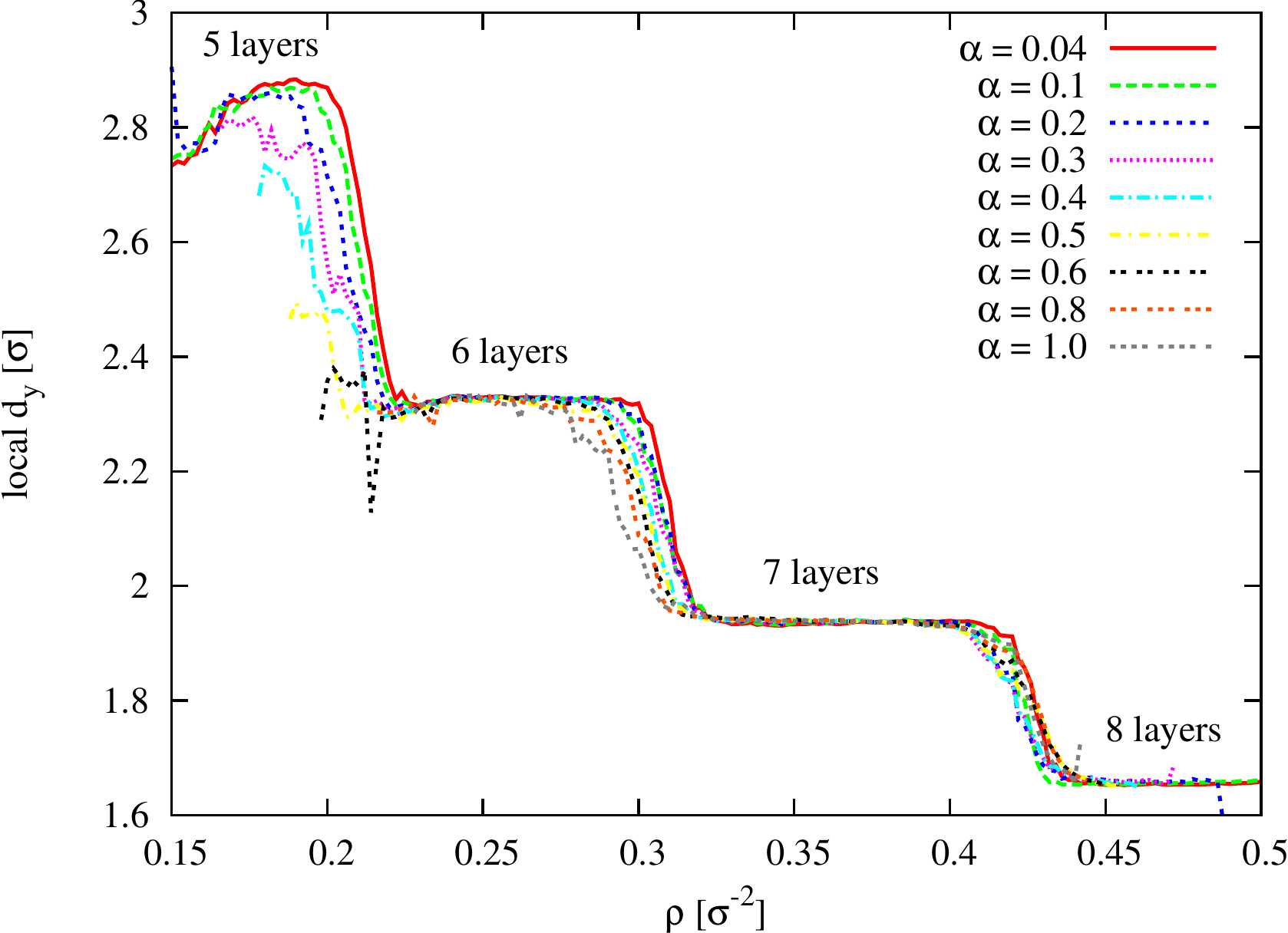}
  \end{center}
  \caption{Simulation: Local lattice constant $d_y$ as function of the local particle
  density $\rho$ for various inclinations $\alpha$.}
  \label{fig:DyRhoAlphaDependency}
\end{figure}
Layer transitions occur at almost identical values of the local particle
density for various inclinations as can be seen in
Fig.~\ref{fig:DyRhoAlphaDependency}. Here the local lattice constant $d_y(x)$
is plotted as a function of the local particle density $\rho(x)$. Transitions
from 8 to 7 layers occur when $\rho(x)$ becomes smaller than $0.42$,
transitions from $7\,\rightarrow\,6$ layers for $\rho(x)<0.3$, and transitions
from $6\,\rightarrow\,5$ layers for $\rho(x)<0.21$.

\subsection*{Static Stretching Analysis}
\begin{figure}[htb]
  \begin{center}
    \includegraphics[width=\columnwidth,clip]{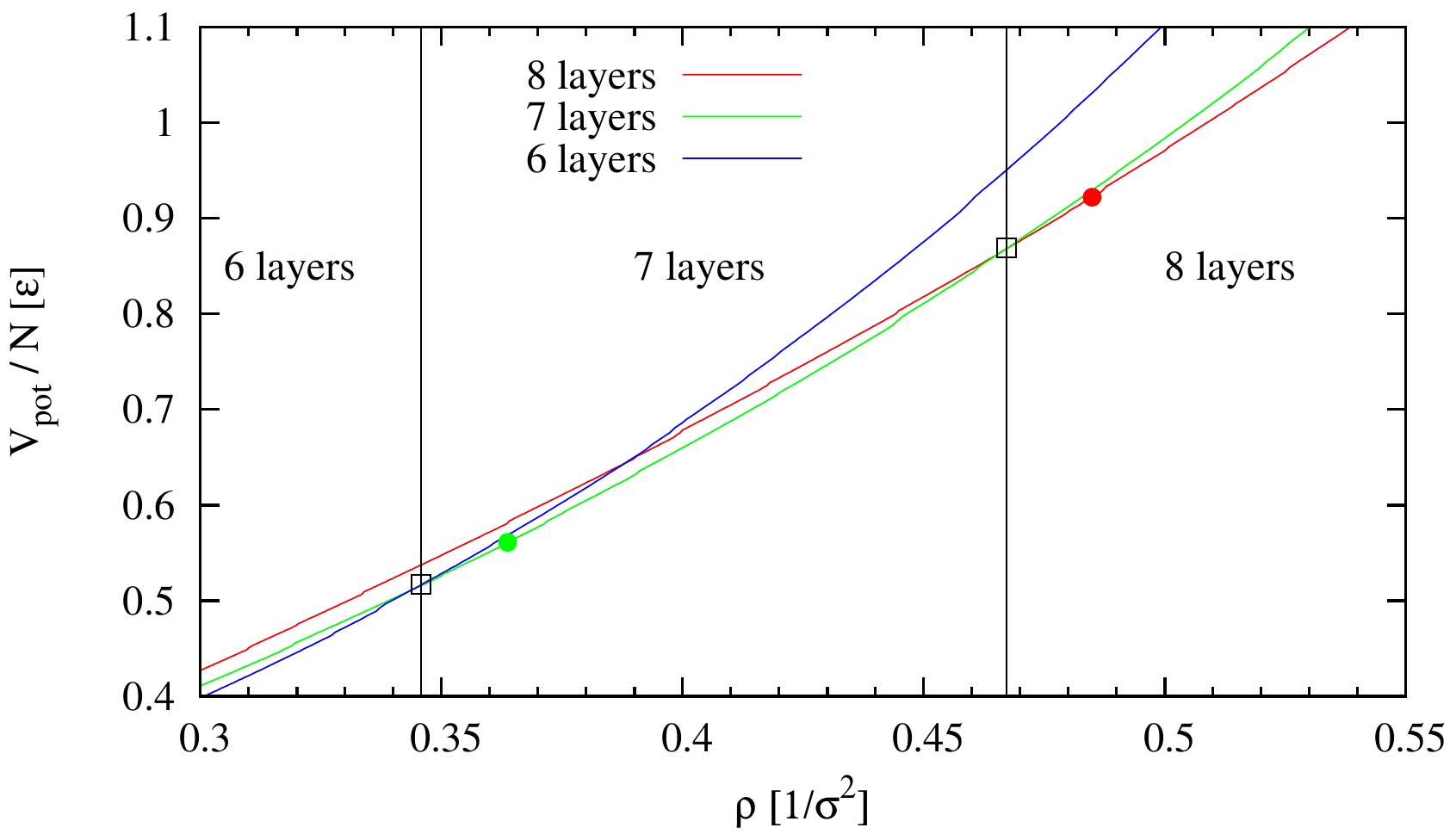}
  \end{center}
  \caption[Result of the stretching analysis of static channel configurations
  of a channel with the width $L_y=10$ and dipolar pair interaction]{Result of
  the stretching analysis of static channel configurations of a channel with
  the width $L_y=10$ and dipolar pair interaction: Shown are the potential
  energies per particle of different layer configurations as function of the
  particle density $\rho$.}
  \label{fig:LayerTransitionDensity}
\end{figure}
The scenario above can be qualitatively confirmed by the following rough
estimation: Starting from an ideal triangular configuration with a given number
of  layers ($n_l$) in a channel of fixed width, we calculate the potential
energy per particle for different particle densities by scaling the channel
length of the static configuration only. Plots of these energies per particle
for different values of $n_l$ as function of the particle density $\rho$ are
shown in Fig.~\ref{fig:LayerTransitionDensity}.  The intersection points, which
are determined from linear approximation of both curves in the region of
intersection, serve as a rough estimate of the density at the layer transition
point. For the given system the values for the transition $8 \rightarrow 7$
layers are: $\rho_{8 \rightarrow 7} \approx 0.467\sigma^{-2}$, and for the
transition $7 \rightarrow 6$ layers: $\rho_{7 \rightarrow 6} \approx
0.345\sigma^{-2}$. The full circles mark the perfect triangular lattices
configurations of the respective number of layers. So, we can conclude that a
given layer structure is stable for up to slightly overstretched perfect
triangular configurations.

They show clear intersection points,
indicating that for a stretched configuration with $n_l$  layers in $x$-direction
it can become energetically more favorable to switch to a compressed
configuration with $(n_l-1)$  layers.

\subsection*{Equilibrium Configurations for Confinement with Non-Parallel Walls}
\begin{figure}[htb]
  \begin{center}
    \includegraphics[angle=270,width=\columnwidth,clip]{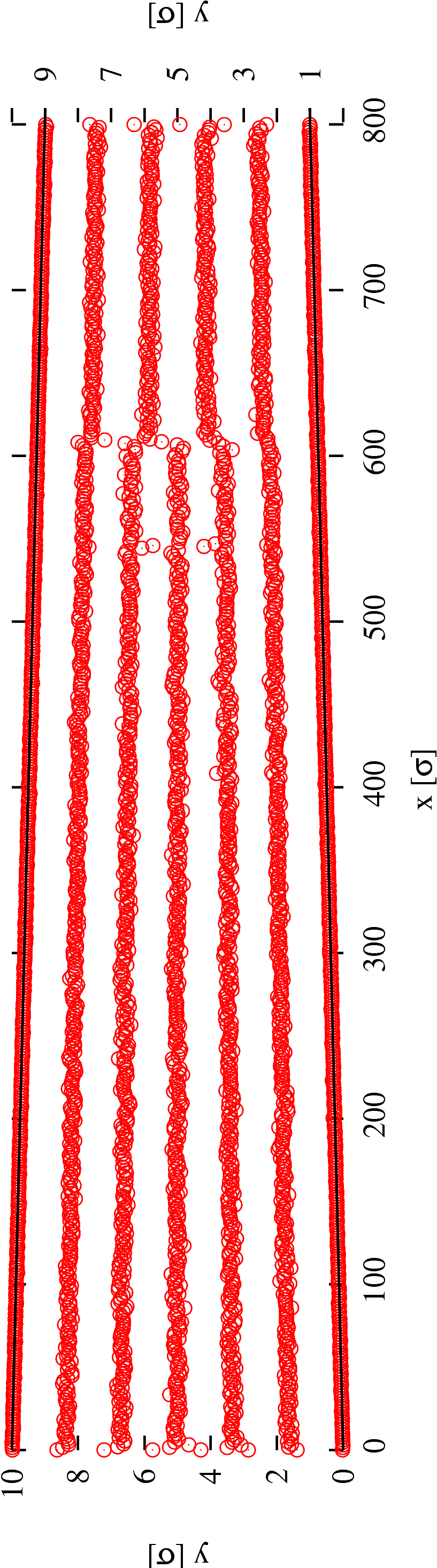}
  \end{center}
  \caption[Equilibrated configuration snapshot of a funnel
  geometry.]{Simulation: Equilibrated configuration snapshot of a funnel geometry with
  opening angle $\alpha_{\mathrm{Hopper}} = 0.143\degree$. All walls are
  modeled as hard walls. No driving force is applied and the particles
  interaction is dipolar ($\Gamma=625.125$).}
  \label{fig:SnapshotHopper}
\end{figure}
Also equilibrium BD simulations, {\it i.e.} simulations with no external
driving force ($\tilde{\mathbf{F}}_{i}^ {\mathrm{ext}} =0$), of closed channels
with non-parallel walls result in a density gradient in the direction of
decreasing channel width. Here, confinement induced arrangement of the
particles into different number of layers takes place.  The particles just
fluctuate about their equilibrium positions. A snapshot of such an equilibrium
configuration is shown in Fig.~\ref{fig:SnapshotHopper} where the confining
funnel has the small opening angle $\alpha_{\mathrm{Funnel}} = 0.143\degree$,
{\it i.e.} over the full channel length of $L_x=800\sigma$ the channel width
decreases by $\Delta L_y=2\sigma$. This kind of layer transition is a purely
geometrical effect, whereas in the case of parallel walls and a constant
longitudinal driving field the occurrence of the density gradient has a
dynamical origin. In both cases the number of layers which form depends on the
value of the local particle density~$\rho(x)$.

\subsection{Comparison with the Experiment}

\begin{figure}[htb]
  \begin{center}
    \includegraphics[width=\columnwidth,clip]{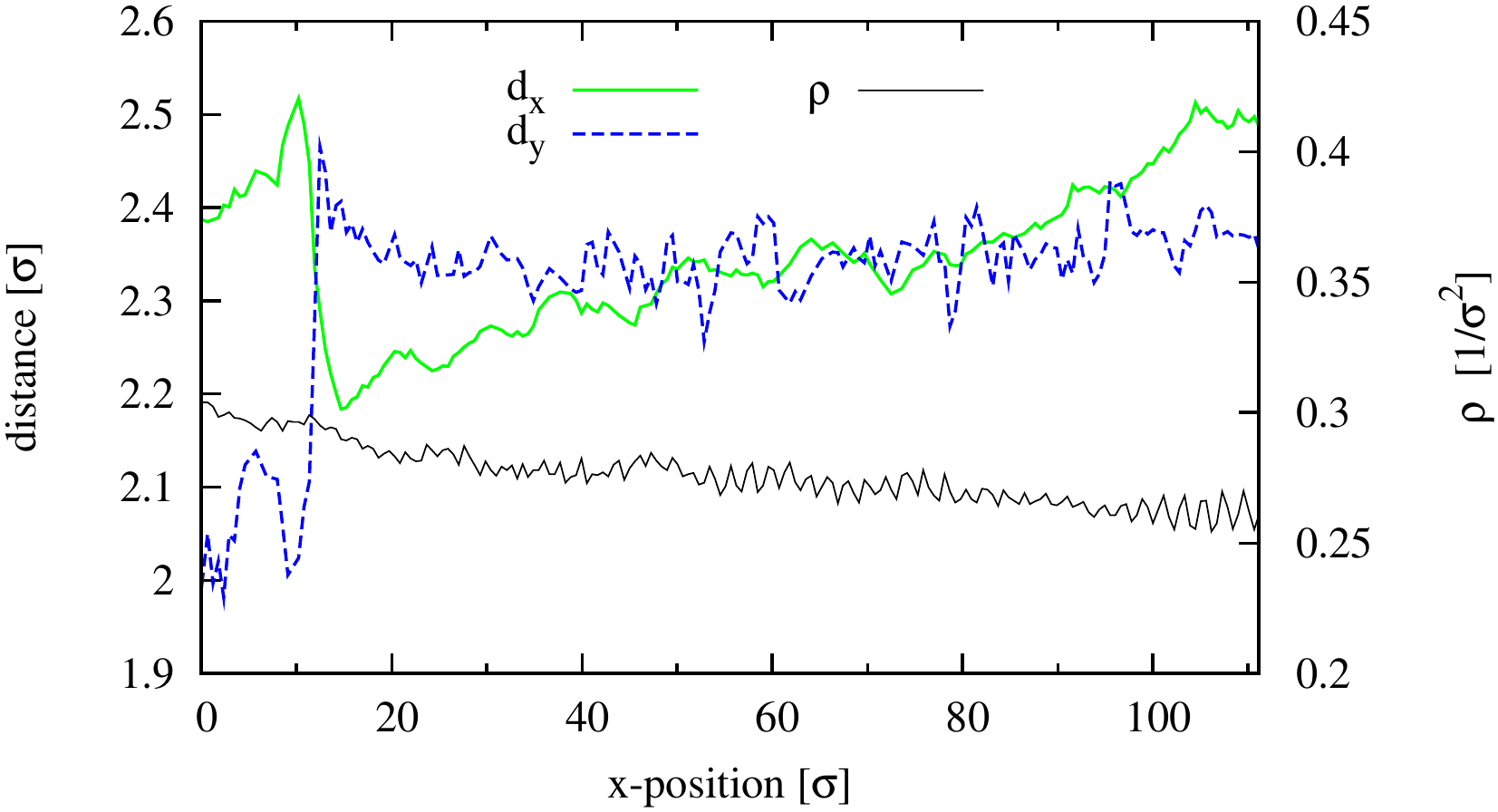}
    \end{center}
    \caption[Local lattice constants $d_x$ and $d_y$ and local particle density
    in the experiment.]{Experiment: Local lattice constants $d_x$ and $d_y$ and
    local particle density. The
    results are obtained for the systems of which is shown in
    Fig.~\hyperref[fig:channelsnapexperiment]{\ref{fig:channelsnapexperiment}(a)}.}
    \label{fig:DistanceDensityExp}
\end{figure}

The experimental result of the density gradient as well as the interparticle
distances are shown in Fig.~\ref{fig:DistanceDensityExp}. The behavior closely
resembles the behavior of the simulated system
(cf.~Fig.~\hyperref[fig:SeperationDensity-jump310-0]{\ref{fig:SeperationDensity-jump310-0}(a)}).
The distance in $x$-direction, $d_x$, is continuously stretched while the
distance in $y$-direction increases in a sharp step at the position of the
layer reduction. The density decreases monotonously along the direction of
motion of the particles by about 20\%. 

\begin{figure}[h!]
  \begin{center}
	\includegraphics[width=\columnwidth]{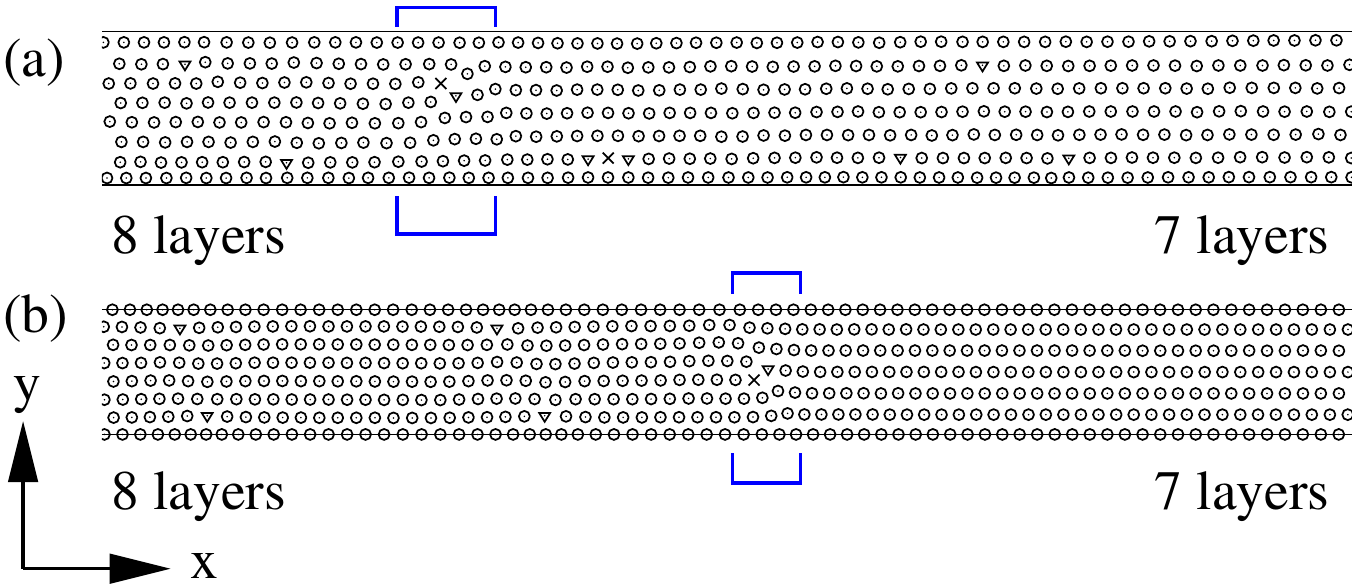}
  \end{center}
  \vspace*{-3ex}
  \caption[Snapshots of defect configurations obtained from a Delaunay
  triangulation of the particles moving in the channel.]{Snapshots of defect
  configurations obtained from a Delaunay triangulation of the particles moving
  in the channel. The particles are coded according to the number of their
  nearest neighbors. Open circles mark the bulk particles with 6 nearest
  neighbors and the edge particles, symbol $\times$ corresponds to a fivefold
  symmetry, and symbol $\triangledown$ to a sevenfold symmetry. {\bf (a)}
  Experiment: In order to minimize the effects of fluctuations on a short time
  scale, 50 images have been averaged. {\bf (b)} BD simulation for a channel
  with parallel walls.}
  \label{fig:voronoi}
\end{figure}
In Fig.~\ref{fig:voronoi} snapshots of non-equilibrium defect configurations
are shown both for the experiment (a) and for the simulation (b). They reveal
that the system is nearly triangular left and right of the point of  layer
reduction. The change is marked by a single defect only.  The number of  layers
is reduced one by one. Reductions of two or more  layers have not been observed
in experiment or in simulation. Naturally, this reduction produces a defect at
the point of the transition. Since the position of the layer reduction  is
mainly determined by the density gradient, its location remains stable with
time on average. A more detailed analysis reveals, however, that the transition
point oscillates back and forth around this average position. At the transition
the driven particles in the bulk  layers have to change the  layer, causing the
transition to move a little bit in direction of the flow. A particle changing
into the edge  layer can neutralize the defect of the transition locally. This
causes a reconfiguration of the ordered structure, which in turn gives rise to
repositioning of the  layer reduction zone back to a region of higher density.

\subsection{Oscillatory Behavior of the Layer Transition}

\begin{figure}[hbt]
  \begin{center}
    \includegraphics[width=\columnwidth,clip]{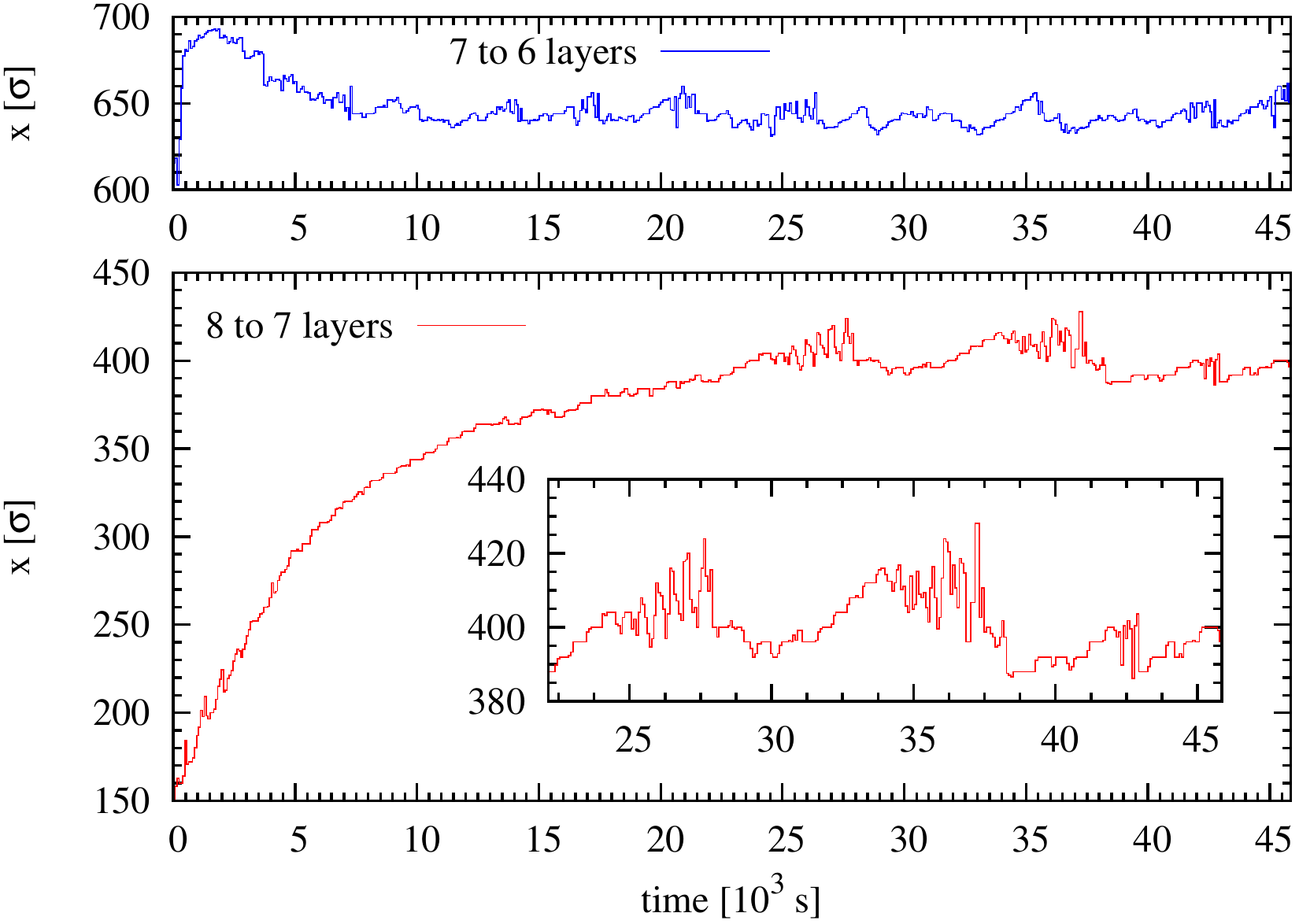}
  \end{center}
  \caption[Movement of the $x$-position of layer transition for the transitions
  $8\rightarrow 7$ layers and $7\rightarrow 6$ layers.]{Simulation: Movement of the
  $x$-position of layer transition for the transitions $8\rightarrow 7$ layers
  and $7\rightarrow 6$ layers. The system parameters are identical with those
  of Fig.~\ref{fig:AllTrajectories}.}
  \label{fig:LayerTransitionMove}
\end{figure}
\begin{figure}[hbt]
  \begin{center}
    \includegraphics[width=\columnwidth,clip]{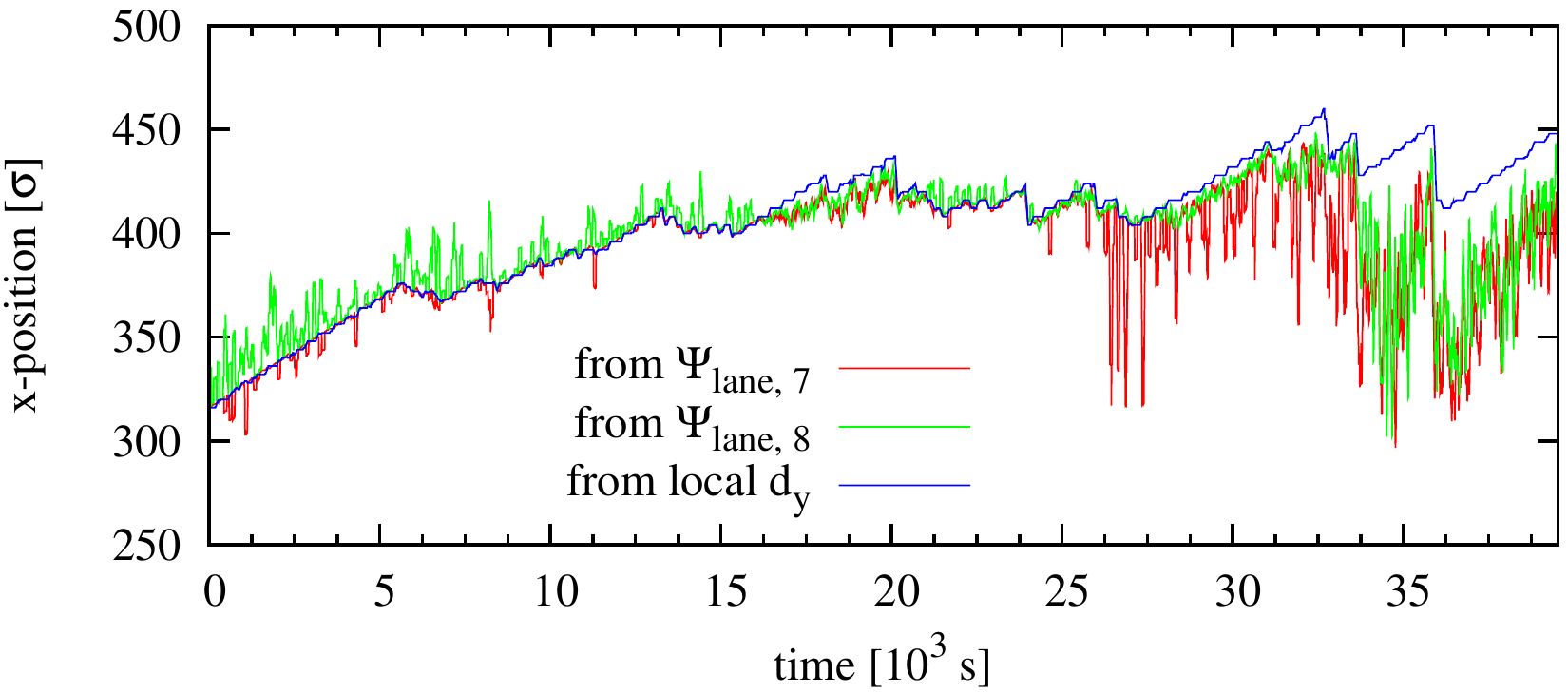}
  \end{center}
  \caption{Simulation: Movement of the $x$-position of layer transition for the transitions
  $8\rightarrow 7$ for an inclination $\alpha=0.2\degree$ and otherwise identical
  parameters as in the previous figure.}
  \label{fig:LayerTransitionMove2}
\end{figure}

There are various ways of numerically localizing the position of the layer
transition. One can either make use of the clear discontinuity of the local
layer order parameters $\Psi_{\mathrm{layer},\,n_l}(x,t)$ (cf.
equation~\ref{eq:LayerOrderParameter}) appropriate for the transition from
$n_l$ to $n_l-1$ layers, or of the location of the discontinuity of the local
lattice constant $d_y(x,t)$.
The local orientational order parameter~$\Psi_6$, which is often used for 2D
systems~\cite{Zahn99}, is not so significant for this system, as it is very
sensitive to any perturbation of the sixfold symmetry. The first three methods
have been used to study the position of the transition from 8 to 7 layers. The
result is given in Fig.~\ref{fig:LayerTransitionMove}. A comparison of all
three methods mentioned is given in Fig.~\ref{fig:LayerTransitionMove2} for
the transition from 8 to 7 layers. The local order parameters are calculated
within bins of size $l_x=2\sigma$ in flow direction limiting the
$x$-resolution. As can be seen all methods give similar results, but special
care needs to be taken in the presence of defects, which occur close to the
layer transition for $t>120\tau_B=\unit{25.2\cdot10^3}{\second}$.

\begin{figure}[hbt]
  \begin{center}
    \includegraphics[width=\columnwidth,clip]{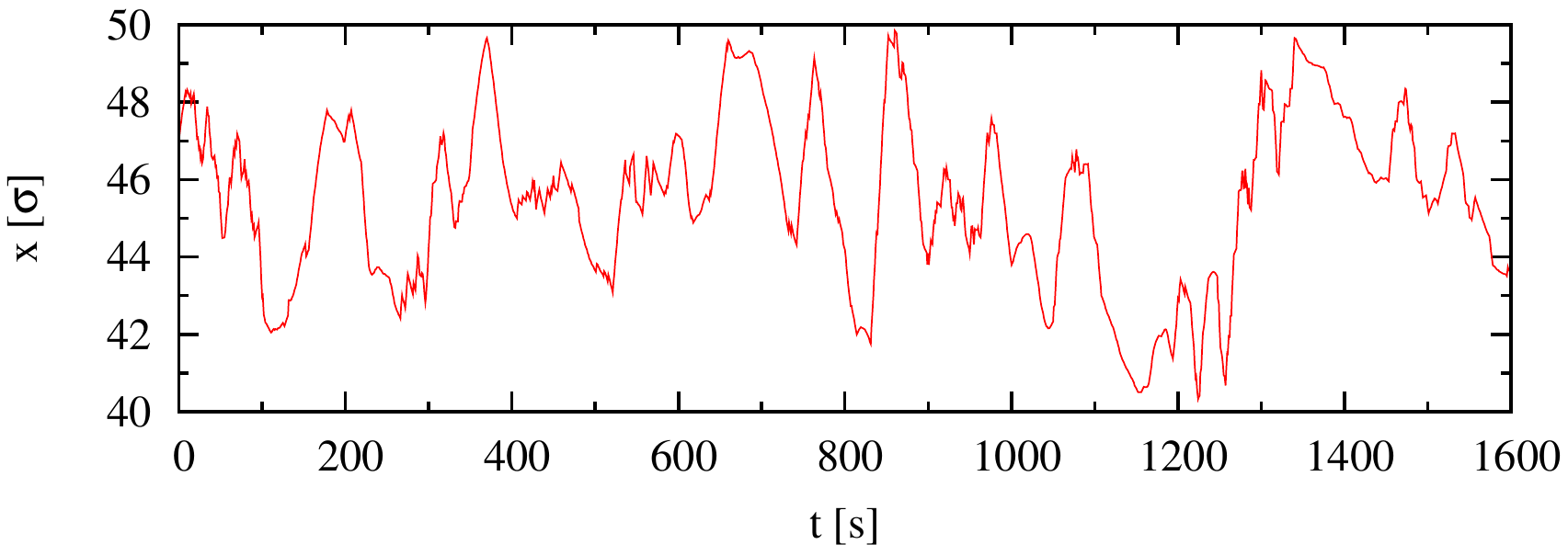}
  \end{center}
  \caption{Experiment: Movement of the $x$-position of layer transition for the transitions
  $8\rightarrow 7$.}
  \label{fig:LayerTransitionMoveExp}
\end{figure}

In the non-equilibrium steady state situation the position of the layer
reduction zone oscillates about a certain $x$-position. This can also
be seen in the experimental data, as shown in
Fig.~\ref{fig:LayerTransitionMoveExp}~(evaluated from the
discontinuity in $d_y$).

% comparison to the experiment

\subsection{Influence of the Particle Interaction Range}

In order to study the influence of the particle interaction range, we
implemented the screened Coulomb (YHC) pair interaction potential 
\begin{equation}
      V_{ij}(r_{ij}) = \left\{ \begin{array}{l@{\quad:\quad}l} \infty &
	r_{ij}<\sigma \\
	V_0\,\,\displaystyle \frac{\exp \left( -\kappa_D \left( r_{ij} - \sigma
	\right) \right)}{r_{ij}} & \sigma \le r_{ij} < r_{\mathrm{cut}} \\
	0 & r_{ij} \ge r_{\mathrm{cut}}
      \end{array} \right.
      \label{eq:YHCInteractionPot}
    \end{equation}
    with the inverse Debye screening length $\kappa_D$ which interpolates the
    potential between the hard core case (for $\kappa_D\rightarrow \infty$) and
    the unscreened Coulomb potential (for $\kappa_D=0$). $V_0$ is the value of
    the pair potential at contact which can be written as
    \begin{equation*}
      \beta V_0 = \frac{Z^2}{\left( 1+\kappa_D\sigma/\,2
      \right)^2}\,\frac{\lambda_B}{\sigma}
      \label{eq:ContactValue}
    \end{equation*}
    where $Z$ is the charge of the colloids and $\lambda_B = e^2/
    (4\pi\epsilon_0 \epsilon_s k_B T)$ is the so-called {\it Bjerrum length} of
    the solvent with permittivity $\epsilon_s$.

\begin{figure}[htb]
  \begin{center}
    \includegraphics[width=\columnwidth,clip]{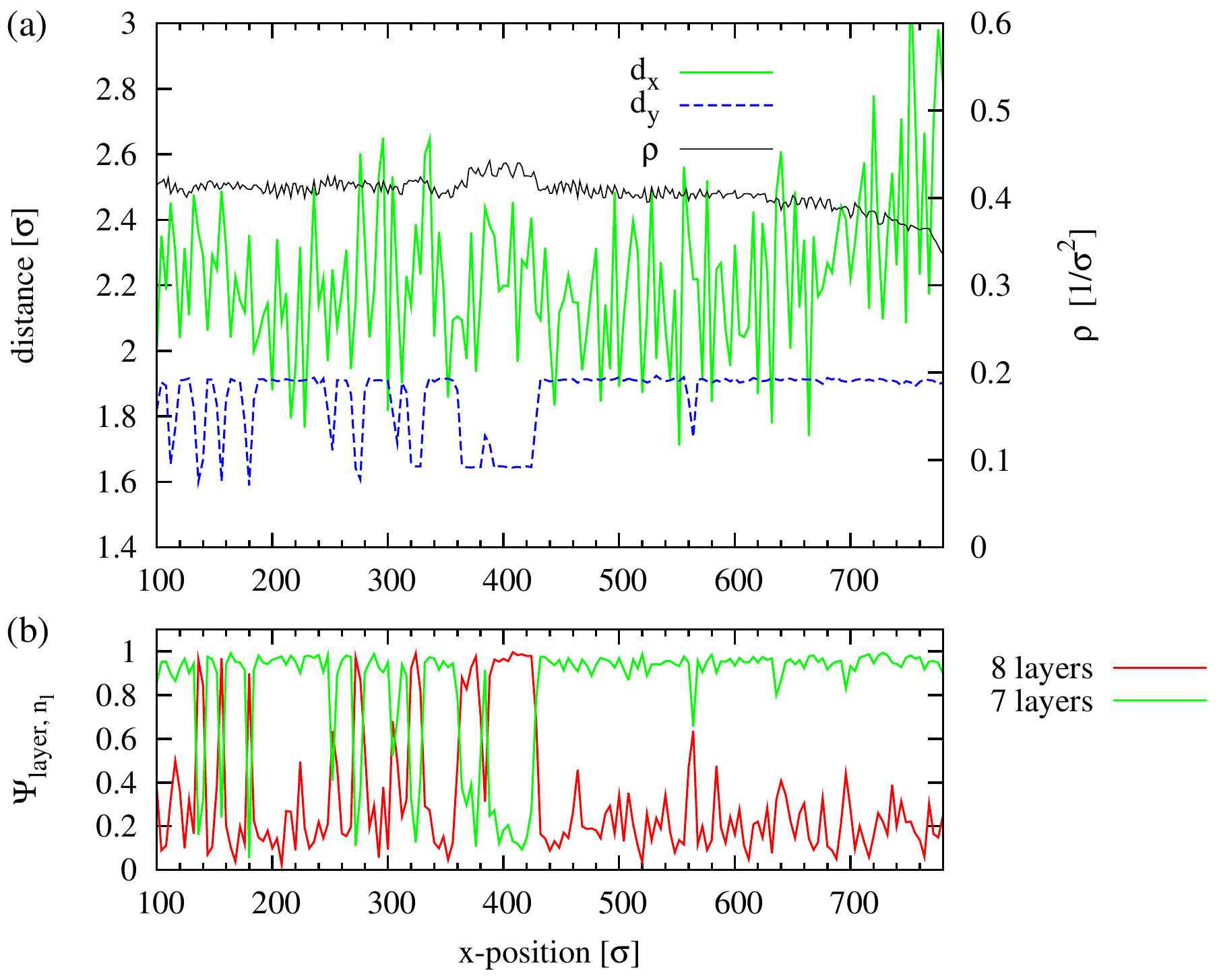}
  \end{center}
  \vspace*{-1.5ex}
  \caption[Local lattice constants, particle density and layer order parameters
  for a system with screened Coulomb interaction.]{Simulation: {\bf (a)} Local lattice
  constants $d_x$ and $d_y$ and local particle density $\rho$ in the BD
  simulation of a  system with screened Coulomb interaction. {\bf (b)}
  Corresponding local layer order parameters $\Psi_{\mathrm{layer},\,n_l}$. The
  system parameters are: $L_x=800\sigma$, $L_y=10\sigma$, $n=0.4\sigma^{-2}$,
  $\beta V_0=400$, $\kappa_D=4.0\sigma^{-1}$, $\Gamma_{\mathrm{YHC}} =
  448.4$, and $\alpha=0.2\degree$.}
  \label{fig:SeperationDensity-jump362-0-YHC}
\end{figure}
Figure~\ref{fig:SeperationDensity-jump362-0-YHC} is the analogous plot to
Fig.~\ref{fig:LayerTransitionDensity} for a system of YHC-particles with the
contact value $\beta V_0=400$ and $\kappa_D=4.0\sigma^{-1}$. Under these
simulation conditions no layer-transition as for the dipolar system is found.
For $x>450\sigma$ the particles are ordered in 7 layers, but for smaller values
only a few islands of particles arranged in layers can be identified from the
local order parameters along the channel in
Fig.~\hyperref[fig:SeperationDensity-jump362-0-YHC]{\ref{fig:SeperationDensity-jump362-0-YHC}(b)}.
The interaction range of a YHC system with $\kappa_D=4.0\sigma^{-1}$ is much
smaller than for the dipolar system, because of the stronger decay of the pair
potential. This decay is also the reason for the large fluctuations of the
local lattice constant $d_x$ in
Fig.~\hyperref[fig:SeperationDensity-jump362-0-YHC]{\ref{fig:SeperationDensity-jump362-0-YHC}(a)}.
A density gradient can not form along the channel, and so no layer transition
is found. The particles need to be strongly coupled with their neighboring
particles to form a density gradient, {\it i.e.} the pair interaction range has
to be at minimum of the order of the average particle spacing.

\section{Alternative Boundary Conditions in Flow Direction}

\begin{figure}[b!ht]
  \begin{center}
    \includegraphics[width=\columnwidth,clip]{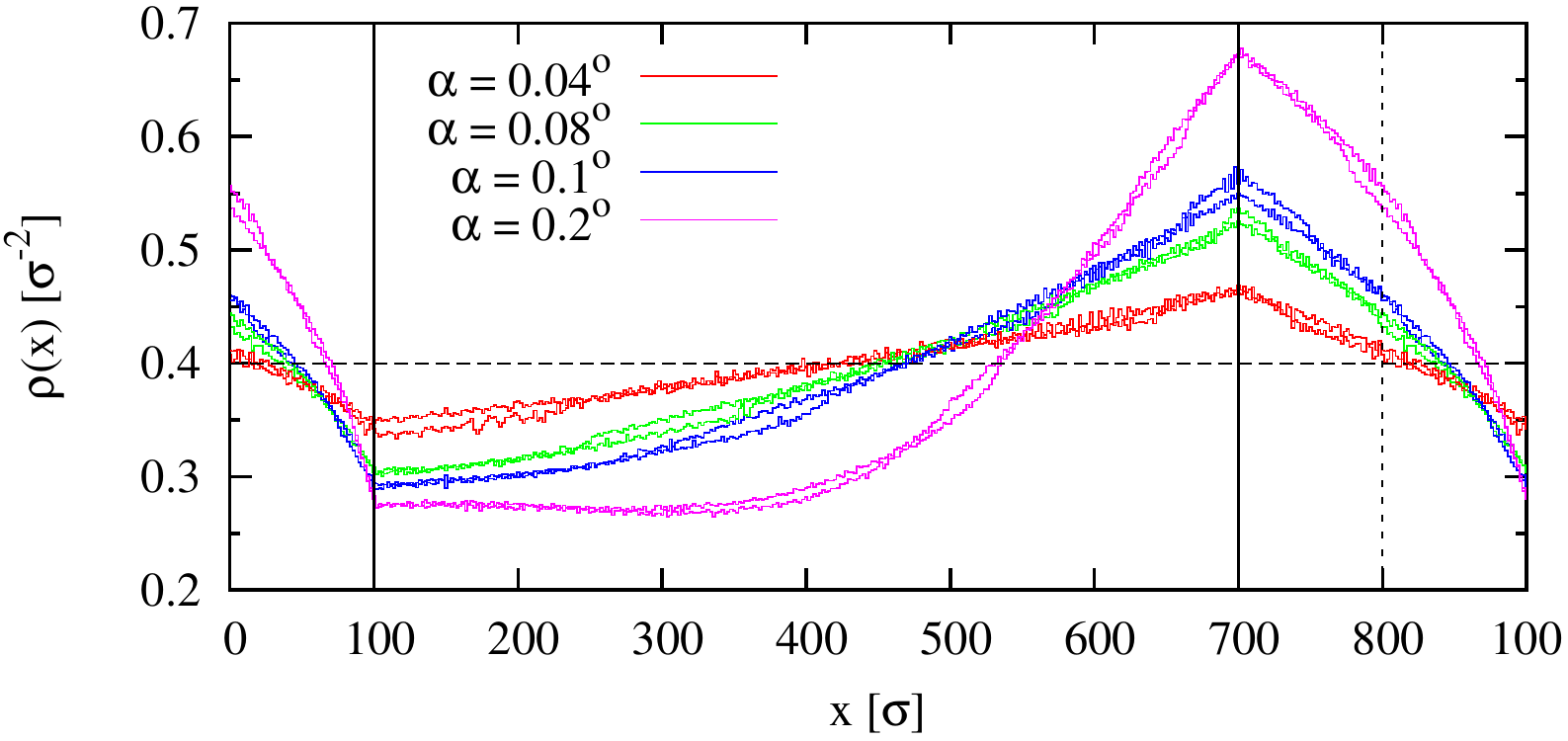}
  \end{center}
  \caption[Density profiles for a selection of inclinations $\alpha$ of a
  system with the driving force applied only within a partition of the
  channel.]{Simulation: Density profiles for a selection of inclinations $\alpha$ of a
  system with the inclination ({\it i.e.} the driving force) applied only
  within the region $x\in[100,\,700]\sigma$. The system is periodic in
  $x$-direction.  Shown are histograms obtained by evaluation of 1000
  configurations of the system having reached a stationary non-equilibrium
  situation (after $\approx 2\cdot10^6$ BD time steps). The applied magnetic
  field strength is $B=\unit{0.25}{\milli\tesla}$, $\Gamma=133.4$, and the
  overall particle density $n=0.4\sigma^{-2}$. For better clarity, we replicate
  the interval $x\in[0,\,100]\sigma$ again on the right hand side of the
  diagram.}
  \label{fig:DensityVglFlowRegion}
\end{figure}

The connection of the channel to the two reservoirs has great influence on the
characteristics of the stationary non-equilibrium density profile along the
channel. Therefore, we performed simulations with an alternative boundary
condition, where the constant external driving force only acts within the
interval $x\in[100,\,700]\sigma$ and a periodic boundary condition is applied
in $x$-direction.
Figure~\ref{fig:DensityVglFlowRegion} shows the resulting stationary
non-equilibrium density profiles after $3\cdot 10^6$ BD time steps for a
selection of inclinations $\alpha$ of a dipolar system. For each inclination
two curves are plotted which correspond to the two channel widths $L_y=8\sigma$
and $L_y=10\sigma$. Obviously, the steady-state density profile along the
channel does not depend on the channel width.

Comparison of these density profiles with those of
Fig.~\ref{fig:densitygradient} highlights the strong influence of the
different realization of the reservoirs. All simulations are started from a
homogeneous particle distribution of local density $\rho=0.4\sigma^{-2}$.
Instead of a density decrease we find in Fig.~\ref{fig:DensityVglFlowRegion}
a buildup of the local density occurring due to the filling of the reservoir at
the channel end. This corresponds to the experimental situation, where the
reservoir at the channel end is filled. For the small inclination
$\alpha=0.04\degree$ a linearly increasing density profile is obtained within
the channel region. Higher inclinations lead to deviation from such a linear
profile. For $\alpha=0.2\degree$ a constant profile with local density
$\rho\approx0.275\sigma^{-2}$ in $x\in[100,\,400]\sigma$ is followed by a sharp
increase of the local density up to $\rho = 0.67\sigma^{-2}$ at the channel end
at $x=700\sigma$.

In the stationary non-equilibrium state the density profile in the reservoirs
can be approximated by a linear gradient. The net flux $J$ in the reservoirs
fulfills Fick's law
\begin{equation}
  J=\frac{k_B T}{2l_0}(\rho_1-\rho_0)
  \label{eq:FickLaw}
\end{equation}
where $\rho_0\equiv\rho(x=100\sigma)$ and $\rho_1\equiv\rho(x=700\sigma)$ are
the local number densities at the channel beginning and end respectively. Due
to the periodic boundary condition in $x$-direction this is equal to the net
flux in the channel region $x\in[100,\,700)\sigma$. Therefore, $J$ may be
approximated by the slope of the linear density profiles in the two reservoir
regions.
\begin{figure}[hbt]
  \begin{center}
    \includegraphics[width=\columnwidth,clip]{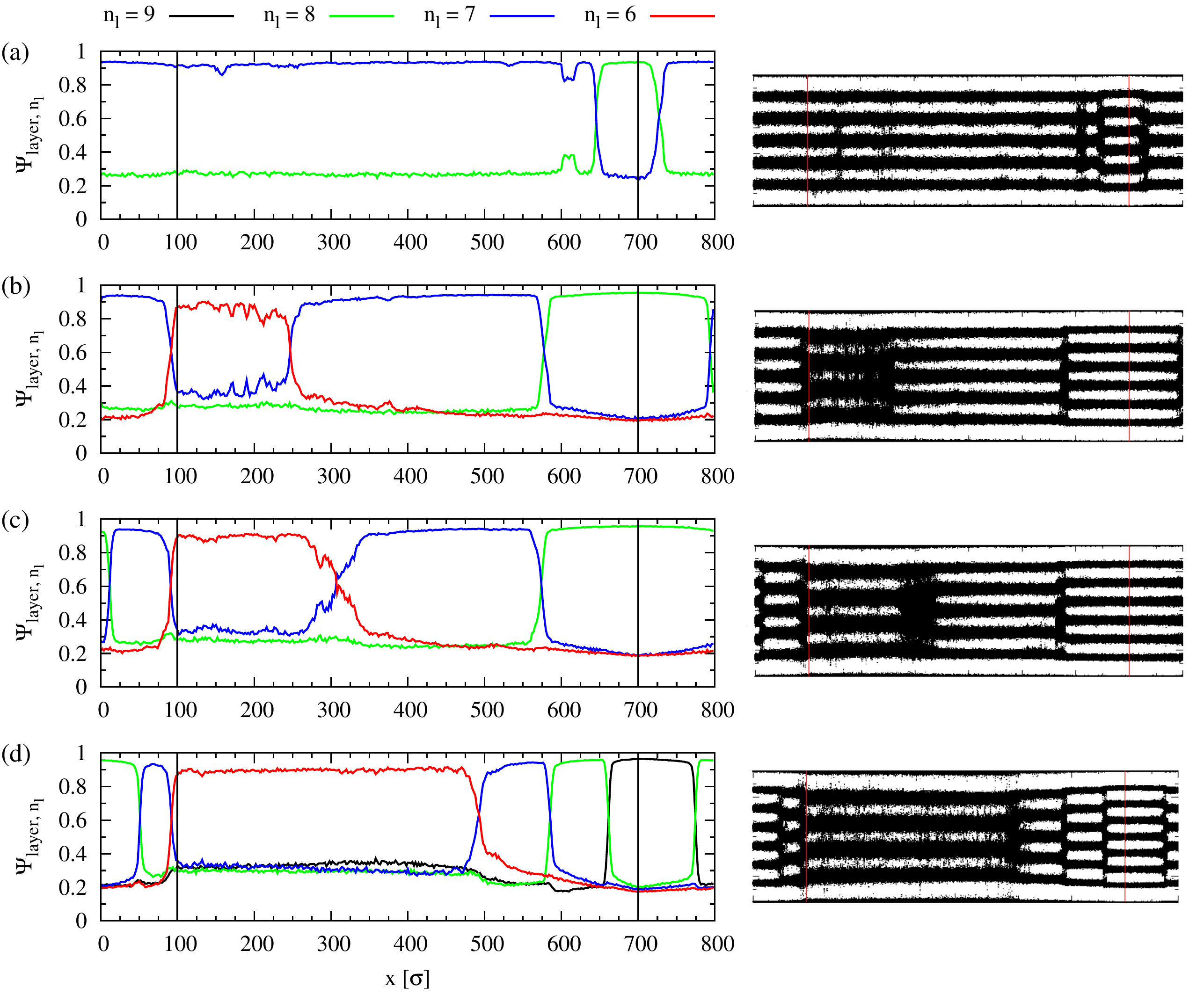}
  \end{center}
  \caption[Stationary non-equilibrium situations of systems where the driving
  force is applied only within part of the channel region for a selection of
  inclinations.]{Simulation: Stationary non-equilibrium situations of systems where the
  driving force is applied only in $x\in[0,\,700]\sigma$ for a selection of
  inclinations: {\bf (a)} $\alpha=0.04\degree$, {\bf (b)}~$\alpha=0.08\degree$,
  {\bf (c)}~\mbox{$\alpha=0.1\degree$}, {\bf (d)} $\alpha=0.2\degree$. For every
  inclination we show the average local layer order parameters
  $\Psi_{\mathrm{layer},\,n_l}$ and the corresponding superposition of 1000
  snapshots. The other simulation parameters are: $L_x=800\sigma$,
  $L_y=10\sigma$, $n=0.4\sigma^{-2}$, $B=\unit{0.25}{\milli\tesla}$, and
  $\Gamma=133.4$.}
  \label{fig:SnapshotsLOPwBC2}
\end{figure}

Figure~\ref{fig:SnapshotsLOPwBC2} shows the layer order parameters
$\Psi_{\mathrm{layer},\,n_l}$ for a selection of inclinations $\alpha$ in
combination with the corresponding superimposed configurations. Clearly, the
layer configuration and the number of layer transitions can be tuned by the
strength of the driving force for the realization of the boundary condition 2
of the flow. Increasing $\alpha$ leads to multiple transitions. Interestingly,
the layer transitions from 7 to 8 layers occur at identical $x$-positions in
the
figures~\hyperref[fig:SnapshotsLOPwBC2]{\ref{fig:SnapshotsLOPwBC2}(b)--(d)}. As
before, the particle flow across the position of the layer transition, which
remains fixed in position.

\subsection*{Systems with Screened Coulomb Interaction}
\begin{figure}[h!bt]
  \begin{center}
    \includegraphics[width=\columnwidth,clip]{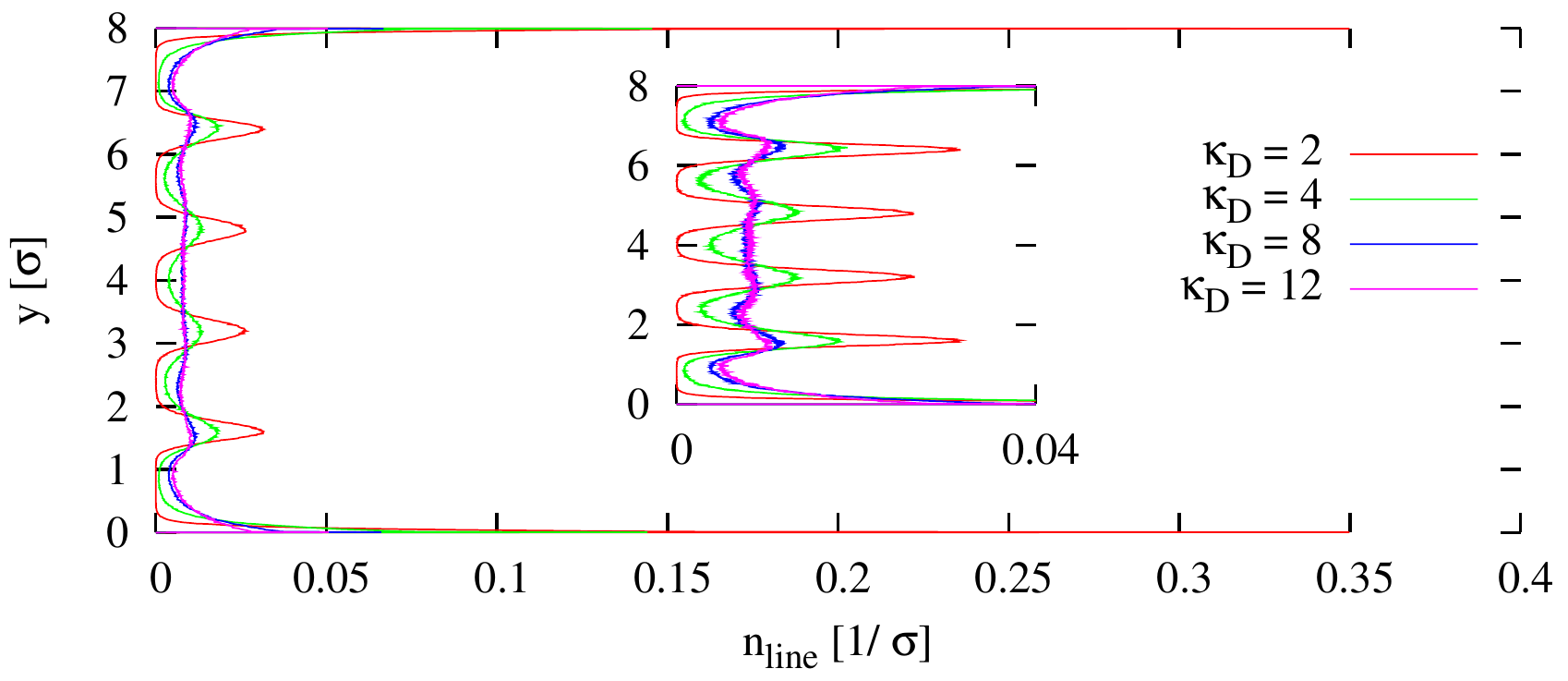}
  \end{center}
  \caption[Full density profiles transverse to the walls for $L_y=8\sigma$ of
  systems with screened Coulomb interaction.]{Simulation: Full density profiles
  transverse to the confining walls for $L_y=8\sigma$ of systems with screened
  Coulomb interaction for a selection of interaction ranges $\kappa_D$. The
  contact value of the potential is $\beta V_0=50$.}
  \label{fig:YHistYHC}
\end{figure}

In Fig.~\ref{fig:YHistYHC} we show the equilibrium density profiles transverse
to the confining walls of a YHC system for a selection of $\kappa_D$ values and
the value at particle contact $\beta V_0=50$. The total particle density is
\mbox{$n=0.45\sigma^{-2}$} which corresponds to a packing fraction
$\eta\approx0.79$. For $\kappa_D=2\sigma^{-1}$ boundary induced layering is
found which becomes less pronounced for increasing $\kappa$, {\it i.e.}
decreasing interaction range. For $\kappa_D>4\sigma^{-1}$ the systems are fluid
in the equilibrium state at this packing fraction, and only a depletion layer
between the edge and the bulk particles can be seen.

The average particle separation of the unbounded system has the value
$R\approx1.38\sigma$.  For $\kappa_D>4\sigma^{-1}$ the characteristic
interaction range is $\sigma+\kappa_D^{-1} < 1.25\sigma$ which is smaller than
$R$.

\begin{figure}[htb]
  \begin{center}
    \includegraphics[width=\columnwidth,clip]{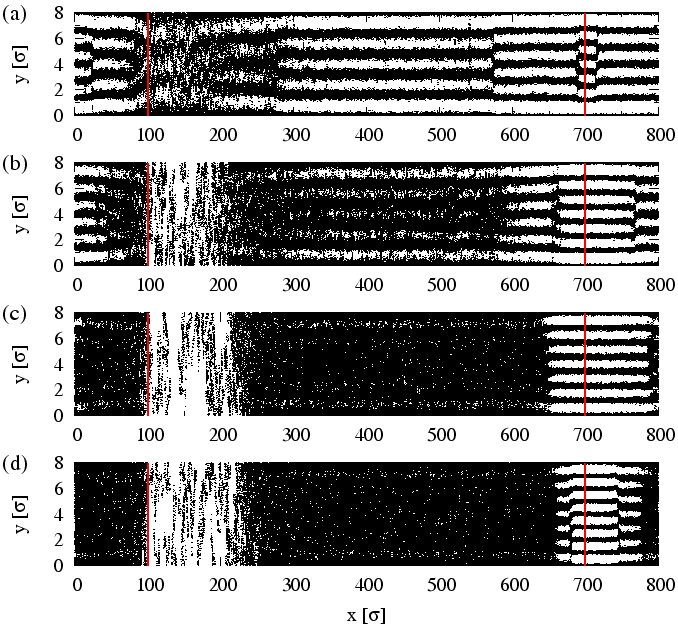}
  \end{center}
  \caption[Superimposed configurations of systems with screened Coulomb pair
  interaction for a selection of inverse screening
  lengths]{Simulation: Superimposed
  configurations of systems with screened Coulomb pair interaction for a
  selection of inverse screening lengths: {\bf (a)}~$\kappa_D=2\sigma^{-1}$,
  {\bf (b)}~$\kappa_D=4\sigma^{-1}$, {\bf (c)}~\mbox{$\kappa_D=8\sigma^{-1}$},
  and {\bf (d)}~$\kappa_D=12\sigma^{-1}$. The particle transport is induced for
  $x\in [0,\,700]\sigma$ by the inclination $\alpha=0.1\degree$. 
  }
  \label{fig:TrajectoryYHCBC2}
\end{figure}
Now, we plot in Fig.~\ref{fig:TrajectoryYHCBC2} the superposition of $100$
configurations with a time separation of $\Delta t=500$ BD steps after
$1.4\cdot10^6$ BD steps for the case of the alternative boundary condition in
flow direction.  The driving force corresponding to an inclination of
$\alpha=0.1\degree$ acts within $x\in[100,\,700]\sigma$. All four superimposed
configurations show the formation of layers near the channel end at
$x=700\sigma$, where the particles enter the reservoir. In
Fig.~\hyperref[fig:TrajectoryYHCBC2]{\ref{fig:TrajectoryYHCBC2}(a)} the
characteristic interaction range of the YHC pair-potential is greater than the
average particle spacing $R$. For this case we find multiple layer transitions
from 5 layers up to 8 layers along the channel. The system behavior is similar
to the situation of the dipolar systems. With increasing values of $\kappa_D$
less layer transitions are observed.
Figures~\hyperref[fig:TrajectoryYHCBC2]{\ref{fig:TrajectoryYHCBC2}(b)--(d)}
show increasing depletion zones at the channel start at $x=100\sigma$. These
depletion zones are followed by regions where the particles are in the liquid
state. Notice, that for $\kappa_D= 8\sigma^{-1}$ and $\kappa_D= 12\sigma^{-1}$
the systems are in the liquid state in equilibrium, too (cf.
Fig.~\ref{fig:TrajectoryYHCBC2}). The corresponding density profiles in
$x$-direction are given in Fig.~\ref{fig:TrajectoryYHCBC2}.

\begin{figure}[hbt]
  \begin{center}
    \includegraphics[width=\columnwidth,clip]{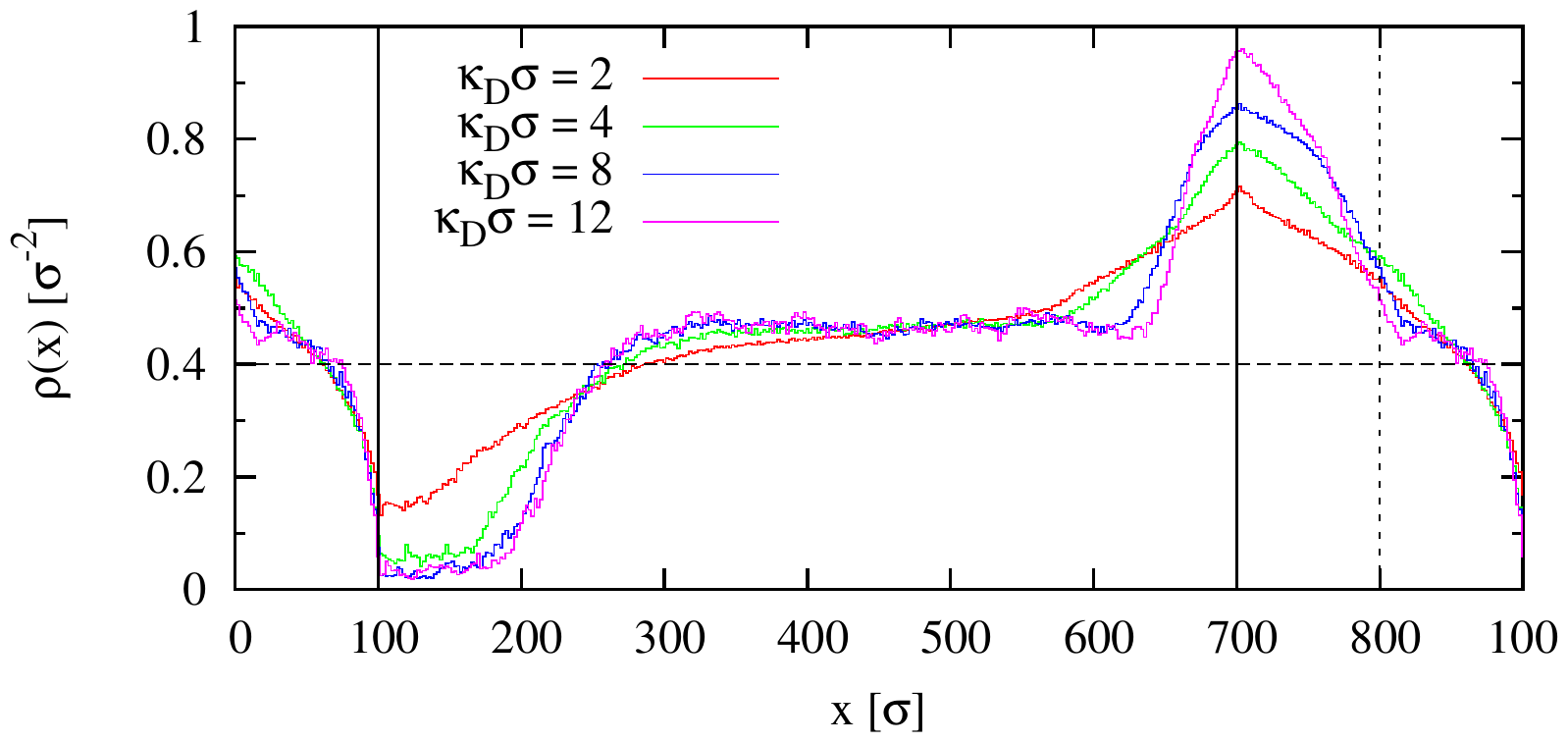}
  \end{center}
  \caption[Density profiles for a selection of Debye screening lengths of a YHC
  system with the driving force applied only within a partition of the
  channel.]{Simulation: Density profiles along the channel for a selection of Debye
  screening lengths $\kappa_D$ of a YHC system ($\beta V_0 = 50$). The driving
  force is applied only within the channel region $x\in[100,\,700]\sigma$ and
  the system is periodic in $x$-direction. These profiles correspond to
  the superimposed configurations of Fig.~\ref{fig:TrajectoryYHCBC2}.}
  \label{fig:DensityVglFlowRegionYHC}
\end{figure}
The systems with $\kappa_D= 8\sigma^{-1}$ and $\kappa_D= 12\sigma^{-1}$ show a
rapid increase of the local density from about $0.5\sigma^{-2}$ up to values
greater than $0.8\sigma^{-2}$ in the interval $x\in[600,\,700]\sigma$. The
particles under the influence of the constant driving force are blocked due to
filling of the reservoir at the channel end. During the simulation run the
particles pile up at the interface between the channel and the reservoir,
because the particles of the channel are pushed into the reservoir but within
the reservoir the particles diffuse almost freely due to the short range of the
YHC interaction (high values of~$\kappa_D$). This leads to a situation where
the influx into the reservoir is greater than the particle drift within the
reservoir being the reason for the sharp density gradients, which lead to the
sudden onset of a layered structure with 8 layers in the
figures~\hyperref[fig:TrajectoryYHCBC2]{\ref{fig:TrajectoryYHCBC2}(c)--(d)}.
For $\kappa_D= 12\sigma^{-1}$ even a layer transition to 9 layers takes place
due to local density values greater than $0.9\sigma^{-2}$ which is not observed
for the other three cases. Alternatively, the particle flux can be blocked in a
controlled fashion by creating so-called laser barriers perpendicular to the
driving field, as we will show in the following section.

%%% Conclusion and Outlook
%\input{LTpaper-Conclusion-v9}
\section{Conclusion}

We have reported on a variety of ordering and transport
phenomena which are induced by the confinement of colloidal particles to
microchannels and by the application of a constant
driving force along the channel. We have analyzed the particle behavior both
under equilibrium and under (stationary) non-equilibrium conditions both in
experiment and by Brownian Dynamics simulations.

First, we have studied the self-assembly of repulsive particles under
equilibrium conditions, {\it i.e.} without a driving force applied.We have
observed a boundary induced formation of a global {\it layered structure} in
the channels. Such a behavior is known for a variety of related
systems~\cite{Glasson01, Teng03, Haghgooie04, Piacente04, Ricci06}.
Systematically, we have analyzed the influence of the channel width $L_y$ and
the influence of the strength of the dipolar particle repulsion. Based
on the order parameter we have calculated the phase diagram of laterally
confined superparamagnetic particles as a function of the channel width $L_y$
within the solid state. We have observed a re-entrant
behavior as a function of $L_y$, where the system behavior oscillates from
solid-like to liquid-like. When the channel width is increased, a periodic
destabilization of the layered structure with $n_l$ layers takes place, and the
system switches to a structure with $n_l+1$ layers. 
The bulk defect concentration $C^b_{\mathrm{defect}}$
shows periodic oscillations as a function of the channel width $L_y$, but not
as a function of the dimensionless interaction parameter~$\Gamma$. The period
of the oscillations is $\sim R$, where $R$ denotes the average distance of two
neighboring particle layers in an unbounded hexagonal system. Such a behavior
previously was reported as a result of both experiments and simulations of a
similar system by Haghgooie~{\it et al.}~\cite{Haghgooie04, Haghgooie05,
Haghgooie06}.  Our data show excellent qualitative and quantitative
agreement with their results.

For very small channel widths $L_y<1\sigma$, where the particles cannot pass
each other, we have observed the well known {\it single file diffusion}
behavior~\cite{Richards77, Fedders77, Wei00}. Our
model system allows for a systematic analysis of the diffusion behavior of
layered structures. In first studies, we have compared the longitudinal and
transversal particle diffusion behavior for the channel width $L_y=10\sigma$,
where the system globally forms 7 layers, to a channel with $L_y=9\sigma$,
where no globally layered structure exists in the solid state. The diffusion
behavior at intermediate time scales is very different for both cases. 
In the presence of global layers, the transversal mean square displacement
$\langle \Delta y^2 \rangle$ has a constant plateau. Additionally, at longer
times a deviation from the Fick diffusion behavior ({\it anomalous diffusion})
is expected by our first simulation results. Evaluation of
the experimental data confirm this deviation from the Fick diffusion
behavior. These effects are absent for the system with
$L_y=9\sigma$. Therefore, it will be interesting, to do a systematic analysis
of this diffusion behavior in the future.

We have predicted and systematically analyzed the phenomenon of particle {\it
layer reduction} under the influence of a constant driving force acting along
the channel.
For small driving forces $F^{\mathrm{ext}}$, where the particles are not yet in
the regime of plug flow the superparamagnetic
particles dynamically re-arrange into different numbers of layers during
transport through the channels.  We have found, that along the channel the
number of layers decreases gradually by steps of one.  The occurrence of the
layer reduction has been confirmed by the experiments. In the experiments, the
massive particles sediment to the bottom of the channel due to gravity, and
there they form a quasi-2D system. After having equilibrated the system, the
whole setup is tilted, so that the colloidal particles are driven through a
lithographically fabricated microchannel under the influence of gravity.

In very good qualitative agreement with the experiments we have shown that the
reduction of layers originates from a density gradient along the channel.
Quantitative differences are expected, because the Stokes diffusion coefficient
$D_0$, which is valid for unbounded systems and is used in the simulations,
differs from the real diffusion coefficient in presence of the confinement
of the experimental setup~\footnote{Haghgooie {\it et al.}~\cite{Haghgooie06}
measured the surface diffusion coefficient of the colloids to be $\sim52\%$ of
the calculated Stokes diffusion coefficient $D_0$ for their system, which is
similar to our system.}.

The reduction of layers takes place for specific values of the local density
$\rho(x)$ and within a distance of only a few particle diameters. We have
explicitly shown that the particles flow across the regions of layer reduction
and thereby dynamically adjust to the local density $\rho(x)$. The origin of
the local density gradient is not fully understood yet. But additional
simulation studies of systems with screened Coulomb particle interaction, where
the interaction range has been varied, have shown that a longitudinal density
gradient and consequently layer transitions occur for particle interaction
ranges which are greater than the average distance of the particles from their
neighbors. For particle pair potentials with smaller interaction ranges than
the average nearest neighbor separation we observe that the layer transition
region smears out, because more particle defects occur due to a smaller density
gradient.  No layer transitions will be observed for the model-case of
hard-core particles. For our choice of boundary conditions we have found, that
the density gradient becomes more pronounced with decreasing inclination
$\alpha$, {\it i.e.} with decreasing driving force.  The density decrease is
maximum at $\alpha=0.0\degree$, because the particle re-insertion scheme, which
we used, induces a pressure difference between both channel ends, even in the
case when no external driving force has been applied.

Generally, we have seen both in simulations and in experiments that the local
density decreases monotonically and continuously along the channel. 
In front of a layer transition the local structure is stretched in longitudinal
direction, whereas after the layer transition the structure is longitudinally
compressed and one layer has disappeared. Therefore, the local lattice constant
$d_x(x)$ in longitudinal direction increases up to the position of the layer
transition, at which it shows a non-continuous decrease. Simultaneously, the
local lattice constant $d_y(x)$ in transversal direction is constant in front
of the layer transition, at which it jumps to the next level according to the
number of layers and remains constant again.  Both effects compensate each
other and thus explain the continuous behavior of the local density along the
channel. 

By a static stretching analysis we have confirmed that a certain layered
structure becomes energetically unstable and thus changes to a structure, where
it has one layer less. The estimated values of the local density, where the
transition takes place, are in quite good agreement with the observation. 
In stationary non-equilibrium the position of the layer transition oscillates
about a fixed position. The amplitude of the oscillations depends on the
strength of the particle interaction. We have shown, that the oscillations of
the layer transition can either be analyzed by the appropriate local layer
order parameters $\Psi_{\mathrm{layer}, n_l}(x)$ or by the local lattice
constant $d_y(x)$. 

Each layer transition is connected to a defect, which is defined by a pair of
particles with five and seven nearest neighbors respectively. Additional
periodic defects have been observed along the channel walls. Due to the purely
repulsive particle interaction the edge particles are pushed against the flat
walls. This leads to very small transverse fluctuations of the edge particles
and a slightly higher line density of the edge particles than of the particles
belonging to the layers in the central region of the channel.

It has been shown, that channel walls made of periodically fixed particles give
rise to shear effects between the particles of the central layers, which move
faster, than the particles, which are in the layer next to the edge particles.
The latter particles show small oscillations about the average drift velocity.

The results shown concern a rather simple classical model system. The
observed phenomena, however, will take place in any systems of long range
interacting particles which are driven through a constriction. Therefore
the results which have been gained from the studies of this system can
be seen as a first step in the understanding of transport processes in
many biological and quantum systems.

%% Thanks for support
We gratefully acknowledge the support of the SFB 513, the SFB TR6 and the NIC,
HLRS, and SSC.

%%% Bibliography
\bibliography{Literature}

\end{document}